\documentclass[]{JHEP3}

\def\cU{{\cal U}}

\def\bR{{\mathbb{R}}}

\def\bZ{{\mathbb{Z}}}

\def\dim{{\rm dim}}

\def\im{{\rm Im}\, }
\def\tr{{\rm tr}}

\def\exp{{\rm exp}}

\def\a{\alpha}
\def\b{\beta}

\def\G{\Gamma}
\def\d{\delta}

\def\k{\kappa}
\def\la{\lambda}

\def\x{\times}

\def\lx{{\hspace{-0.04cm}\ltimes\hspace{-0.05cm}}}

\def\ker{\textrm{ker}}
\def\im{\textrm{im}}

\newcommand{\lb}{\left(}
\newcommand{\rb}{\right)}

\newcommand{\sfi}{{\mathsf i}}
\newcommand{\sfk}{{\mathsf k}}

\newcommand{\beq}{\begin{equation}}
\newcommand{\eeq}{\end{equation}}
\newcommand{\beqa}{\begin{eqnarray}}
\newcommand{\eeqa}{\end{eqnarray}}
\newcommand{\bgt}{\begin{gather}}
\newcommand{\bal}{\begin{align}}
\newcommand{\eal}{\end{align}}

\newcommand{\barr}{\begin{array}}
\newcommand{\earr}{\end{array}}
\newcommand{\ben}{\begin{enumerate}}
\newcommand{\een}{\end{enumerate}}
\newcommand{\bit}{\begin{itemize}}
\newcommand{\eit}{\end{itemize}}
\newcommand{\tx}[1]{\textrm{#1}} 
 
\newcommand{\gt}[1]{\mathfrak{#1}}

\def\1{\mathbb{I}}
\def\2{[2]_q}

\newcommand{\non}{\nonumber\\}

\newcommand{\nl}{\newline}

\newcommand{\un}{\underline}

\newcommand{\Ng}{{\bf g}}
\newcommand{\Nt}{{\bf t}}

\newcommand{\qq}{\begin{eqnarray}}

\newcommand{\ee}{{\rm e}}

\newcommand{\qqq}{\end{eqnarray}}

\newcommand{\CA}{{\cal A}}

\newcommand{\CD}{{\cal D}}

\newcommand{\CS}{{\cal S}}

\newcommand{\CU}{{\cal U}}

\newcommand{\wrt}{with respect to }

\def\qU2{\cU_q (\mathfrak{su}(2))}

\def\too{\longrightarrow}

\def\resu2{\textrm{REA}_q(\mathfrak{su}(2))}

\usepackage{latexsym,amsfonts,amsmath,amssymb,mathrsfs,xypic}
\usepackage{bm,epsfig}
\usepackage{indentfirst}

\input xy

\xyoption{all}

\title{WZW orientifolds and finite group cohomology}

\author{Krzysztof Gaw\c{e}dzki, {\it\,CNRS, Laboratoire de Physique,
ENS-Lyon, 46 All\'ee d'Italie, F-69364 Lyon, France}}
\author{Rafa\l ~R.~Suszek, {\it\,Laboratoire de Physique,
ENS-Lyon, 46 All\'ee d'Italie, F-69364 Lyon, France}} 
\author{Konrad Waldorf, {\it\,Fachbereich Mathematik, Universit\"at Hamburg,
Bundesstrasse 55, D-20146 Hamburg, Germany}}

\abstract{\,The simplest orientifolds of the WZW models
are obtained by gauging a $\,\bZ_2\,$ symmetry group generated
by a combined involution of the target Lie group $\,G\,$ and of
the worldsheet. The action of the involution on the target is 
by a twisted inversion $\,g\mapsto(\zeta g)^{-1}$, where $\,\zeta\,$ 
is an element of the center of $\,G$. \,It reverses the sign of 
the Kalb-Ramond torsion field $\,H\,$ given by a bi-invariant closed 
3-form on $\,G$. \,The action on the worldsheet reverses its orientation.
An unambiguous definition of Feynman amplitudes of the orientifold 
theory requires a choice of a gerbe with curvature $\,H\,$ on the 
target group $\,G$, \,together with a so-called Jandl structure 
introduced in \cite{SSW}. More generally, one may gauge orientifold 
symmetry groups $\,\Gamma=\bZ_2\lx Z\,$ that combine the $\,\bZ_2$-action
described above with the target symmetry induced by a subgroup $\,Z\,$ 
of the center of $\,G$. \,To define the orientifold theory in such
a situation, one needs a gerbe on $\,G\,$ with a $\,Z$-equivariant Jandl
structure. \,We reduce the study of the existence of such structures 
and of their inequivalent choices to a problem in group-$\Gamma\,$ 
cohomology that we solve for all simple simply-connected compact Lie 
groups $\,G\,$ and all orientifold groups $\,\Gamma=\bZ_2\lx Z$.
\vskip 0.3cm}


\begin{document}

\section{Introduction}\label{sec:intro}
\noindent Unoriented string theory, both in the closed and in 
the open sector, has a long history \cite{Schwz,PS}. From the 
two-dimensional point of view, it involves conformal field theory 
defined on unoriented worldsheets. Such a theory may be viewed as 
an ``orientifold'' obtained from a conformal field model defined 
on oriented surfaces by gauging a discrete symmetry containing 
transformations reversing the worldsheet orientation. If the 
conformal theory is a sigma model whose target space carries a 
background Kalb-Ramond 2-form field $\,B$, \,the worldsheet 
orientation-changing transformations have to be combined with 
target-space transformations that change the sign of $\,B\,$ so
that the $\,B$-field contribution to the sigma model action 
functional stays invariant. This leads to subtle issues if the 
$\,B$-field is topologically non-trivial, like the one present in the 
Wess-Zumino-Witten (WZW) sigma models with Lie group targets 
\cite{Witt}. In those models, only the closed torsion 3-form 
$\,H=dB$, \,a right-left invariant 3-form 
on the group manifold, is globally defined. Orientifolds
of the WZW models have been studied intensively within 
the algebraic approach, following the pioneering
work of the Rome group \cite{PSS95a,PSS95b}. The main tool
in this approach was the use of sewing and modular 
duality constraints in order to find consistent expressions 
for the crosscap states encoding the action of the orientation 
inversion in the closed string sector. The algebraic approach
was further developed in the context of more general 
orientifolds combining simple-current orbifolds and orientation
reversal in \cite{HSS1,HSS2,HS,FHSSW,BH}. It gave rise to an
abstract formulation of the relevant topological structures
in the language of tensor categories \cite{FFRS}. The interpretation 
of the results of the algebraic approach in terms of the target 
geometry was the subject of papers \cite{BCW} 
and \cite{B} that studied orientifolds of the $\,SU(2)\,$ 
and $\,SO(3)\,$ WZW theories. 
\vskip 0.05cm

In general, one may expect that the intricacies appearing in 
the algebraic studies of WZW orientifolds have their source 
in the classical target geometry, more precisely in non-triviality 
of the $\,B$-field background, similarly to the ones involved in the 
simple-current orbifolds of the WZW models. In the latter case, 
it was argued in \cite{top} that the proper treatment of the 
non-trivial $\,B$-field background in the closed string sector may 
be achieved by employing the third (real) Deligne cohomology. This approach 
lay behind the classification of the WZW models on non-simply 
connected simple compact groups obtained in \cite{FGK}. The third 
Deligne cohomology classifies geometric structures called bundle
gerbes with connections introduced in \cite{Murr,MurrS}.
The latter are in a similar relation to the closed 3-forms $\,H\,$
as line bundles with connection are to their curvature 2-forms
$\,F$. \,Consequently, the 3-form $\,H\,$ corresponding to a gerbe
is called its curvature. The geometric language of bundle gerbes 
is sometimes more convenient than the cohomological one of Deligne 
cohomology. 
For general simple groups, in particular, it appeared to be easier 
to construct the bundle gerbes with the curvature given by a bi-invariant 
3-form $\,H\,$ than the corresponding Deligne cohomology 
classes. Such a construction was accomplished for the simply connected 
groups in \cite{Meinr} and was generalized to the non-simply connected 
ones in \cite{GR2}. An extension of the geometric analysis including 
open strings and $\,D$-branes required studying gerbe-modules 
carrying Chan-Paton gauge fields twisted by the gerbe \cite{Kapust}. 
In the algebraic language, the WZW models with non-simply connected 
target groups are simple-current orbifolds of the models with simply 
connected targets. The geometric analysis of \cite{GR1,G3}, employing 
(bundle) gerbes and gerbe modules, permitted a systematic classification 
of symmetric $D$-branes in the WZW models and exposed the classical origin 
of the finite group cohomology that appeared in the algebraic analysis 
of the simple-current orbifolds. Indeed, the relevant cohomological 
aspects pass undeformed to the quantum theory that is obtained 
by geometric quantization of the classical one \cite{G3}. 
\vskip 0.05cm

The recent paper \cite{SSW} introduced additional data, called a Jandl 
structure on a gerbe, that are required to define Feynman amplitudes 
for closed unoriented worldsheets in the presence of a topologically 
non-trivial $\,B$-field. A Jandl structure may be viewed as a symmetry
of the gerbe under a transformation of the underlying space that changes 
the sign of the curvature 3-form. In this paper, we classify such structures
on all gerbes on simple compact groups with the gerbe curvature equal to 
a bi-invariant torsion 3-form $\,H$. \,More precisely, 
on the simply connected group targets $\,G$, \,we consider the action 
of orientifold groups $\,\G=\bZ_2\lx Z$. \,This action combines
the involutive twisted inversion $\,g\mapsto(\zeta g)^{-1}$, \,where 
the twist element $\,\zeta\,$ belongs to center $\,Z(G)\,$ of $\,G$, 
\,with the multiplication by elements of the ``orbifold'' subgroup 
$\,Z\subset Z(G)$. \,The action of $\,Z\,$ preserves the 
bi-invariant 3-form $\,H$, \,whereas the action of the twisted inversion 
changes its sign. We introduce the notion of a $\,\G$-equivariant 
structure on the gerbe with curvature $\,H\,$ on group $\,G$. \,Such 
a structure may be regarded as a $\,Z$-equivariant Jandl structure on 
that gerbe. It determines a genuine Jandl structure 
on the quotient gerbe on the non-simply connected group $\,G/Z\,$
and enables to define unambiguously the contribution of the $\,B$-field 
to Feynman amplitudes of unoriented string world histories 
represented by maps from unoriented closed surfaces to the target 
$\,G/\G$.  
\vskip 0.05cm

We show that obstructions to existence of $\,\G$-equivariant
structures are contained in the cohomology group $\,H^3(\G,U(1)_\epsilon)$, 
\,where the subscript $\,\epsilon\,$ indicates that $\,U(1)\,$ is considered
with the action $\,\lambda\mapsto\lambda^{-1}\,$ of the elements of 
$\,\Gamma\,$ that reverse the sign of $\,H$. \,If the obstruction class 
vanishes, non-equivalent 
$\,\Gamma$-equivariant structures may be labeled by elements of the 
cohomology group $\,H^2(\G,U(1)_\epsilon)$. \,Each choice gives a 
different (closed-string) orientifold theory. Let us recall
that obstructions to existence of the quotient gerbe on 
$\,G/Z\,$ (and of the $\,Z$-orbifold theory) lie in $\,H^3(Z,U(1))\,$
and that ambiguities in its construction (the ``discrete torsion'' of
\cite{Vafa}) take values in $\,H^2(Z,U(1))$, \,see \cite{G3}.  
The present paper is devoted to the study of obstruction
3-cocycles for all simple simply connected groups $\,G\,$ of the Cartan
series and all choices of the orientifold groups $\,\G=\bZ_2\lx Z$. 
\,We find the conditions under which the obstruction
cocycles are coboundaries, i.e. the obstruction cohomology class
is trivial. This provides an extension of the work of \cite{GR2}
from the orbifold to the orientifold case. Similarly as in the orbifold 
case analyzed in \cite{GR2,G3}, the cochains trivializing the 
obstruction cocycles enter directly the construction of  
$\,\G$-equivarient structures on the gerbes on groups $\,G\,$ and 
the analysis of the symmetric $\,D$-branes in the WZW orientifolds. 
These topics, involving more geometric considerations as well as 
a discussion of the relation between our approach and the algebraic 
one of \cite{BH}, are postponed to a later publication \cite{GSW2}. 
In the present paper, we shall avoid geometry by sticking to a 
local description of gerbes, staying close to the Deligne cohomology 
approach of \cite{top}. 
\vskip 0.05cm

The paper is organized as follows.     
In Sect.\,\ref{sec:bgo}, we summarize the description of
gerbes by local data and the relation of gerbes on discrete
quotients to finite group cohomology. The
application to gerbes on simple simply connected 
compact groups $\,G\,$ and their non-simply connected
quotients $\,G/Z\,$ is recalled from \cite{GR2}. Finally, we
extend the construction to the case of quotients
by orientifold groups $\,\G\,$ and describe a 3-cocycle whose 
cohomology class obstructs existence of $\,\G$-equivariant 
structures on gerbes on the simply connected groups $\,G\,$
for $\,\G=\bZ_2\lx Z$. 
\,In Sect.\,\ref{sec:cohocon}, we study the relevant cohomology groups: 
the one containing the obstruction classes: $\,H^3(\G,U(1)_\epsilon)\,$ 
and the one classifying non-equivalent $\,\G$-equivariant structures: 
$\,H^2(\G,U(1)_\epsilon)$. They are more difficult to calculate 
than the corresponding orbifold cohomologies but information 
about those groups may be obtained from the Lyndon-Hochschild-Serre 
spectral sequence that we discuss in some detail. In particular, we 
are able to calculate the classifying group $\,H^2(\G,U(1)_\epsilon)\,$ 
in all relevant cases. Sect.\,\ref{sec:case} is the most technical part 
of the paper. It analyzes the obstruction 3-cocycles for all simple
groups $\,G\,$ of the Cartan series and all choices of the
twisted orientifold group actions and finds cohomologically 
inequivalent trivializing cochains whenever 
the obstruction cohomology class is trivial. The results are
tabulated in Appendix. In Sect.\,\ref{sec:concl}, we collect 
our conclusions.
\vskip 0.2cm

\noindent{\bf Acknowledgements}. \ K.G and R.R.S. acknowledge the 
support of the European Commission under the contract 
EUCLID/HPRN-CT-2002-00325 and the funding by the Agence National
de Recherche grant ANR-05-BLAN-0029-03. K.W. was partly supported 
by Rudolf und Erika Koch-Stiftung.

\section{Bundle gerbes and orientifolds}\label{sec:bgo}
\subsection{Local description of bundle gerbes}\label{sec:local}
\noindent (Bundle) {\bf gerbes} (with hermitian structure 
and unitary connection) are geometric structures that allow to define 
the contribution of the Kalb-Ramond torsion 3-form $\,H\,$ to closed-string 
Feynman amplitudes. A simple, although not always convenient, way to 
present a gerbe on a manifold $\,M\,$ is via its local data.  
In this paper, we shall stick to such a local description of
bundle gerbes that reduces the geometric structures to the cohomological 
ones described already in \cite{top}. A discussion, in relation 
to orientifolds, of the geometric structures underlying the notion 
of bundle gerbes \cite{Murr,MurrS} is postponed to \cite{GSW2}. 
\vskip 0.2cm

Gerbe local data subordinate to a good open covering\footnote{In a good 
open covering, the sets $\,O_i\,$ and all their (non-empty) intersections
$\,O_{i_1i_2\dots i_k}=O_{i_1}\cap O_{i_2}\cap\cdots\cap O_{i_k}\,$ are 
contractible.} $\,(O_i)\,$ of $\,M\,$ are 
a collection $\,(B_i,A_{ij},g_{ijk})\,$ where $\,B_i\,$ are 2-forms 
on the sets $\,O_i$, $\,A_{ij}=-A_{ji}\,$ are 1-forms on $\,O_{ij}\,$ and 
$\,g_{ijk}=g_{jik}^{-1}=g_{jki}=g_{ikj}^{-1}\,$ are $\,U(1)$-valued 
functions on $\,O_{ijk}$ such that the following descent equations 
hold: 
\qq
&&\hbox to 6cm{$B_j-B_i\,=\,dA_{ij}$\hfill} {\rm on}
\ \ \,O_{ij},\label{dceq2}\cr
&&\hbox to 6cm{$A_{ij}-A_{ik}+A_{jk}\,=\,\sfi\,g_{ijk}^{-1}dg_{ijk}$\hfill}
{\rm on}\ \ \,O_{ijk},\label{dceq3}\cr
&&\hbox to 6cm{$g_{ijk}\,g_{ijl}^{-1}g_{ikl}\,g_{jkl}^{-1}\,=\,1$\hfill} 
{\rm on}\ \ \,O_{ijkl}.
\nonumber
\qqq
The global closed 3-form $\,H\,$ equal to $\,dB_i\,$ on the sets $\,O_i\,$ 
is called the curvature of the gerbe. 
The necessary and sufficient condition for existence of a gerbe
with curvature $\,H\,$ (and of the corresponding local data) is that 
the periods of the 3-form $\,{1\over2\pi}H\,$ be integers. The local 
data $\,(B'_i,A'_{ij},g'_{ijk})\,$ and $\,(B_i,A_{ij},g_{ijk})\,$ are 
considered equivalent if there exist 1-forms $\,\Pi_i\,$ on $\,O_i\,$ 
and $\,U(1)$-valued functions $\,\chi_{ij}=\chi_{ji}^{-1}\,$ on $\,O_{ij}\,$
such that
\qq
&&B'_i\,=\,B_i+d\Pi_i,
\cr
&&A'_{ij}\,=\,A_{ij}+\Pi_j-\Pi_i-\sfi\,
\chi_{ij}^{-1}d\chi_{ij},\label{equiv2}\cr
&&g'_{ijk}\,=\,g_{ijk}\,\chi_{ij}^{-1}\chi_{ik}\,\chi_{jk}^{-1}.
\nonumber
\qqq
Equivalent local data correspond to gerbes that are called stably 
isomorphic \cite{MurrS}. Clearly, such gerbes have the same curvature 
3-form $\,H$. \,In general, two gerbes with the same curvature differ by 
a flat gerbe with vanishing curvature. Up to equivalence, the local data of 
a flat gerbe are of the form $\,(0,0,u_{ijk})\,$ with 
$\,u_{ijk}\in U(1)\,$ \cite{top}. \,Their equivalence classes 
are in a one-to-one correspondence with elements of the cohomology group 
$\,H^2(M,U(1))$. \,In particular, if $\,H^2(M,U(1))\,$ is trivial 
then all gerbes with the same curvature are stably isomorphic.
If there is no torsion in $\,H^3(M,\bZ)\,$ then one may 
also put the flat gerbe local data into an equivalent form $\,(B,0,1)\,$ 
where $\,B\,$ is a global closed 2-form.
\vskip 0.2cm 

If $\,\Sigma\,$ is an oriented closed connected surface and $\,X\,$ maps 
$\,\Sigma\,$ to $\,M\,$ then, pulling back the gerbe 
by $\,X\,$ to $\,\Sigma$, \,one obtains a 
flat gerbe on $\,\Sigma\,$ which, up to a stable isomorphism, is characterized 
by a cohomology class in $\,H^2(\Sigma,U(1))=U(1)$. \,The corresponding 
number in $\,U(1)\,$ is called the holonomy of the gerbe on $\,M\,$ along 
$\,X$. \,If the local data for the pulled-back gerbe are taken in the form
$\,(B,0,1)\,$ then the holonomy along $\,X\,$ is given by 
$\,\exp[\hspace{0.025cm}\sfi\int_\Sigma B]$. \,It enters as 
a factor in the Feynman amplitude of the closed-string world 
history $\,X$. 
\vskip 0.2cm

It is convenient to use the cohomological language to describe
gerbe local data and their equivalence classes \cite{top}. 
We shall denote by $\,\check{C}^p(\CS)\,$ the Abelian group of 
$\check{\rm C}$ech $\,p$-cochains with values in an (Abelian) sheaf $\,\CS$. 
\,An element $\,c\in\check{C}^p(\CS)\,$ is a collection of sections 
$\,c_{i_0\cdots i_p}\,$ of $\,\CS\,$ over the sets $\,O_{i_0\cdots i_p}\,$ 
that is antisymmetric in the indices $\,i_0,\dots,i_p$.
The groups $\,\check{C}^p(\CS)\,$ form the $\check{\rm C}$ech 
complex $\,\check{C}(S)$,
\qq
0\ \longrightarrow\ \check{C}^0(\CS)\ \mathop{\longrightarrow}
\limits^{\check{\delta}}\ \check{C}^1(\CS)\ 
\mathop{\longrightarrow}\limits^{\check{\delta}}\ \check{C}^2(\CS)\ 
\mathop{\longrightarrow}\limits^{\check{\delta}}\ \ \cdots\,,
\label{Cech}
\qqq
where the $\check{\rm C}$ech coboundary $\,\check{\delta}\,$
is defined by
\qq
(\check{\delta}c)_{i_0\cdots i_{p+1}}\,=\,\sum\limits_{j=0}^{p+1}(-1)^j
c_{i_0\cdots i_{j-1}i_{j+1}\cdots i_{p+1}}\,.
\nonumber
\qqq
The $\check{\rm C}$ech cohomology groups $\,H^p(M,\CS)\,$ are composed
of $\check{\rm C}$ech $\,p$-cocycles modulo $\,p$-coboundaries.
In particular,  taking the sheaf of locally constant $\,U(1)$-valued
functions, one obtains the cohomology groups $\,H^p(M,U(1))$.
\,Given a complex $\,\CD\,$ of sheaves
\qq
0\ \longrightarrow\ \CS^0 \mathop{\longrightarrow}
\limits^{d_0}\ \CS^1\ \mathop{\longrightarrow}\limits^{d_1}
\ \CS^2\ \mathop{\longrightarrow}\limits^{d_2}\ \ \cdots\,,
\label{comp00}
\qqq
one may build a double complex
\qq
&&\hspace{1.7cm}\downarrow\check{\delta}\hspace{2cm}\downarrow\check{\delta}
\hspace{2cm}\downarrow\check{\delta}\cr
&&0\ \longrightarrow\ \ \check{C}^p(\CS^0)\ \ \ \mathop{\longrightarrow}
\limits^{d_0}\ \ \check{C}^p(\CS^1)\ \ \ \mathop{\longrightarrow}\limits^{d_1}
\ \ \check{C}^p(\CS^2)\ \ \ \mathop{\longrightarrow}\limits^{d_2}\ \ \cdots\cr
&&\hspace{1.7cm}\downarrow\check{\delta}\hspace{2cm}\downarrow\check{\delta}
\hspace{2cm}\downarrow\check{\delta}\cr
&&0\ \longrightarrow\ \check{C}^{p+1}(\CS^0)\ \mathop{\longrightarrow}
\limits^{d_0}\ \check{C}^{p+1}(\CS^1)\ \mathop{\longrightarrow}\limits^{d_1}
\ \check{C}^{p+1}(\CS^2)\ \mathop{\longrightarrow}\limits^{d_2}\ \ \cdots\cr
&&\hspace{1.7cm}\downarrow\check{\delta}\hspace{2cm}\downarrow\check{\delta}
\hspace{2cm}\downarrow\check{\delta}
\nonumber
\qqq
The hypercohomology groups $\,\mathbb{H}^s(M,\CD)\,$ of the complex $\,\CD\,$  
are defined as the cohomology groups of the diagonal complex $\,K(\CD)\,$
\qq
0\ \longrightarrow\ A^0 \mathop{\longrightarrow}
\limits^{D_0}\ A^1 \mathop{\longrightarrow}
\limits^{D_1}\ A^2\ \mathop{\longrightarrow}
\limits^{D_2}\ A^3 \mathop{\longrightarrow}
\limits^{D_3}\ \ \cdots
\label{compCA}
\qqq
where
\qq
A^s\,=\,\mathop{\oplus}\limits_{p+q=s}\check{C}^p(\CS^q)
\label{CAr}
\qqq
and $\,\,D_s=(-1)^{q+1}\check{\delta}+d_q\,\,$ on $\,\check{C}^p(\CS^q)$.
\vskip 0.2cm

We shall denote by $\,\CU\,$  the sheaf of local (smooth) $\,U(1)$-valued 
functions on $\,M\,$ and by $\,\Lambda^q\,$ the sheaves of (smooth) $q$-forms 
on $\,M$. \,For the complex $\,\CD(2)$,
\qq
0\ \longrightarrow\ \CU \mathop{\longrightarrow}
\limits^{\frac{1}{\sfi}\,d\,\log}\ \Lambda^1\ 
\mathop{\longrightarrow}\limits^{d}\ \Lambda^2\,
\label{comp0}
\qqq
where $\,d\,$ is the exterior derivative, the groups $\,A^s\,$ 
of (\ref{CAr}) are: 
\qq
&&A^0\ =\ \check{C}^0(\CU)\ =\ \{\,(f_i)\,\}\,,\label{a01}\\ 
&&A^1\ =\ \check{C}^0(\Lambda^1)\oplus 
\check{C}^1(\CU)\ =\ \{\,(\Pi_i,\chi_{ij})\,\}\,,
\label{a11}\\
&&A^2\ =\ \check{C}^0(\Lambda^2)\oplus\check{C}^1(\Lambda^1)
\oplus\check{C}^2(\CU) 
\ =\ 
\{\,(B_i,A_{ij},g_{ijk})\,\}\,,\label{a21}\\
&&A^3\ =\ \check{C}^1(\Lambda^2)\oplus\check{C}^2(\Lambda^1)\oplus 
\check{C}^3(\CU)
\ =\ 
\{\,(F_{ij},D_{ijk},\sigma_{ijkl})\,\}\label{a31}
\nonumber
\qqq
where $\,f_i$, $\,\chi_{ij}$, $\,g_{ijk}$, $\,\sigma_{ijkl}\,$ 
are $\,U(1)$-valued functions on $\,O_i$, $\,O_{ij}$, $\,O_{ijk}\,$ 
and $\,O_{ijkl}$, \,respectively, $\,\Pi_i$, $\,A_{ij}\,$ and $\,D_{ijk}\,$ 
are 1-forms on $\,O_i$, $\,O_{ij}\,$ and $\,O_{ijk}\,$ and $\,B_i$, $\,F_{ij}\,$
are 2-forms on $\,O_i\,$ and $\,O_{ij}$. The differentials $\,D_i\,$ 
combine the exterior derivative with the $\check{\rm C}$ech coboundary:
\qq
&&D_0(f_i)\ =\ (-\sfi\,f_i^{-1}df_i,\,f_j^{-1}f_i),
\cr
&&D_1(\Pi_i,\chi_{ij})\ =\ (d\Pi_i,
\,-\sfi\,\chi_{ij}^{-1}d\chi_{ij}+\Pi_j-\Pi_i,\,\chi^{\,-1}_{jk}\chi_{ik} 
\chi^{\,-1}_{ij})\,
\cr
&&D_2(B_i,A_{ij},g_{ijk})\ =\ (dA_{ij}-B_j+B_i,\,
-\sfi\,g_{ijk}^{\,-1}dg_{ijk}+A_{jk}-A_{ik}+A_{ij},\cr
&&\hspace{9cm}g_{jkl}^{\,-1} 
g_{ikl}\,g_{ijl}^{\,-1}g_{ijk})\,.
\nonumber
\qqq
The hypercohomology of the complex $\,\CD(2)\,$ 
of (\ref{comp0}), i.e. the cohomology of the complex $\,K(\CD(2))$, 
\,see (\ref{compCA}), is
\qq
&&\mathbb{H}^0(M,\CD(2))\,=\,{\rm ker}\,D_0\,\,\cong\,H^0(M,U(1))\,,\cr\cr
&&\mathbb{H}^1(M,\CD(2))\,=\,{{\rm ker}\,D_1
\over{\rm im}\,D_0}\,\cong\,H^1(M,U(1))
\label{cohA}
\qqq
and, in the second degree,
\qq
\,\mathbb{H}^2(M,\CD(2))={{\rm ker}\,D_2\over{\rm im}\,D_1}\,.
\nonumber
\qqq
$H^0(M,U(1))\,$ is the group of constant $\,U(1)$-valued
functions on $\,M$, $\,H^1(M,U(1))\,$ is the one 
of the isomorphism classes of flat line bundles on $\,M\,$ and 
$\,H^2(M,\CD(2))\,$ is the third real Deligne cohomology group
\cite{Bryl,Gajer}. The local data of a gerbe 
$\,c=(B_i,A_{ij},g_{ijk})\,$ satisfy the cocycle 
condition\footnote{We use the additive notation for the Abelian groups 
$\,A^n$.} $\,D_2c=0\,$ and equivalent local data differ by 
a coboundary $\,D_1\beta\,$ with $\,\beta=(\Pi_i,\chi_{ij})\,$ so that the 
elements of the hypercohomology group $\,\mathbb{H}^2(M,\CD(2))\,$ are 
in a one-to-one correspondence with stable isomorphism 
classes of gerbes. 
\vskip 0.2cm

\subsection{Gerbes on orbifolds and group cohomology}\label{sec:orbi}
\noindent Suppose now that a discrete group $\,\Gamma\,$ acts
on $\,M\,$ preserving the closed 3-form $\,H$. \,Let us assume that the open 
covering $\,(O_i)\,$ is such that $\,\gamma(O_i)
=O_{\gamma i}\,$ for an action 
$\,(\gamma,i) \mapsto\gamma i\,$ of $\,\Gamma\,$ on the index set.
We shall call $\,\Gamma\,$ the orbifold group. In a natural way, we may 
lift its action to the Abelian groups $\,A^n\,$ of (\ref{a01})-(\ref{a31}) 
by defining
\qq
\gamma f_i\,=\,{\gamma^{-1}}^\ast f_{\gamma^{-1}i}\,,\qquad
\gamma\Pi_i\,=\,{\gamma^{-1}}^\ast\Pi_{\gamma^{-1}i}\,,
\label{act}
\qqq
etc. This turns the complex $\,K(\CD(2))\,$ of (\ref{compCA}) induced 
from the sheaf complex (\ref{comp0}) into one of $\,\Gamma$-modules. 
\vskip 0.2cm

Below, we shall employ the language of the group $\,\Gamma\,$ cohomology,
\,see e.g. \cite{Brown} or Appendix A of \cite{GR1}, \,defining 
$\,p$-cochains on $\,\Gamma\,$ with values in a $\,\Gamma$-module $\,N\,$ 
as maps from $\,\Gamma^p\,$ to $\,N$, \,and the coboundary operator
$\,\delta\,$ by
\qq
(\delta n)_{\gamma,\gamma',\dots,\gamma^{(p)}}\,=\,
\gamma n_{\gamma',\dots,\gamma^{(p)}}\,-\,n_{\gamma\gamma',\gamma'',
\dots,\gamma^{(p)}}\,+\,\cdots\,+\,(-1)^pn_{\gamma,\dots,\gamma^{(p-1)}
\gamma^{p}}\cr
+\,(-1)^{p+1}n_{\gamma,\gamma',\dots,\gamma^{(p-1)}}.
\nonumber
\qqq
The Abelian groups $\,C^p(N)\,$ of $\,p$-cochains on $\,\Gamma\,$
form the complex $\,C(N)$,
\qq
0\ \longrightarrow\ C^0(N)\ \mathop{\longrightarrow}\limits^\delta\ 
C^1(N)\ \mathop{\longrightarrow}\limits^\delta\ C^2(N)
\ \mathop{\longrightarrow}\limits^\delta\ \ \cdots\,.
\nonumber
\qqq
The cohomology groups $\,H^p(\Gamma,N)\,$ are composed of 
$\,p$-cocycles on $\,\Gamma\,$ modulo $\,p$-coboundaries.
\,Given a complex $\,K\,$ of $\,\Gamma$-modules
\qq
0\ \longrightarrow\ N^0 \mathop{\longrightarrow}
\limits^{d_0}\ N^1\ \mathop{\longrightarrow}\limits^{d_1}
\ N^2\ \mathop{\longrightarrow}\limits^{d_2}\ \ \cdots\,,
\label{comp00g}
\qqq
we may consider again a double complex formed from the groups
$\,C^p(N^q)\,$ and the induced diagonal complex.
The cohomology groups of the latter define the hypercohomology
groups $\,\mathbb{H}^s(\Gamma,K)$.  
\vskip 0.2cm

We shall be interested in gerbes on $\,M\,$ with 
$\,\Gamma$-{\bf equivariant structures} ($\Gamma$-{\bf gerbes} for short)
that permit to define the contribution of the torsion field $\,H\,$ to 
Feynman amplitudes of closed strings moving in the orbifold $\,M/\Gamma$. 
\,$\Gamma$-gerbes may be presented by their local 
data $\,(c,b_\gamma,a_{\gamma,\gamma'})$,
\,where $\,c=(B_i,A_{ij},g_{ijk})\in A^2$, $\,b_\gamma
=(\Pi^\gamma_i,\chi^\gamma_{ij})\in A^1\,$ and $\,a_{\gamma,\gamma'}
=(f^{\gamma,\gamma'}_i)\in A^0\,$
satisfy the relations
\qq
&&D_2c\,=\,0,\label{gg0}\\
&&(\delta c)_\gamma\,\equiv\,\gamma c\,-\,c\,=\,D_1b_\gamma,\label{gg1}\\
&&(\delta b)_{\gamma,\gamma'}\,\equiv\,
\gamma b_{\gamma'}\,-\,b_{\gamma\gamma'}\,
+\,b_{\gamma}\,=\,-D_0a_{\gamma,\gamma'},\label{gg2}\\
&&(\delta a)_{\gamma,\gamma',\gamma''}\,\equiv\,
\gamma a_{\gamma',\gamma''}\,-\,a_{\gamma\gamma',\gamma''}\,+\,
a_{\gamma,\gamma'\gamma''}\,-\,a_{\gamma,\gamma'}\,=\,0.
\label{gg3}
\qqq
The $\,\G$-gerbe local data $\,(c',b'_\gamma,a'_{\gamma,\gamma'})\,$
and $\,(c,b_\gamma,a_{\gamma,\gamma'})\,$ will be considered
equivalent if there exist $\,\beta\in A^1\,$ and $\,\phi_\gamma\in A^0\,$
such that
\qq
&&c'\,=\,c\,+\,D_1\beta,\label{equiv21}\\
&&b'_\gamma\,=\,b_\gamma\,+\,\gamma\beta\,-\,\beta\,+\,D_0\phi_\gamma\,
\equiv\,b_\gamma\,+\,(\delta\beta)_\gamma\,+\,D_0\phi_\gamma,\label{equiv22}\\
&&a'_{\gamma,\gamma'}\,=\,a_{\gamma,\gamma'}\,-\,\gamma\phi_{\gamma'}
\,+\,\phi_{\gamma\gamma'}\,-\,\phi_\gamma\,\equiv\,a_{\gamma,\gamma'}\,-\,
(\delta\phi)_{\gamma,\gamma'}.
\label{equiv23}
\qqq
In particular, $\,c'\,$ and $\,c\,$ are equivalent local data for gerbes
on $\,M$. $\,\Gamma$-gerbes with equivalent local data will be called stably
isomorphic. Equivalence classes of local data 
$\,(c,b_\gamma,a_{\gamma,\gamma'})\,$ form the hypercohomology group
$\,\mathbb{H}^2(\Gamma,K(\CD(2)))$.  
\vskip 0.2cm

It is easy to see that, up to equivalence, the local data 
$\,(c,b_\gamma,a_{\gamma,\gamma'})\,$ of a flat $\,\Gamma$-gerbe 
are of the form
\qq
c\,=\,(0,0,u_{ijk}),\qquad
b_\gamma\,=\,(0,v_{\gamma;ij}^{-1}),\qquad
a_{\gamma,\gamma'}\,=\,(w_{\gamma,\gamma';i})\,,
\label{fgg1}
\qqq
where $\,u_{ijk},v_{\gamma;ij},w_{\gamma,\gamma';i}\in U(1)$. 
\,This form is preserved by the transformations 
(\ref{equiv21})-(\ref{equiv23}) with 
\qq
\beta\,=\,(0,v_{ij}^{-1}),\qquad\phi_\gamma\,=\,(w_{\gamma;i})
\label{fgg2}
\qqq
with $\,v_{ij},w_{\gamma;i}\in U(1)$. \,The equivalence classes
of local data for a flat $\,\Gamma$-gerbe form the hypercohomology
group $\,\mathbb{H}^2(\Gamma,\check{C}(U(1)))\,$ where $\,\check{C}(U(1))\,$
is the $\check{\rm C}$ech complex (\ref{Cech}) for the sheaf
of locally constant $\,U(1)$-valued functions (viewed as
a complex of $\,\Gamma$-modules).
\vskip 0.2cm

In general, there are obstructions to existence of a 
$\,\Gamma$-equivariant structure on a gerbe with local
data $\,c$. \,First, existence of $\,b_\gamma\in\CA^1\,$ such 
that (\ref{gg1}) holds requires that the equivalence class
of the flat-gerbe local data
\qq
[\gamma c-c]\,\in\,H^2(M,U(1))
\nonumber
\qqq
be trivial (or, geometrically, that the pullback of the gerbe
by $\,\gamma\,$ stays in the same stable isomorphism class). This is
automatically assured if $\,H^2(M,U(1))=0$. \,Suppose then that
$\,\gamma c-c=D_1b_\gamma\,$ for $\,b_\gamma\in A^1$.
\,It follows that $\,D_1(\delta b)_{\gamma,\gamma'}=0\,$ so that 
$\,(\delta b)_{\gamma,\gamma'}\,$ defines a 2-cocycle 
$\,r_{\gamma,\gamma'}\,$ on $\,\Gamma\,$ with values in 
$\,H^1(M,U(1))\equiv H^1$, 
\,see (\ref{cohA}). Its cohomology class 
\qq
[r_{\gamma,\gamma'}]\,\in\,H^2(\Gamma,H^1)
\nonumber
\qqq
is the next obstruction to existence of a $\,\Gamma$-equivariant 
structure. If it is trivial, which holds automatically if $\,H^1=0$, \,then 
there exist $\,e_\gamma\in A^1\,$ with $\,D_1e_\gamma=0\,$ and 
$\,a_{\gamma,\gamma'}\in A^0\,$ such that  
\qq
(\delta b)_{\gamma,\gamma'}\,=\,(\delta e)_{\gamma,\gamma'}\,-\,D_0
a_{\gamma,\gamma'}.
\nonumber
\qqq
Note that $\,D_0(\delta a)_{\gamma,\gamma',\gamma''}=0$. \,Hence
$\,u_{\gamma,\gamma',\gamma''}=(\delta a)_{\gamma,\gamma',\gamma''}\,$
is a 3-cocycle on $\,\Gamma\,$ with values in 
$\,{\rm ker}\,D_0=H^0(M,U(1))\equiv H^0$.
\,Its cohomology class 
\qq
[u_{\gamma,\gamma',\gamma''}]\,\in\,H^3(\Gamma,H^0)
\label{obstr}
\qqq
is the last obstruction to existence of a $\,\Gamma$-equivariant 
structure. If it is trivial, i.e. if $\,u_{\gamma,\gamma',\gamma''}=
(\delta v)_{\gamma,\gamma',\gamma''}\,$ for some $\,v_{\gamma,\gamma'}\in H^0$,
\,then taking 
$\,b_\gamma-e_\gamma\,$ as a new $\,b_\gamma\,$ and $\,a_{\gamma,\gamma'}
-v_{\gamma,\gamma'}\,$ as a new $\,a_{\gamma,\gamma'}$, \,we obtain the 
relations (\ref{gg1})-(\ref{gg3}). Note that in (\ref{obstr}), the group
$\,H^0\,$ of locally constant $\,U(1)$-valued functions $\,f\,$ should be 
viewed as a $\,\Gamma$-module with $\,\gamma f={\gamma^{-1}}^\ast f$. 
\,If $\,M\,$ is connected then $\,H^0=U(1)\,$ with
the trivial action of $\,\Gamma$.
\vskip 0.2cm

An important question arises as to how many inequivalent 
$\,\Gamma$-equivariant
structures exist on a gerbe on $\,M\,$ if all obstructions vanish.
Two sets of local data for a $\,\Gamma$-gerbe with the same
underlying gerbe local data $\,c\,$ differ by 
$\,(b_\gamma,a_{\gamma,\gamma'})\,$ such that 
\qq
D_1b_\gamma\,=\,0,\qquad
(\delta b)_{\gamma,\gamma'}\,=\,-D_0a_{\gamma,\gamma'},
\qquad
(\delta a)_{\gamma,\gamma',\gamma''}\,=\,0.\label{db}
\qqq
The equivalence classes of $\,(b_\gamma,a_{\gamma,\gamma'})\,$
satisfying (\ref{db}) modulo 
\qq
((\delta\beta)_\gamma+D_0\phi_\gamma,\,-(\delta\phi)_{\gamma,\gamma'})
\nonumber
\qqq
with $\,D_1\beta=0\,$ label then inequivalent $\,\Gamma$-equivariant 
structures on the gerbe with local data $\,c$. \,Note that $\,b_\gamma\,$
and $\,a_{\gamma,\gamma'}\,$ above may be taken in the form (\ref{fgg1})
and $\,\beta\,$ and $\,\phi_\gamma\,$ in the form (\ref{fgg2}).
The set of equivalence classes $\,[b_\gamma,a_{\gamma,\gamma'}]\,$ 
forms an Abelian group that we shall denote $\,H_\Gamma$. \,It may
be interpreted as the hypercohomology group $\,\mathbb{H}^2(\Gamma,K(U(1)))\,$
where $\,K(U(1))\,$ is the complex 
\qq 
0\ \longrightarrow\ \check{C}^0(U(1))\ \mathop{\longrightarrow}
\limits^{\check{\delta}}\ \check{Z}^1(U(1))\cr 
\nonumber
\qqq
of $\,\Gamma$-modules with $\,\check{Z}^1(U(1))={\rm ker}
\,\check{\delta}|_{\check{C}^1(U(1))}$.
\,There is a natural map from $\,H_\Gamma\,$ to 
$\,H^1(\Gamma,H^1)\,$ that assigns to 
$\,[b_\gamma,a_{\gamma,\gamma'}]\,$ the cohomology class $\,[b_\gamma]\,$ 
of the image of $\,b_\gamma\,$ in $\,H^1$. 
\vskip 0.2cm

If $\,H^1(\Gamma,H^1)=0$, \,e.g. if $\,H^1=0$, \,then 
$\,[b_\gamma]=0\,$ and there exist $\,(\beta,\phi_\gamma)\,$ such that 
$\,D_1\beta=0\,$ and $\,b_\gamma=(\delta\beta)_\gamma+D_0\phi_\gamma$.
\,For $\,\alpha_{\gamma,\gamma'}=a_{\gamma,\gamma'}
+(\delta\phi)_{\gamma,\gamma'}$, \,one has the relation 
$\,D_0\alpha_{\gamma,\gamma'}=0$. \,It follows that
$\,\phi_\gamma\,$ may be modified so that 
$\,a_{\gamma,\gamma'}=-(\delta\phi)_\gamma\,$ if and only if
the cohomology class $\,[\alpha_{\gamma,\gamma'}]\in H^2(\Gamma,H^0)\,$
is trivial. This results in the isomorphism 
\qq
H_\Gamma\,\ni\,[b_\gamma,a_{\gamma,\gamma'}]\ 
\longmapsto\ [\alpha_{\gamma,\gamma'}]\,\in\,H^2(\Gamma,H^0)
\nonumber
\qqq
of Abelian groups. We infer this way that if 
$\,(c,b_\gamma,a_{\gamma,\gamma'})\,$ 
are local data for a $\,\Gamma$-gerbe then, for 2-cocycles 
$\,v_{\gamma,\gamma'}\,$ on $\,\Gamma\,$ with values in $\,H^0$, 
\qq
(c,b_\gamma,a_{\gamma,\gamma'}+v_{\gamma,\gamma'})
\nonumber
\qqq
are also local data for a $\,\Gamma$-gerbe and, up to
equivalence, all $\,\Gamma$-gerbe local data with the same gerbe 
local data $\,c\,$ are obtained in such a way.
\,The local data $\,(c,b_\gamma,a_{\gamma,\gamma'})\,$ and 
$\,(c,b_\gamma,a_{\gamma,\gamma'}+v_{\gamma,\gamma'})\,$
are equivalent if and only if $\,v_{\gamma,\gamma'}=
(\delta w)_{\gamma,\gamma'}\,$ for $\,w_\gamma\in H^0$.
\,Hence elements of $\,H^2(\Gamma,H^0)\,$ label
inequivalent $\,\Gamma$-structures on a gerbe on 
$\,M\,$ provided that $\,H^1(\Gamma,H^1)=0$. 
\vskip 0.2cm

Suppose now that $\,\Gamma\,$ acts on $\,M\,$ without fixed points 
and that $\,M/\Gamma\equiv M'\,$ is a manifold. Under the assumption that 
the open covering $\,(O_i)\,$ of $\,M\,$ is such that 
$\,O_{i(\gamma i)}\not=\emptyset\,$ only if $\,\gamma=1$, \,the sets 
$\,O'_i=\pi(O_i)$, \,where $\,\pi:M\rightarrow M'\,$ is the canonical 
projection, form a good covering of $\,M'\,$  and 
\qq
O'_{ij'}\equiv O'_i\cap O'_{j'}\,=\mathop{\sqcup}\limits_{j=\gamma j'}
\pi(O_{ij}),\qquad O'_{ij'k'}\,=\mathop{\sqcup}\limits_{j=\gamma j'
\atop k=\gamma\gamma'k'}\pi(O_{ijk})
\nonumber
\qqq
etc. In that situation, a $\,\Gamma$-gerbe with local data 
$\,(c,b_\gamma,a_{\gamma,\gamma'})$, \,where $\,c=(B_i,A_{ij},g_{ijk})$, 
$\,b_\gamma=(\Pi^\gamma_i,\chi^\gamma_{ij})\,$ 
and $\,a_{\gamma,\gamma'}=(f^{\gamma,\gamma'}_i)$, \,induces in a canonical
way a gerbe on $\,M'\,$ with local data $\,(B'_i,A'_{ij'},g'_{ij'k'})\,$
given by the relations \cite{Nuno}:
\qq
&&\hbox to 6cm{$\pi^\ast B'_i\,=\,B_i$\hfill}{\rm on}\,\ \ O_i,\cr
&&\hbox to 6cm{$\pi^\ast A'_{ij'}\,=\,A_{ij}+\Pi^\gamma_j$\hfill}
{\rm on}\,\ \ O_{ij}\ \ {\rm for}\ \ j=\gamma j',\cr
&&\hbox to 6cm{$\pi^\ast g'_{ij'k'}\,=\,g_{ijk}(\chi^\gamma_{jk}
f^{\gamma,\gamma'}_k)^{-1}$\hfill}{\rm on}\,\ \ O_{ijk}\ \ {\rm for}
\ \ j=\gamma j',\ k=\gamma\gamma'k'.
\nonumber
\qqq
Equivalent $\,\Gamma$-gerbe local data on $\,M\,$
are associated with equivalent gerbe local data on $\,M'$. \,Note that
the latter correspond to the curvature 3-form $\,H'\,$ such that
$\,\pi^\ast H'=H$. \,In the more general context where 
$\,\Gamma\,$ acts on $\,M\,$ with fixed points, we shall sometimes talk, 
by an abuse of language, of $\,\Gamma$-gerbes on $\,M\,$ 
as gerbes on the orbifold $\,M/\Gamma$. \,A more sophisticated 
approach to gerbes on orbifolds may be found in \cite{LuUr}.
\vskip 0.2cm

Let $\,\Sigma=\tilde\Sigma/\pi_1\,$ be an oriented closed connected surface 
with $\,\pi_1\,$ its fundamental group and $\,\tilde\Sigma\,$ its
universal covering space. The maps $\,X:\tilde\Sigma\rightarrow M\,$
such that there exists a homomorphism $\,x:\pi_1\rightarrow\Gamma\,$
for which
\qq
X(a\tilde\sigma)\,=\,x(a)X(\tilde\sigma)
\nonumber
\qqq
if $\,a\in\pi_1\,$ and $\,\tilde\sigma\in\tilde\Sigma\,$ describe world 
histories of the closed string moving in the orbifold 
$\,M/\Gamma$. 
\,The pullback by $\,X\,$of the local data for a $\,\Gamma$-gerbe on $\,M\,$
defines local data for a flat $\,\pi_1$-gerbe on $\,\tilde\Sigma$. 
\,Those, in turn, determine local data for a flat gerbe on $\,\Sigma\,$ 
by the construction described above and an element in 
$\,H^2(\Sigma,U(1))=U(1)\,$ called the holonomy along $\,X\,$ that 
represents the contribution of the Kalb-Ramond field to the Feynman 
amplitude of $\,X$.

\subsection{Gerbes on simple compact Lie groups}\label{sec:gcLg}
\noindent Gerbes on Lie groups have been studied in the context of
the Wess-Zumino-Witten (WZW) models \cite{Witt} of conformal field 
theory describing the motion of strings in group manifolds. Let
$\,G\,$ be a connected and simply connected compact simple
Lie group and let $\,H_{\sfk}\,$ be the bi-invariant closed 3-form 
on $\,G$,
\qq
H_{\sfk}\,=\,{_\sfk\over^{12\pi}}\,\tr\,(g^{-1}dg)^3.
\nonumber
\qqq
Here, $\,\tr\,$ denotes the $\,ad$-invariant positive bilinear
symmetric (Killing) form on the Lie algebra $\,\Ng$, \,normalized so 
that the 3-form $\,{1\over2\pi}H_\sfk\,$ has integer periods if and
only if $\,\sfk\,$ (called the level) is an integer. For
such $\,\sfk$, \,there exists a gerbe on $\,G\,$ with curvature
$\,H_\sfk\,$ and it is unique up to stable isomorphisms
since $\,H^2(G,U(1))=0$. \,We shall call it the level $\,\sfk\,$ 
gerbe on $\,G$. \,An explicit construction
of such gerbes was given in \cite{top} for $\,G=SU(2)$, 
in \cite{ChatH} for $\,G=SU(N)\,$ and in \cite{Meinr} for all 
simple simply-connected compact Lie groups. In the last two cases, 
the construction used a more geometric description of gerbes along
the lines of \cite{Murr,MurrS} rather than the one employing local data. 
\vskip 0.2cm

Let $\,Z(G)\,$ be the center of the simply connected group $\,G\,$
and let $\,\Gamma=Z\subset Z(G)\,$ be its subgroup. The case of non-simply 
connected quotients $\,G/Z\equiv G'\,$ was studied in \cite{GR1} for 
$\,G=SU(N)\,$ and in \cite{GR2} for other groups $\,G$. \,In 
those references, gerbes on groups $\,G'\,$ with curvature
$\,H'_\sfk\,$ were explicitly constructed whenever possible. 
Equivalently, the construction provides $\,Z$-equivariant
structures on the level $\sfk\,$ gerbe on $\,G$.
\,Since the groups $\,H^2(G,U(1))\,$ and $\,H^1(G,U(1))\,$ are trivial 
and $\,H^0(G,U(1))=U(1)\,$ with the trivial action of $\,Z$, 
\,the only obstruction to existence 
of such $\,Z$-equivariant structures is the cohomology class 
$\,[u_{z,z',z''}]\in H^3(Z,U(1))$, \,see (\ref{obstr}).
The main part of the construction of \cite{GR1,GR2} consisted 
in analyzing the cohomological equation
\qq
u_{z,z',z''}\,=\,(\delta v)_{z,z',z''}
\label{cohoeq}
\qqq
and finding its solutions for all levels $\,\sfk\,$ for which the 
obstruction cohomology class (\ref{obstr}) is trivial. In agreement
with the analysis of the last subsection, solutions
$\,v_{z,z'}\,$ differing by non-cohomologous 2-cocycles
gave rise to inequivalent $\,Z$-equivariant structures
and hence to stably non-isomorphic gerbes on $\,G'\,$ with curvature
$\,H'_\sfk$. \,The levels $\,\sfk\,$ for which the obstruction class 
is trivial are the ones for which the 3-form $\,{1\over2\pi}H'_\sfk\,$ 
on $\,G'\,$ has integer periods. They were identified for the first
time in \cite{FGK}. 
\vskip 0.2cm

Let us recall here the form of the obstruction 3-cocycle
$\,u_{z,z',z''}\,$ obtained in \cite{GR2}. 
The cocycle was related to the action of the center $\,Z(G)\,$
on the set of conjugacy classes of $\,G$. \,Each conjugacy 
class has a single representative of the form $\,\ee^{2\pi\sfi\tau}\,$ 
where $\,\tau\,$ belongs to the positive Weyl alcove $\,\CA$, 
\,a simplex in the Cartan algebra $\,\Nt\subset\Ng\,$ with the vertices
\qq
\tau_0=0,\qquad\tau_i={_1\over^{k_i}}\lambda_i^\vee\quad\ {\rm for}\quad
i=1,\dots,r
\nonumber
\qqq
where $\,r=\dim\Nt\,$ is the rank of $\,G$, \,$\lambda_i^\vee\,$
are the simple coweights in $\,\Nt\,$ and $\,k_i\,$ are the corresponding
Coxeter labels. The latter are defined by the relations
\qq
\tr\,\lambda_i^\vee\alpha_j=\delta_{ij}\,,\qquad\phi=\sum\limits_{i=1}^r
k_i\alpha_i
\nonumber
\qqq
where $\,\alpha_i,\ i=1,\dots,r$, \,are the simple roots and $\,\phi\,$
is the highest root of the Lie algebra $\,\Ng$. \,Multiplication 
by an element $\,z\in Z(G)\,$ sends conjugacy classes into conjugacy 
classes and induces an affine map $\,\tau\mapsto\gamma\tau\,$ of the 
positive Weyl alcove. More exactly,
\qq
z\,\ee^{2\pi\sfi\tau}\,=\,w_z^{-1}\,\ee^{2\pi\sfi(z\tau)}\,w_z
\label{zact}
\qqq
for  some $\,w_z\,$ belonging to the normalizer $\,N(T)\,$ of the
Cartan subgroup $\,T\subset G$. \,For the vertices of $\,\CA$,
\,we have
\qq
z\tau_i\,=\,\tau_{zi}\qquad{\rm for}\quad i=0,\dots,r\,.
\nonumber
\qqq
Upon identification of the set of indices $\,i=0,1,\dots,r\,$ 
with the set of nodes of the extended Dynkin diagram of  
$\,\Ng$, \,the action $\,i\mapsto zi\,$ induces a symmetry of the 
diagram. The group elements $\,w_z\,$ are 
determined up to left multiplication by elements of $\,T\,$
and, in general, cannot be chosen to depend multiplicatively
on $\,z\,$ but $\,w_zw_{z'}w_{zz'}^{-1}\in T$. \,Let 
$\,b_{z,z'}\in\Nt\,$ be such that 
\qq
w_zw_{z'}w_{zz'}^{-1}\,=\,\ee^{2\pi\sfi\,b_{z,z'}}.
\nonumber
\qqq
For $\,z\in Z(G)$, \,the vertex $\,\tau_{z^{-1}0}\,$ of $\,\CA\,$
is a simple coweight such that $\,z=\ee^{-2\pi\sfi\,\tau_{z^{-1}0}}$. 
\,The formula
\qq
u_{z,z',z''}\,=\,\ee^{-2\pi\sfi\sfk\,\tr\,\tau_{z^{-1}0}\,b_{z',z''}}
\label{z3cocyc}
\qqq
defines a 3-cocycle on $\,Z(G)\,$ whose
cohomology class does not depend on the choices made in 
the definition\footnote{The 3-cocycle 
analyzed in \cite{GR2} differed by a coboundary from the one 
of (\ref{z3cocyc}).}. \,The restriction
of $\,u_{z,z',z''}\,$ to $\,z,z',z''\in Z\subset Z(G)\,$
gives the 3-cocycle whose cohomology class in $\,H^3(Z,U(1))\,$ 
is the obstruction 
(\ref{obstr}) to existence of a $\,Z$-equivariant structure
on the level $\,\sfk\,$ gerbe on $\,G$. \,The cohomological 
equation (\ref{cohoeq}) was discussed case by case 
in \cite{GR2}.
\vskip 0.2cm

\subsection{Gerbes on orientifolds}\label{sec:gonor} 
\noindent 
A simple generalization of the notion of a $\,\Gamma$-gerbe 
developed in Sect.\,\ref{sec:orbi} is to admit a more general action 
of the discrete group $\,\Gamma\,$ on $\,M\,$ such that 
$\,\gamma^*H=\epsilon(\gamma)H\,$
for a homomorphism $\,\epsilon:\Gamma\rightarrow\{\pm1\}$. 
\,The only modification required
is in the definition (\ref{act}) of the action of $\,\Gamma\,$ 
on the groups $\,A^n\,$ of (\ref{a01})-(\ref{a31}) that should read:
\qq
\gamma f_i\,=\,({\gamma^{-1}}^\ast 
f_{\gamma^{-1}i})^{\epsilon(\gamma)}\,,
\qquad\gamma\Pi_i\,=\,\epsilon(\gamma)\,{\gamma^{-1}}^\ast
\Pi_{\gamma^{-1}i}\,,
\nonumber
\qqq
etc. The change assures, for example, that if $\,c\in\CA^2\,$ with 
$\,D_2c=0\,$ gives local data for a gerbe with curvature $\,H\,$ then 
so does $\,\gamma c$. 
\,Let $\,\Gamma_0\subset\Gamma\,$ denote the kernel of 
$\,\epsilon\,$ so that one has the exact sequence of groups:
\qq
1\ \longrightarrow\ \Gamma_0\ \longrightarrow\ \Gamma
\ \mathop{\longrightarrow}^\epsilon\ \bZ_2\ \longrightarrow\ 1\,.
\nonumber
\qqq 
We shall call $\,\Gamma\,$ an orientifold group if $\,\Gamma_0\not=\Gamma$. 
\,The whole discussion of Sect.\,\ref{sec:orbi} except for the two 
end paragraphs about gerbes on non-singular quotients and about 
the holonomy extends to the case of orientifold group actions 
generalizing the notions of $\,\Gamma$-equivariant structures and 
of $\,\Gamma$-gerbes to that case. We shall loosely talk of
$\,\Gamma$-gerbes for $\,\Gamma\,$ an orientifold group as gerbes
on the orientifold $\,M/\Gamma$. \,As before, if 
$\,H^1(M,U(1))=0=H^2(M,U(1))$ then the only obstruction 
to existence of a $\,\Gamma$-equivariant structure on the gerbe 
with local data $\,c\,$ is the class (\ref{obstr}), where the group 
$\,H^0\equiv H^0(M,U(1))\,$ of locally constant $\,U(1)$-valued functions 
is viewed now as a $\,\Gamma$-module with 
$\,\gamma f=({\gamma^{-1}}^\ast f)^{\epsilon(\gamma)}$. \,For 
$\,M\,$ connected, $\,H^0=U(1)\,$ with the action 
$\,\gamma \lambda=\lambda^{\epsilon(\gamma)}$. \,If the obstruction 
(\ref{obstr}) 
is trivial then inequivalent $\,\Gamma$-equivariant structures 
are labeled by elements of $\,H^2(\Gamma,H^0_\epsilon)$, \,where
the subscript $\,\epsilon\,$ indicates that $\,H^0\,$ is taken
with the $\,\Gamma$-module structure just described.
\vskip 0.2cm

The simplest example is that of the {\,\bf inversion group}
$\,\Gamma=\{\pm1\}\equiv\bZ_2\,$ with $\,\epsilon(\pm 1)=\pm1$. 
\,$\Gamma$-equivariant structures on a gerbe on $\,M\,$ 
for such $\,\G\,$ were introduced (in an equivalent formulation) 
in \cite{SSW} under the name of \,{\bf Jandl structures}.
A particular case is when $\,M\,$ is the orientation double 
$\,\hat\Sigma\,$ of an unoriented closed connected surface
$\,\Sigma=\hat\Sigma/\bZ_2$. \,$\hat\Sigma\,$ is an oriented 
closed surface (connected if $\,\Sigma\,$ is non-orientable 
and with two components otherwise). The group of stable isomorphism 
classes of $\,\bZ_2$-gerbes on $\,\hat\Sigma\,$ is 
$\,\mathbb{H}^2(\bZ_2,\check{C}(U(1)))$, \,as described in 
Sect.\,\ref{sec:orbi}. In \cite{SSW}, a natural group homomorphism 
$\,\iota\,$ was constructed that renders the diagram
\vskip -0.3cm
\qq
\mathbb{H}^2(\bZ_2,C(U(1)))\quad
&\mathop{\longrightarrow}\limits^{\iota}&\quad U(1)\cr
\hspace*{1.3cm}\searrow&\qquad\qquad&\swarrow{\rm sq}\label{diag}\\
&H^2(\hat\Sigma,U(1))=U(1)&
\nonumber
\qqq
commutative. In the diagram, the south-east arrow 
is induced by forgetting the $\,\bZ_2$-equivariant structure and 
the south-west one by $\,{\rm sq}(\lambda)=\lambda^2$. 
\,Let $\,X:\hat\Sigma\rightarrow M\,$ be such that
\qq
X(-1\cdot\hat\sigma)\,=\,-1\cdot X(\hat\sigma). 
\nonumber
\qqq
for $\,\hat\sigma\in\hat\Sigma$. \,Such maps $\,X$, \,invariant under the 
combined worldsheet orientation reversal and a target map that changes
the sign of the torsion field, describe world histories of the closed 
unoriented string moving in the orientifold $\,M/\bZ_2$. \,The pullback 
by $\,X\,$ of a $\,\bZ_2$-gerbe on $\,M\,$ to $\,\hat\Sigma\,$ defines
a $\,\bZ_2$-gerbe on $\,\hat\Sigma$. \,The number in $\,U(1)\,$ associated 
to the stable isomorphism class of the latter by the homomorphism 
$\,\iota\,$ is called the holonomy of the $\,\bZ_2$-gerbe on $\,M\,$ 
along $\,X$. \,It represents the contribution of the Kalb-Ramond field 
to the Feynman amplitude of the world history $\,X$.
\vskip 0.2cm

For more general orientifold groups $\,\G$, \,the restriction of the 
$\,\G$-equivariant structure to the $\,\G_0$-equivariant one
may be used to define a quotient gerbe on $\,M'=M/\Gamma_0\,$ if 
$\,\G_0\,$ acts without fixed points. The full $\,\Gamma$-equivariant
structure induces then a Jandl structure on the quotient gerbe, 
see \cite{GSW2}. We could, more correctly, call a $\,\Gamma$-equivariant 
structure  a $\,\Gamma_0$-{\bf equivariant Jandl structure}, 
\,but we shall stick in what follows to the former name. 
The construction of holonomy of gerbes with Jandl structures
described above may be extended to the equivariant case. Let
$\,\Sigma\,$ be an unoriented closed connected surface, $\,\hat\Sigma\,$
its orientation double, $\,\hat\pi_1\,$ the fundamental group of
$\,\hat\Sigma$, \,and $\,\tilde\Sigma\,$ the universal covering 
of $\,\hat\Sigma$. \,There is a natural group $\,\tilde\pi\,$ 
entering the exact sequence of groups
\qq
1\ \longrightarrow\ \hat\pi_1\ \longrightarrow\ \tilde\pi
\ \mathop{\longrightarrow}^{\tilde\epsilon}\ \bZ_2\ \longrightarrow\ 1
\nonumber
\qqq
and acting on $\,\tilde\Sigma\,$ without fixed points in a way that
extends the action of $\,\hat\pi_1$, projects to the natural action 
of $\,\bZ_2\,$ on $\,\hat\Sigma\,$ and to the identity on $\,\Sigma$. 
\,Suppose that $\,X:\tilde\Sigma\rightarrow M\,$
is such that, for a homomorphism $\,x:\tilde\pi\rightarrow\Gamma\,$
with $\,\epsilon(x(\tilde a))=\tilde\epsilon(\tilde a)$, \,one has:
\qq
X(\hat a\tilde\sigma)\,=\,x(\tilde a)X(\tilde\sigma)  
\nonumber
\qqq
for $\,\tilde a\in\tilde\pi\,$ and $\,\tilde\sigma\in\tilde\Sigma$.
\,Such maps $\,X$, \,covariant with respect to the action of the
orientifold groups, describe world histories of the closed unoriented
string moving in the orientifold $\,M/\Gamma$. \,The pullback 
of a $\,\Gamma$-gerbe on 
$\,M\,$ by $\,X\,$ defines a flat $\,\tilde\pi$-gerbe on $\,\tilde\Sigma$.
\,Using the restriction of the $\,\tilde\pi$-equivariant structure
to $\,\hat\pi_1\subset\tilde\pi$, \,one may then obtain a flat gerbe on
$\,\hat\Sigma$. \,The $\,\tilde\pi$-equivariant structure descends
to a Jandl structure on it. Applying the homomorphism 
$\,\iota\,$ of (\ref{diag}) to the $\,\bZ_2$-gerbe on $\,\hat\Sigma\,$
obtained this way, one finds then the holonomy of the $\,\Gamma$-gerbe 
along $\,X\,$ that contributes to the Feynman amplitude 
of the closed unoriented string moving in the orientifold 
$\,M/\Gamma$.
\vskip 0.2cm

\subsection{Orientifolds of simple compact Lie groups}\label{sec:orcLg}
\noindent We may consider the inversion group 
$\,\Gamma=\{\pm1\}\cong\bZ_2\,$ with $\,\epsilon(\pm1)=\pm1\,$ 
and $\,-1\equiv z_0\,$ acting on a connected simply connected 
simple compact Lie group $\,G\,$ by
\qq
G\ni g\,\rightarrow\,z_0g=(\zeta g)^{-1}\in G
\label{tw}
\qqq
for $\,\zeta\in Z(G)\,$ that we shall call the {\bf twist element}. 
\,The action of $\,z_0\,$ changes, indeed, the sign of $\,H_\sfk\,$ 
so that the relation $\,\gamma^*H_\sfk=\epsilon(\gamma)H_\sfk\,$ holds. 
\,More generally, we shall consider orientifold groups 
$\,\Gamma=\bZ_2\lx Z\,$ for $\,Z\subset Z(G)\,$ with 
the multiplication table 
\qq
&&\hbox to 6.5cm{$(1,z)\cdot(1,z')=(1,zz'),$\hfill}(-1,z)\cdot(1,z')
=(-1,zz'),\cr
&&\hbox to 6.5cm{$(1,z)\cdot(-1,z')=(-1,z^{-1}z'),$\hfill}
(-1,z)\cdot(-1,z')=(1,z^{-1}z')
\nonumber
\qqq
and $\,\epsilon(\pm1,z)=\pm1\,$ so that $\,\Gamma_0=Z$. \,Note 
that $\,\G\,$ is a non-Abelian
group if $\,Z\,$ is non-trivial and different from $\,\bZ_2\,$ or
from a direct product of $\,\bZ_2\,$ factors. To simplify the notation, 
we shall write $\,(1,z)\equiv z\,$ and $\,(-1,z)\equiv z_0z$. \,For
the action of $\,\G\,$ on $\,G\,$ we shall take the one that combines
(\ref{tw}) with the action of $\,Z\,$ by  multiplication so that 
$\,z_0zg=(\zeta zg)^{-1}$. \,Note that if $\,h_{\zeta'}\,$
for $\,\zeta'\in Z\,$ denotes the automorphism of 
$\,\G=\bZ_2\lx Z\,$ defined by the relations
\qq
h_{\zeta'}(z)\,=\,z\,,\qquad h_{\zeta'}(z_0z)\,=\,z_0\zeta'z
\label{twh}
\qqq
then the composition of the action of $\,\G\,$ on $\,G\,$ with 
$\,h_{\zeta'}\,$ induces the change $\,\zeta\mapsto\zeta\zeta'\,$ 
of the twist element.
Hence twist elements in the same coset of $\,Z(G)/Z\,$
give rise to equivalent orientifold group actions. This is in 
agreement with the observation \cite{GSW2} that 
the restriction of a $\,\Gamma$-equivariant structure on a 
gerbe on $\,G\,$ to the $\,Z$-equivariant structure induces
a gerbe on the non-simply connected group $\,G'=G/Z\,$
and the full $\,\Gamma$-equivariant structure gives rise
to a Jandl structure on that gerbe. Indeed, actions of $\,\G\,$ 
on $\,G\,$ corresponding to twist elements in the same coset 
of $\,Z(G)/Z\,$ induce the same action of $\,\bZ_2\,$ on $\,G'$. 
\vskip 0.2cm

As discussed in the previous subsection, the sole obstruction
to existence of a $\,\Gamma$-equivariant structure on the level
$\,\sfk\,$ gerbe on $\,G\,$ is given by the
cohomology class $\,[u_{\gamma,\gamma',\gamma''}]\in 
H^3(\Gamma,U(1)_\epsilon)$, \,where the subscript $\,\epsilon\,$
indicates that $\,U(1)\,$ is taken with the action 
$\,\gamma\lambda=\lambda^{\epsilon(\gamma)}\,$ of $\,\Gamma$. \,The 
3-cocycle $\,u_{\gamma,\gamma',\gamma''}\,$ may be found by a 
straightforward generalization of the work done in \cite{GR2} where 
the case of orbifold groups $\,Z\,$ was treated, see Sect.\,\ref{sec:gcLg} 
above. We shall only describe the result here, postponing a more detailed 
exposition to \cite{GSW2}. 
\vskip 0.2cm

First, let us observe that the inversion map 
$\,g\mathop{\mapsto}\limits^\kappa g^{-1}\,$ sends
conjugacy classes to conjugacy classes. Upon identification
of the set of conjugacy classes in $\,G\,$ with the positive Weyl 
alcove $\,\CA\subset\Nt$, \,see Sect.\,\ref{sec:gcLg}, it induces an 
affine map $\,\tau\mapsto\kappa\tau\,$ on $\,\CA\,$ such that
\qq
\kappa\tau_i\,=\tau_{\kappa i}
\nonumber
\qqq
for $\,\kappa0=0\,$ and $\,i\mapsto\kappa i\,$ for $\,i=1,\dots,r$,
\,giving rise to a symmetry of the (unextended) Dynkin diagram 
of $\,\Ng$. \,More precisely,
\qq
\ee^{-2\pi\sfi\tau}\,=\,w_\kappa^{-1}\,\ee^{2\pi\sfi(\kappa\tau)}\,w_\kappa
\label{kact}
\qqq
for  some $\,w_\kappa\,$ belonging to the normalizer $\,N(T)\,$ of the
Cartan subgroup $\,T\subset G$. \,The element $\,w_\kappa\,$ is determined 
up to left multiplication by elements of $\,T$. \,Combining the 
relations (\ref{kact}) and (\ref{zact}), we infer that for any 
$\,\gamma\in\Gamma=\bZ_2\lx Z\,$ there exist an affine map 
$\,\tau\mapsto\gamma\tau\,$ of the Weyl alcove $\,\CA\,$ 
and $\,w_\gamma\in N(T)\,$ such that 
\qq
\gamma\,\ee^{2\pi\sfi\tau}\,=\,w_\gamma^{-1}\,\ee^{2\pi\sfi(\gamma\tau)}
\,w_\gamma\,.
\label{gact}
\qqq
One has $\,\gamma\tau=z\tau\,$ for $\,\gamma=z\,$ and 
$\,\gamma\tau=\kappa\zeta z\tau\,$ for $\,\gamma=z_0z\,$ and one may 
take
\qq
w_\gamma\,=\,w_z\quad\tx{for}\quad\gamma=w\,,\qquad
w_\gamma\,=\,w_\kappa w_\zeta w_z\quad\tx{for}\quad\gamma=z_0z\,.
\label{wg}
\qqq 
The action of $\,\Gamma\,$ on the vertices of $\,\CA$,
\qq
\gamma\tau_i\,=\,\tau_{\gamma i}\qquad{\rm for}\quad i=0,\dots,r
\label{gi}
\qqq
induces a symmetry $\,i\mapsto\gamma i\,$ of the extended Dynkin diagram of  
$\,\Ng$. \,This symmetry preserves the Coxeter labels: $\,k_{\gamma i}=k_i\,$ 
if one sets $\,k_0=1$. \,From the relations (\ref{gact})
and (\ref{gi}), one obtains the formula:
\qq
\gamma\tau\,=\,\epsilon(\gamma)\,w_\gamma\tau w_\gamma^{-1}\,
+\,\tau_{\gamma0}
\nonumber
\qqq
 for the action of $\,\gamma\,$ on $\,\CA$. \,As before, it is easy to 
see that $\,w_\gamma w_{\gamma'}w_{\gamma\gamma'}^{-1}\in T\,$ so that one 
may choose $\,b_{\gamma,\gamma'}\in\Nt\,$ such that 
\qq
w_\gamma w_{\gamma'}w_{\gamma\gamma'}^{-1}\,=\,
\ee^{2\pi\sfi\,b_{\gamma,\gamma'}}.
\label{gggb}
\qqq
The 3-cocycle on $\,\Gamma$, \,whose cohomology class defines the obstruction 
to existence of a $\,\Gamma$-equivariant structure on the level
$\,\sfk\,$ gerbe on $\,G$, \,takes the form:
\qq
u_{\gamma,\gamma',\gamma''}\,=\,\ee^{-2\pi\sfi\sfk\,\epsilon(\gamma)\,\tr\,
\tau_{\gamma^{-1}0}\,b_{\gamma',\gamma''}}\,.
\label{g3cocyc}
\qqq
The cocycle condition means that
\qq
(\delta u)_{\gamma,\gamma',\gamma'',\gamma'''}\,\equiv\,
u_{\gamma',\gamma'',\gamma'''}^{\epsilon(\gamma)}\,
u_{\gamma\gamma',\gamma'',\gamma'''}^{-1}
u_{\gamma,\gamma'\gamma'',\gamma'''}\,
u_{\gamma,\gamma',\gamma''\gamma'''}^{-1}\,
u_{\gamma,\gamma',\gamma''}\ =\ 1\,.
\nonumber
\qqq
It may be verified by a direct calculation.
The cohomology class of $\,u_{\gamma,\gamma',\gamma''}\,$ is independent 
of the choices made in its definition.
Note that the 3-cocycle (\ref{g3cocyc}) on $\,\Gamma\,$ restricts 
to the 3-cocycle (\ref{z3cocyc}) on $\,Z=\Gamma_0$.
\vskip 0.2cm

Let us finally remark that, since the orientifold action (\ref{tw})
of $\,\G=\bZ_2\lx Z\,$ with the twist element $\,\zeta\zeta'\,$ for 
$\,\zeta'\in Z\,$ may be obtained from that with the twist
element $\,\zeta\,$ by composing with the automorphism (\ref{twh}) 
of $\,\G$, \,the cocycle $\,u_{\gamma,\gamma',\gamma''}\,$ for the new 
action defines the same cohomology class in $\,H^3(\G,U(1)_\epsilon)\,$ 
as the one for the original action composed with the automorphism 
$\,h_{\zeta'}$. \,The composition with an automorphism of $\,\Gamma\,$ 
that leaves the homomorphism $\,\epsilon\,$ invariant commutes 
with the coboundary $\,\delta\,$ and induces an automorphism of the 
cohomology groups $\,H^n(\G,U(1)_\epsilon)$. 
\vskip 0.2cm

\section{Lyndon-Hochschild-Serre spectral sequence}\label{sec:cohocon}
\noindent Recall that the cohomology class
\qq
[u_{\gamma,\gamma',\gamma''}]\ \in\ H^3(\G,U(1)_\epsilon)
\nonumber
\qqq 
is the obstruction to existence of a $\,\Gamma$-equivariant
structure on the level $\,\sfk\,$ gerbe on the simply connected
group $\,G$. \,The purpose of the present paper is to discuss in detail 
the cohomological equation
\qq
u_{\gamma,\gamma',\gamma''}\ =\ 
v_{\gamma',\gamma''}^{\epsilon(\gamma)}\,v_{\gamma\gamma',\gamma''}^{-1}
\,v_{\gamma,\gamma'\gamma''}\,v_{\gamma,\gamma'}^{-1}\ \equiv\ 
(\delta v)_{\gamma,\gamma',\gamma''}.
\label{gcohoeq}
\qqq
which is solvable if and only if the cohomology class 
$\,[u_{\gamma,\gamma',\gamma''}]\,$ is trivial. Knowledge of
the general structure of the cohomology group $\,H^3(\G,U(1)_\epsilon)\,$ 
will be useful in checking the latter condition. \,In what follows, we shall 
call $\,u_{\gamma,\gamma',\gamma''}\,$ the \textbf{obstruction cocycle} 
and $\,v_{\gamma,\gamma'}\,$ a \textbf{trivializing cochain}. As will 
be shown in \cite{GSW2}, trivializing cochains enter directly
the construction of a $\,\Gamma$-equivariant structure on the
level $\,\sfk\,$ gerbe on $\,G$, \,similarly as in the case of orbifold
groups that was discussed in \cite{GR2}. The classification 
of inequivalent $\,\Gamma$-gerbes may, likewise, 
be formulated in the cohomological language, with inequivalent
$\,\Gamma$-gerbes corresponding to trivializing cochains differing 
by non-cohomologous 2-cocycles $\,v_{\gamma,\gamma'}$, 
\beq
[v_{\gamma,\gamma'}]\ \in\ H^2(\G,U(1)_\epsilon)\,.
\eeq
This way $\,H^2(\G,U(1)_\epsilon)\,$ plays the role of  
the classifying group for inequivalent $\,\Gamma$-gerbes on $\,G$. 
\,Its structure will provide valuable insights into certain algebraic 
properties and the origin of trivializing cochains prior to entering
the straightforward yet tedious computations of Sect.\ref{sec:case}. 
It should be stressed at this point that while obstructions to 
existence of orientifold gerbes do not, in general, exhaust 
the obstruction cohomology group $\,H^3(\G,U(1)_\epsilon)$, \,it 
is the entire classifying group $H^2(\G,U(1)_\epsilon)$ that captures 
inequivalent orientifold gerbe structures. 
\vskip 0.2cm

In consequence of the semi-direct product nature
of the orientifold group $\,\Gamma=\bZ_2\lx Z$, \,the main tool 
which will be used in exploring the $U(1)_\epsilon$-valued cohomology 
of $\,\G\,$ is the Lyndon-Hochschild-Serre (LHS) spectral sequence \cite{weib}
\beq
E_r^{p,q} \Longrightarrow H^{p+q}(\G,U(1)_\epsilon)
\label{spconv}
\eeq
associated with the short exact sequence of groups:
\beq
1 \too Z \too \G \too \bZ_2 \too 1.
\eeq
Recall \cite{specseq} that the $r^{\rm th}$ page of a spectral
sequence with $\,r\geq0\,$  is a collection of Abelian groups 
$\,E_r^{p,q}\,$ \,vanishing for negative 
$\,p\,$ or $\,q$, \,together with the  coboundary homomorphisms 
$\,d^{p,q}_r:E^{p,q}_r\rightarrow
E^{p+r,q-r+1}_r\,$ such that $\,d^{p+r,q-r+1}_rd^{p,q}_r=0$.
\,The groups of the next page are defined by setting
$\,E^{p,q}_{r+1}=\ker d^{p,q}_r/\im\,d^{p-r,q+r-1}$. 
\,The second page of the LHS spectral sequence is composed of the 
groups
\beq
E_2^{p,q} = H^p(\bZ_2,H^q(Z,U(1))_\epsilon)\,,
\eeq
with the action of $\,\bZ_2\,$ on $\,H^q(Z,U(1))\,$ induced by
the one on the $q$-cochains on $\,Z$:
\qq
(-1\cdot c)_{z,z',\dots,z^{(q)}}\,=\,c_{z^{-1},{z'}^{-1},\dots,{z^{(q)}}^{-1}}^{-1}\,.
\label{orac}
\qqq
The relation (\ref{spconv}) of the LHS sequence to the cohomology 
groups $\,H^n(\Gamma,U(1)_\epsilon)\,$ is established with the help 
of a filtration
\qq
0=H^n_{n+1}\,\subset\,H^n_n\,\subset\ \cdots\ \subset H^n_1\,\subset\,H^n_0
=H^n(\Gamma,U(1)_\epsilon)
\nonumber
\qqq
such that
\qq
H^n_p/H^n_{p+1}\,\cong\,E^{p,n-p}_\infty
\nonumber
\qqq
where $\,E^{p,q}_\infty\,$ denotes the group at which $\,E^{p,q}_r\,$
stabilize for $\,(p,q)\,$ fixed and $\,r\,$ sufficiently large.
\vskip 0.2cm

Already the second page of the LHS spectral sequence provides a great 
deal of information on the possible structure of the cohomology groups 
$H^n(\G,U(1)_\epsilon)$, \,at least for $\,Z\,$ cyclic to which case we 
shall specialize first, taking \,$Z = \bZ_m\,$
with $\,m>0$. \,The cyclic group cohomology is well known, see \cite{Brown}:
\qq
H^q(\bZ_m,U(1))\ =\ \left\{ \barr{ccl} 
U(1) & \qquad \tx{if} & \quad q = 0, \\
0 & \qquad \tx{if} & \quad q>0\ \ \tx{is\ \,even},\\
\bZ_m & \qquad \tx{if} & \quad q\ \ \tx{is\ \,odd}
\earr \right.
\nonumber
\qqq
for the trivial action of the orbifold group $\,\bZ_m\,$ on $\,U(1)$.
\,The action of the generator $\,-1\,$ of the orientifold group $\,\bZ_2\,$ 
on $\,H^q(\bZ_m,U(1))\,$ induced by (\ref{orac}) reduces to the
inversion for $\,q\,$ even and to the trivial action for $\,q\,$ odd.
One further has: 
\beq
H^p(\bZ_2,U(1)_\epsilon)\ =\ \left\{ \barr{ccl}
\bZ_2 & \qquad \tx{if} & \quad p\ \ \tx{is\ \,even},\\
0 & \qquad \tx{if} & \quad p\ \ \tx{is\ \,odd}
\earr \right.
\label{peps}
\eeq
for the action of $\,-1\,$ on $\,U(1)\,$ by inversion, and
\beq
H^p(\bZ_2,\bZ_m)\ =\ \left\{ \barr{ccl}
\bZ_m & \qquad \tx{if} & \quad p = 0, \\
\bZ_2 & \qquad \tx{if} & \quad p>0\ \ \tx{and}\ \ m\ \ \tx{is\ \,even},\\ 
0 & \qquad \tx{if} & \quad p>0\ \ \tx{and}\ \ m\ \ \tx{is\ \,odd}
\earr \right.
\eeq
for the trivial action of $\,\bZ_2\,$ on $\,\bZ_m$.
\,This gives for the second page of the spectral sequence: \nl
\begin{center}
$E_2^{p,q} \,: \quad$ \begin{tabular}{|c}
\xymatrix{q \uparrow \\
\vdots & \vdots & \vdots & \vdots & \vdots \\
\bZ_m \ar@{.>}[dddrrrr]_{d_4^{0,3}}& 0 & 0 &  0 & 0 & \cdots \\
0  & 0 & 0 &  0 & 0 & \cdots \\
\bZ_m \ar[drr]_{d_2^{0,1}} & 0 & 0 &  0 & 0 & \cdots \\
E_2^{0,0} = \bZ_2 & 0 & \bZ_2 & 0 & \bZ_2 & \cdots &\xrightarrow{p}} \\ \hline
\end{tabular}
\end{center}
~\nl
for $\,m\,$ odd, and 
\begin{center}
\mbox{$E_2^{p,q} \,: \quad$ \begin{tabular}{|c}
\xymatrix{q \uparrow \\
\vdots & \vdots & \vdots & \vdots & \vdots & \vdots \\
\bZ_m   \ar@{.>}[dddrrrr]_{d_4^{0,3}} \ar@{-->}[ddrrr]^{d_3^{0,3}}& \bZ_2
& \bZ_2 & \bZ_2 & \bZ_2 & \bZ_2  & \cdots \\ 
0  & 0 & 0 &  0 & 0 & 0  & \cdots \\
\bZ_m \ar[drr]_{d_2^{0,1}} & \bZ_2 \ar[drr]_{d_2^{1,1}} & \bZ_2 \ar[drr]_{d_2^{2,1}}
& \bZ_2 \ar[drr]^{d_2^{4,1}} & \bZ_2 & \bZ_2  & \cdots \\
E_2^{0,0} = \bZ_2 & 0 & \bZ_2 & 0 & \bZ_2 & 0 & \cdots & \xrightarrow{p}}
\\ \hline
\end{tabular}}
\end{center}
~\nl
for $\,m\,$ even. The images of the coboundary homomorphisms $\,d^{p,q}_r\,$
for the second page (the continuous lines) and of the higher ones (the dotted
and dashed lines), together with the definition of the groups
entering next pages, lead us to the conclusion that the LHS
spectral sequence stabilizes quickly for the cohomology groups
of interest: the classifying group $\,H^2(\G,U(1)_\epsilon)$, \,and the 
obstruction group $\,H^3(\G,U(1)_\epsilon)$. 
\vskip 0.2cm

For $\,m\,$ odd, taking into account that there are no non-trivial 
homomorphisms from 
$\,\bZ_m\,$ to $\,\bZ_2$, \,we conclude that the sequence collapses
to the second page giving
\qq
H^n(\bZ_2\lx \bZ_m,U(1)_\epsilon)\ =\ \left\{ \barr{ccl}
\bZ_2 & \quad \tx{if} & \quad n\ \ \tx{is\ \,even},\\
\bZ_{m} & \quad \tx{if} & \quad n\ \ \tx{is\ \,odd}.
\earr \right.
\nonumber
\qqq
The case of $\,m\,$ even is somewhat more complicated. We shall argue
that $\,d_2^{0,1}=0\,$ also in this case. It is shown in
\cite{sah} that for $\,\G=\bZ_2\lx Z\,$ there exists a $\,7$-term 
exact sequence :
\vskip 0.6cm

\xymatrix{\quad 0 \ar[r] & H^1(\bZ_2,H^0(Z,U(1))_\epsilon) 
\ar[r] & H^1(\G,U(1)_\epsilon)
\ar[r]^-{\rho} & H^0(\bZ_2,H^1(Z,U(1))_\epsilon) \ar@(d,u)[dll]^{d_2^{0,1}}
\\
& H^2(\bZ_2,H^0(Z,U(1))_\epsilon) \ar[r] & H^2(\G,U(1)_\epsilon)_1 
\ar[r]
& H^1(\bZ_2,H^1(Z,U(1))_\epsilon) \ar@(d,u)[dll]^{d_2^{1,1}} \\
& H^3(\bZ_2,H^0(Z,U(1))_\epsilon)}

\vskip 0.6cm
\noindent with $\,\rho\,$ denoting the restriction map and 
$\,H^2(\G,U(1)_\epsilon)_1\,$
entering the exact sequence
\qq
0\,\longrightarrow\,H^2(\G,U(1)_\epsilon)_1\,\longrightarrow\,
H^2(\G,U(1)_\epsilon)
\,\longrightarrow\,H^2(Z,U(1))^{\bZ_2}
\nonumber
\qqq  
where the last group is the subgroup of $\,\bZ_2$-invariant elements
of $\,H^2(Z,U(1))$. \,Since every 1-cocycle $\,w_z\,$ on $\,Z\,$ with values 
in $\,U(1)\,$ (i.e. a character of $\,Z$) \,extends to a 1-cocycle
on $\,\G\,$ with values in $\,U(1)_\epsilon\,$ upon setting $\,w_{z_0z}
=w_z^{-1}$, \,the restriction map is surjective. Besides, $\,H^2(Z,U(1))=0\,$
for $\,Z\,$ cyclic  
so that $\,H^2(\G,U(1)_\epsilon)_1\cong H^2(\G,U(1)_\epsilon)\,$
in this case. \,If 
$\,Z=\bZ_m\,$ with $\,m\,$ even then 
\qq
&&H^1(\bZ_2,H^0(Z,U(1))_\epsilon)\,=\,H^1(\bZ_2,U(1)_\epsilon)\,=0\,,\cr
&&H^0(\bZ_2,H^1(Z,U(1))_\epsilon)\,=\,H^0(\bZ_2,Z)\,=\,Z\,,\cr
&&H^2(\bZ_2,H^0(Z,U(1))_\epsilon)\,=\,H^2(\bZ_2,U(1)_\epsilon)\,=\,\bZ_2\,,\cr
&&H^1(\bZ_2,H^1(Z,U(1))_\epsilon)\,=\,H^1(\bZ_2,Z)\,=\,\bZ_2\,,\cr
&&H^3(\bZ_2,H^0(Z,U(1))_\epsilon)\,=\,H^3(\bZ_2,U(1)_\epsilon)\,=\,0
\nonumber
\qqq
so that the $\,7$-term exact sequence reduces to
\qq
0 \,\longrightarrow\, H^1(\Gamma,U(1)_\epsilon) 
\mathop{\,\longrightarrow\,}\limits^{\rho} Z 
\mathop{\,\longrightarrow\,}\limits^{d_2^{0,1}} \bZ_2 \,\longrightarrow\,
 H^2(\G,U(1)_\epsilon)) \,\longrightarrow\, \bZ_2
\mathop{\,\longrightarrow\,}\limits^{d_2^{1,1}} 0\,.
\nonumber
\qqq
It follows, in particular, that $\,\rho\,$ is an isomorphism and $\,d_2^{0,1}=0$. 
\,Finally, using this information 
in the LHS spectral sequence, we infer that for $\,m\,$ even,   
\bgt\label{cobstrodd}
H^n(\bZ_2\lx \bZ_m, U(1)_\epsilon) = \left\{ \barr{ccl}
\bZ_2 & \quad \tx{if} & \quad n = 0\,, \\
\bZ_{m} & \quad \tx{if} & \quad n = 1\,, \\
\bZ_4 \quad \tx{or} \quad \bZ_2 \times \bZ_2 & \quad \tx{if} & \quad n =
2\,, \\ 
 \bZ_{2k}\ \ \tx{or}\quad \bZ_2 \times \bZ_k\,, \quad 
\frac{m}{k}=1,2,4, & \quad \tx{if} & \quad n = 3\,.
\earr \right.
\end{gather}
The ambiguity in \eqref{cobstrodd} can actually be resolved for the 
group $\,H^2(\bZ_2\lx \bZ_{m},U(1)_\epsilon)$. \,Indeed, consider
its element defined by the cocycle
\beq
v^{(1)}_{z_0^{n}z,z_0^{n'}z'} = (-1)^{nn'}
\label{1stcoc}
\eeq
for $\,z,z'\in\bZ_m$. \,Suppose 
that $\,v^{(1)}\,$ is a coboundary, 
$\,v^{(1)}_\gamma=(\d w)_\gamma$, \,from which it would follow, in particular, 
that
\beq
w_{z_0z'}\,(w_{z_0z^{-1}z'})^{-1}w_{z}
\,=\,1 \qquad {\rm and} \qquad (w_{z_0z'})^{-1}(w_{z^{-1}z'})^{-1}
w_{z_0z}\,=\,-1.
\eeq
The two conditions are, however, contradictory as can be verified
by replacing $\,z\,$ by $\,z^{-1}z'\,$ in the second one.
Hence the class $\,[v^{(1)}_{\gamma,\gamma'}]\,$ 
generates a $\bZ_2$ subgroup of
$\,H^2(\bZ_2\lx \bZ_{m},U(1)_\epsilon)$. \,For $\,m\,$ odd,
this is the whole group but for $\,m\,$ even we may repeat the same 
reasoning \wrt the $2$-cocycle
\beq
v^{(2)}_{z_0^{n}z,z_0^{n'}z'} = \begin{cases}\ 1\ \quad \tx{if}
\quad (z')^{m/2}=1\,,\cr
\ (-1)^n \quad \tx{if} \quad (z')^{m/2}\not=1 
\end{cases}
\label{2ndcoc}
\eeq
and $\,v^{(2)}_{\gamma,\gamma'}(v^{(1)}_{\gamma,\gamma'})^{-1}$, 
\,establishing that the class $[v^{(2)}_{\gamma,\gamma'}]$ in 
$\,H^2(\bZ_2\lx \bZ_{m},U(1)_\epsilon)\,$ is non-trivial and
different from $\,[v^{(1)}_{\gamma,\gamma'}]$. 
\,This immediately implies that
\beq\label{H2Z2}
H^2(\bZ_2\lx \bZ_{m},U(1)_\epsilon)\ =\ \bZ_2\times\bZ_2
\eeq
for $\,m\,$ even and it is generated by the cohomology classes
$\,[v^{(1)}_{\gamma,\gamma'}]\,$ and $\,[v^{(2)}_{\gamma,\gamma'}]$. 
\vskip 0.2cm

Finally, we give, for the sake of completeness, the classifying cohomology
group for the case of the non-cyclic orbifold subgroup $\,\bZ_2 \x \bZ_2\,$
that will be encountered in the study of the Cartan series $\,D_{2s}\,$ 
of simple groups. Since
\qq
H^q(\bZ_2\times\bZ_2,U(1))\,=\,\begin{cases}\,\hbox to 2cm{$U(1)$\hfill}
\tx{if}\quad q=0,\cr
\,\hbox to 2cm{$\bZ_2\times\bZ_2$\hfill}\tx{if}\quad q=1,\cr
\,\hbox to 2cm{$\bZ_2$\hfill}\tx{if}\quad q=2,\end{cases}
\nonumber
\qqq
see \cite{GR2}, in the LHS sequence
\qq
E^{0,2}\,=\,\bZ_2\,,\qquad E^{1,1}\,=\,\bZ_2\times\bZ_2\,,
\qquad E^{2,0}\,=\,\bZ_2\,.
\nonumber
\qqq
It follows that $\,H^2(\bZ_2\hspace{-0.03cm}\ltimes\hspace{-0.03cm}
(\bZ_2 \x \bZ_2),U(1)_\epsilon)\,$ has rank smaller or equal to 16.
In Sect.\,\ref{sec:Dre}, we shall exhibit 16 cohomologically inequivalent 
2-cocycles on $\,\bZ_2\hspace{-0.03cm}\ltimes\hspace{-0.03cm}
(\bZ_2 \x \bZ_2)\,$ taking values $\,\pm1$. \,This will establish 
the equality
\beq\label{H2Z2Z2}
H^2(\bZ_2\hspace{-0.03cm}\ltimes\hspace{-0.03cm}
(\bZ_2 \x \bZ_2),U(1)_\epsilon)\ =\ \bZ_2\times\bZ_2\times\bZ_2\times\bZ_2\,.
\eeq
\vskip 0.2cm

Prior to refining the tools of analysis of the obstruction cocycles,
let us make one general comment. In the cyclic case $\,Z=\bZ_m\,$, 
the (possibly non-factorizing) component of the obstruction group 
$\,H^3(\Gamma,U(1)_\epsilon)\,$ coming from  $\,H^0(\bZ_2,H^3(\bZ_m,
U(1))_\epsilon)\,$ is determined completely by the orbifold subgroup. 
After imposing conditions that allow a trivialization 
of the obstruction cocycle $\,u_{z,z',z''}\,$ restricted to $\,Z$, 
we are left with at most a sign obstruction. We shall encounter 
such sign obstructions for $\,m\,$ even in the cases considered below. 
On the other hand, for $\,m\,$ odd, the restriction to $\,Z\,$ is 
clearly the sole source of obstruction. In particular, there are 
no obstructions to the trivializability of $\,u_{\gamma,\gamma,\gamma''}\,$ 
for trivial $\,Z\,$ and, consequently, no obstruction to existence 
of Jandl structures on the gerbes on simply connected groups $\,G$.

\section{Case-by-case analysis}\label{sec:case}
\noindent Trivializability of the obstruction 3-cocycle 
$\,u_{\gamma,\gamma',\gamma''}\,$ given by (\ref{g3cocyc})
constrains the admissible values of the level $\,\sfk\,$ in terms 
of the other elements such as the structure of the group $\,G$, 
the choice of the orbifold subgroup $\,Z\subset Z(G)$, \,and that of 
the twist element $\,\zeta\in Z\,$ entering the action (\ref{tw}) 
on $\,G\,$ of the $\,\bZ_2\,$ component of the orientifold
group $\,\Gamma=\bZ_2\lx Z$. \,Below, we shall calculate the cocycles
$\,u_{\gamma,\gamma',\gamma''}\,$ on $\,\G\,$ and classify 
the cases when they may be trivialized, giving also an explicit form
of trivializing cochains. The latter provide the main input
in the explicit construction of $\,\Gamma$-equivariant structures
on the level $\,\sfk\,$ gerbe on $\,G\,$ that will be
described in \cite{GSW2}. Cohomologically inequivalent 
trivializing cochains give rise to inequivalent 
$\,\Gamma$-equivariant structures. The construction of 
\cite{GSW2} is a direct generalization of the one 
of gerbes on non-simply connected groups $\,G/Z\,$ described 
in \cite{GR2}. 
\vskip 0.2cm

Below, we shall denote by $\,z\,$ a fixed generator of $\,Z(G)\,$
for the groups $\,G\,$ with cyclic center $\,Z(G)$.  
\,The essential input in the calculations of the obstruction cocycle
$\,u_{\gamma,\gamma',\gamma''}\,$ is the identification of the
elements $\,w_z\,$ and $\,w_\kappa\,$ in the normalizer $\,N(T)\,$
of the Cartan subgroup of $\,T\subset G\,$ and of the maps
$\,\tau\mapsto z\tau\,$ and $\,\tau\mapsto\kappa z\,$ of the positive
Weyl alcove that satisfy (\ref{zact}) and (\ref{kact}).
To simplify the notation, we shall abbreviate $\,z^n\equiv n\,$ 
and $\,z_0z^n\equiv\un{n}$, \,where $\,n=0,\dots,|Z(G)|-1\,$ for 
the elements of the maximal orientifold group $\,\G=\bZ_2\lx Z(G)$. 
\,For any integer $\,n$, \,we shall denote by $\,[n]\,$ the number equal 
to $\,n\,$ modulo $\,|Z(G)|\,$ and such that $\,0\leq[n]<|Z(G)|$. 
\,In accordance with (\ref{wg}), for the general elements 
$\,\gamma=z^n,z_0z^n\,$ of $\,\Gamma\,$ with $\,n=0,\dots,|Z(G)|-1$, 
\,one may set:
\qq
w_n\equiv w_{z^n}\,=\,w_z^n,\qquad w_{\un{n}}\equiv w_{z_0z^n}\,=\,
w_\kappa w_z^{n_0} w_z^n
\label{wnn}
\qqq
if the twist element $\,\zeta\,$ entering the action 
(\ref{tw}) of $\,z_0\,$ on $\,G\,$ is equal to $\,z^{n_0}\equiv n_0$. 
\,With these choices of $\,w_\gamma$, \,the calculation of the
obstruction cocycle $\,u_{\gamma,\gamma',\gamma''}\,$
will follow (\ref{gggb}) and (\ref{g3cocyc}). 
For smaller orientifold groups $\,\G=\bZ_2\lx Z\subset\bZ_2\lx Z(G)\,$ 
with $\,Z\subset Z(G)$, \,the obstruction cocycle will be obtained
by restriction of the one for the maximal $\,\G$.
\vskip 0.2cm

Obstructions to existence of a trivializing cochain
coming from the orbifold subgroup $\,Z\subset\Gamma\,$ were
analyzed in \cite{GR2}. To look for a further obstruction associated
with the subgroup $\,\bZ_2\lx\bZ_2\subset\Gamma\,$ 
for $\,Z(G)\,$ cyclic with $\,|Z(G)|\,$ even, we shall consider, 
for $\,n={1\over2}|Z(G)|$, \,the combination
\qq
X\,=\,u_{\un{n},n,\un{0}}^{-2}\,
u_{0,\un{n},\un{n}}^2\,
u_{\un{0},0,\un{0}}^2\,
u_{0,\un{0},\un{0}}^{-2}\,
u_{\un{n},\un{n},\un{n}}^{-1}\,
u_{\un{n},0,\un{n}}^{-1}\,
u_{n,\un{0},n}u_{\un{0},n,n}\,
u_{n,n,\un{0}}^{-1}\,
u_{\un{0},\un{0},\un{0}}
\label{loeq}
\qqq
By direct substitution, one may check that $\,X=1\,$ if
$\,u_{\gamma,\gamma',\gamma''}\,$ satisfies (\ref{cohoeq})
(or its restriction to $\,\bZ_2\lx\bZ_2$). 
In a few cases, this equality will impose further non-trivial conditions
on existence of trivializing cochains.  
\vskip 0.2cm

\,Recall from Sect.\,\ref{sec:cohocon} that, for $\,Z\,$ cyclic, the 
cohomology groups $\,H^2(\bZ_2\lx Z)=\bZ_2\,$ if the rank $\,|Z|\,$ 
of $\,Z\,$ is odd, and $\,H^2(\bZ_2\lx Z)=\bZ_2\times\bZ_2\,$ if $\,|Z|\,$ 
is even. If a 2-cochain $\,v_{\gamma,\gamma'}\,$ solves the cohomological
equation (\ref{cohoeq}) then 
\qq
v_{\gamma,\gamma'}\quad{\rm and}\quad 
v_{\gamma,\gamma'}\,v^{(1)}_{\gamma,\gamma'}
\nonumber
\qqq 
give two cohomologically inequivalent solutions if the rank 
of $\,|Z|\,$ is odd and 
\qq
v_{\gamma,\gamma'},\quad v_{\gamma,\gamma'}\,v^{(1)}_{\gamma,\gamma'},
\quad v_{\gamma,\gamma}\,v^{(2)}_{\gamma,\gamma'}\quad
v_{\gamma,\gamma'}\,v^{(1)}_{\gamma,\gamma'}\,v^{(2)}_{\gamma,\gamma'}
\nonumber
\qqq 
give four cohomologically inequivalent solutions if 
$\,|Z|\,$ is even. All other solutions of (\ref{cohoeq}) 
differ from those by 2-coboundaries. In the notation
introduced above, the 2-cocycles $\,v^{(1)}_{\gamma,\gamma'}\,$ 
and $\,v^{(2)}_{\gamma,\gamma'}\,$ of (\ref{1stcoc}) and (\ref{2ndcoc})
read
\qq
&&v^{(1)}_{n,n'}\,=\,v^{(1)}_{n,\un{n}'}\,=\,v^{(1)}_{\un{n},n'}\,=\,1,
\quad v^{(1)}_{\un{n},\un{n}'}\,=\,-1,\label{v11}\\
&&v^{(2)}_{n,n'}\,=\,v^{(2)}_{n,\un{n}'}\,=\,1,\quad v^{(2)}_{\un{n},n'}\,
=\,v^{(2)}_{\un{n},\un{n}'}\,=\,(-1)^{n'|Z|/|Z(G)|}.\label{v22}
\qqq 
\vskip 0.1cm

\subsection{The case of $\,G=A_r={SU}(r+1)$}\label{sec:Ar}
\noindent The Lie algebra $\,\Ng=\gt{su}(r+1)\,$ is composed of 
the hermitian traceless $\,(r+1)\times(r+1)$-matrices. 
The Cartan algebra $\,\Nt\subset\Ng\,$ is chosen in the standard
way as composed from the diagonal matrices in $\,\Ng$.
We shall denote by $\,e_i$, $i=1,\dots,r+1$, \,the diagonal 
matrices with the matrix elements $\,(e_i)_{j,j'}
=\delta_{i,j}\delta_{i,j'}$. \,The scalar product 
$\,\tr\,e_ie_{i'}=\delta_{i,i'}\,$ defines the Killing form on
$\,\Nt\,$ with the required normalization. 
\,The center $\,Z(G)\cong\bZ_{r+1}\,$ is generated by the element 
$\,z=\ee^{-2\pi\sfi\,\lambda_r^\vee}=\ee^{-{2\pi\sfi\over r+1}}$, \,with
the simple (co)weights 
\qq
\lambda_i^\vee\,=\,\sum\limits_{j=1}^i e_j-{_i\over^{r+1}}
\sum\limits_{j=1}^{r+1}e_j
\nonumber
\qqq
corresponding to the standard choice 
of the simple (co)roots\footnote{We shall always identify the 
Cartan algebra $\,\Nt\,$ with its dual using the Killing form $\,\tr$.} 
$\,\alpha_i=e_i-e_{i+1}$.
\,The positive Weyl alcove $\,\CA\,$ is the simplex in $\,\Nt\,$
with the vertices $\,\tau_0=0\,$ and $\,\tau_i=\lambda_i^\vee$.
\,For $\,\tau\in\CA$, \,the relations (\ref{zact}) and (\ref{kact}) 
hold for
\qq
(w_z)_{j,j'}\,=\,\ee^{\frac{\pi\sfi\,r}{r+1}}\,\delta_{j-1,[j']}
\,,\qquad 
(w_\kappa)_{j,j'}\,=\,\ee^{\frac{\pi\sfi\,r}{2}}\,\delta_{j,r+2-j'}
\nonumber
\qqq
and for the transformations of the positive Weyl alcove acting on the 
vertices of $\,\CA\,$ by
\qq
z\tau_i\,=\,\tau_{[i+1]}\,,\qquad\kappa\tau_i\,=\,\tau_{[r-i+1]}\,.
\nonumber
\qqq
The corresponding index transformations induce, respectively, the 
symmetry of the extended Dynkin diagram and the symmetry of the 
unextended one that are depicted in Fig.\ref{fig:Ar} 
and Fig.\ref{fig:ArK}.

\FIGURE[!h]{
\epsfxsize = 3in
\epsfysize = 1.2in
\centerline{
\vspace{0.5cm}
\epsfbox{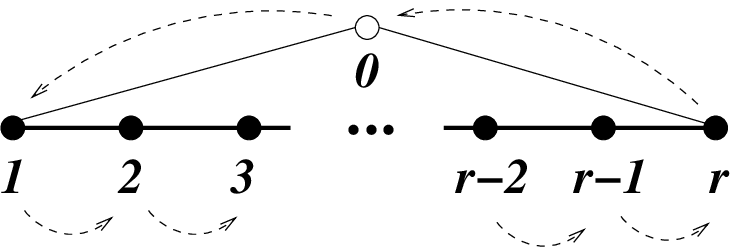}}
\vspace{-0.2cm}
\caption{{\sf The rotation of the extended Dynkin diagram of $A_r$ under
$z$.}}
\label{fig:Ar}}
\FIGURE[!h]{
\epsfxsize = 3in
\epsfysize = 1.5in
\centerline{
\epsfbox{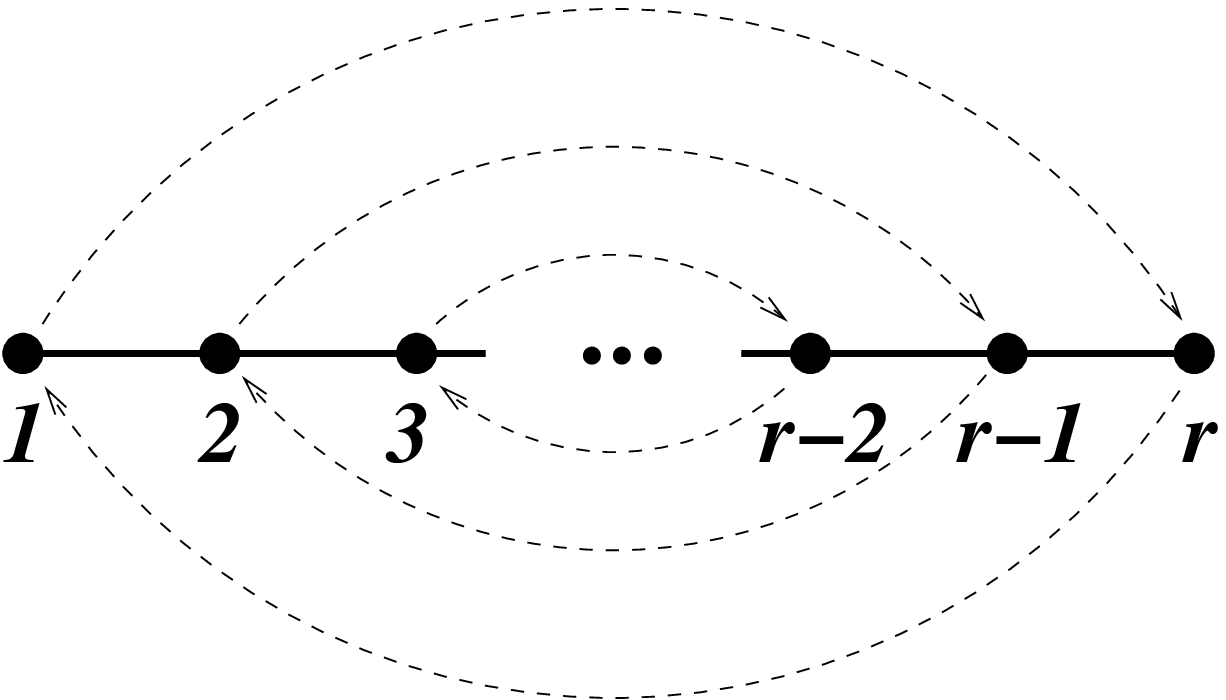}}
\vspace{-0.2cm}
\caption{{\sf The reflection of the Dynkin diagram of $A_r$ under 
$\kappa$.}}
\label{fig:ArK}}
\noindent For the maximal orientifold group 
$\,\Gamma=\bZ_2\lx\bZ_{r+1}\,$ with 
the action of the generator $\,z_0\,$ of $\,\bZ_2\,$ given by
(\ref{tw}), we shall define $\,w_n\,$ and $\,w_{\un{n}}\,$
according to (\ref{wnn}).
To satisfy the relation (\ref{gggb}) for $\,\gamma,\gamma'=n,{\un n}$,
\,one may take
\qq
&&b_{n,n'}\ =\ b_{\un{n},n'}\ =\ \left\{ \barr{lcl}\hbox to 2.92cm{$
0$\hfill}\tx{if} \quad  n + n' < r + 1\,, \\
\hbox to 2.92cm{$\frac{r(r+1)}{2} \la^\vee_r$\hfill}\tx{if} 
\quad  n + n' \geq r + 1\,,\earr \right.
\\ \non
&&b_{n,\un{n'}}\ =\ \left\{ \barr{lcl}
\hbox to 4.5cm{$r n \la^\vee_r$\hfill}\tx{if} \quad  n' \geq n\,, \\
\hbox to 4.5cm{$r \lb n + \frac{r+1}{2} \rb \la^\vee_r$\hfill}\tx{if} 
\quad  n' < n\,,\earr \right. \\ \non
&&b_{\un{n},\un{n'}}\ =\ \left\{ \barr{lcl}\hbox to 4.5cm{$
r \lb n + n_0 + \frac{r+1}{2} \rb \la^\vee_r$\hfill}\tx{if} \quad  n'
\geq n\,, \\ 
\hbox to 4.5cm{$r \lb n + n_0 \rb \la^\vee_r$\hfill}\tx{if} 
\quad  n' < n\,.\earr \right. 
\nonumber
\qqq
Using the identity
\beq
\tr \lb \la^\vee_i \la^\vee_r \rb
= \frac{i}{r+1}\,,
\eeq
one obtains from (\ref{g3cocyc}) the explicit form of the
obstruction cocycle on the group
$\,\G=\bZ_2\lx\bZ_{r+1}$: 
\qq
&&u_{n,n',n''}\ =\ \Phi^{\,n\,\frac{n'+n'' -
[n'+n'']}{r+1}}\ =\ u_{n,\un{n}',n''}\,, \label{u1}\\ \non  
&&u_{\un{n},n',n''}\ =\ 
\Phi^{\,(n_0+n)\,\frac{n'+n'' -
[n'+n'']}{r+1}}\ =\ u_{\un{n},\un{n}',n''}\,, \\ \non
&&u_{n,n',\un{n}''}\ =\ 
\Phi^{\,n\,\frac{n''-n' -
[n''-n']}{r+1}}\,\Psi^{-nn'}\,, \\ \non
&&u_{\un{n},n',\un{n}''}\ =\ 
\Phi^{\,(n_0+n)\,\frac{n''-n' -
[n''-n']}{r+1}}\,\Psi^{(n_0+n)n'}\,, \\ \non 
&&u_{n,\un{n}',\un{n}''}\ =\ 
\Phi^{\,n\,\left(1+\frac{n''-n' -
[n''-n']}{r+1}\right)}\,\Psi^{-n(n_0+n')}\,, \\ \non
&&u_{\un{n},\un{n}',\un{n}''}\ =\ 
\Phi^{(n_0+n)\,\left(1+\frac{n''-n' -
[n''-n']}{r+1}\right)}\,\Psi^{(n_0+n)(n_0+n')}\,, \label{u2}
\qqq
where $\,\Phi\equiv(-1)^{\sfk r}\,$ and 
$\,\Psi\equiv\ee^{\frac{2\pi\sfi\sfk}{r+1}}$.
\,For the case when $\,\Gamma=\bZ_2\lx\bZ_m\,$ for a proper
subgroup $\,\bZ_m\subset\bZ_{r+1}\,$ composed 
of elements $\,n\,$ that are multiples of $\,\frac{r+1}{m}$, 
\,the obstruction cocycle is obtained by restriction 
of $\,n,n'\,$ and $\,n''\,$ to such values.
\vskip 0.2cm

A necessary condition for the solvability 
of the cohomological equation (\ref{gcohoeq}) is the triviality of
the cohomology class $\,[u_{n,n',n''}]\in H^3(Z,U(1))$. \,This
condition was analyzed in \cite{GR1,GR2} where it was shown
that it implies that
\qq
\sfk\quad\tx{is \,even}\quad\tx{if}\quad m\quad\tx{is \,even}\quad
\tx{and}\quad\frac{_{r+1}}{{^m}}\quad\tx{is \,odd}.
\label{orbcond}
\qqq
The latter restriction means that $\,\sfk r\frac{r+1}{m}\,$ is even
so that the factors $\,\Phi^n\,$ in the expression
for $\,u_{\gamma,\gamma',\gamma''}\,$ may be replaced 
by $\,1\,$ and that $\,u_{n,n',n''}\equiv1$, \,in particular. 
\vskip 0.2cm

Another necessary condition for the solvability
of (\ref{gcohoeq}) is the trivializability of the
restriction of $\,u_{\gamma,\gamma',\gamma''}\,$ 
to $\,\bZ_2\subset\G$,
\,i.e. to $\,\gamma,\gamma',\gamma''=0,\un{0}$.
\,This, however, always holds because of the triviality of the
cohomology group $\,H^3(\bZ_2,U(1)_\epsilon)$, 
\,see (\ref{peps}). The 2-cochain on 
$\,\bZ_2\,$ which trivializes the restricted 3-cocycle is
\qq
\tilde v_{0,0}\,=\,\tilde v_{\un{0},0}\,=\,\tilde v_{0,\un{0}}\,=\,1\,,\qquad 
\tilde v_{\un{0},\un{0}}\,
=\,\pm\,\Psi^{-\frac{1}{2}n_0\left(n_0+\frac{r(r+1)}{2}\right)}\,,
\nonumber
\qqq
with the two signs giving cohomologically inequivalent 2-cochains. All
other trivializing 2-cochains differ from them by 2-coboundaries 
(recall that $\,H^2(\bZ_2,U(1)_\epsilon)=\bZ_2$). \,Note again that
the triviality of $\,H^3(\bZ_2,U(1)_\epsilon)\,$ implies 
that if the orbifold subgroup $\,Z\,$ is trivial then the cohomological 
equation (\ref{gcohoeq}) is always solvable.  
\vskip 0.2cm

Returning to the case of non-trivial $\,Z\cong\bZ_{m}$, \,further 
simplification of the 3-cocycle (\ref{u1})-(\ref{u2}) may
be achieved by extracting from it the coboundary 
$\,(\delta v')_{\gamma,\gamma',\gamma''}\,$ for
\qq
&&\hbox to 4cm{$v'_{n,n'}\,=\,\Psi^{nn'}\,,\hfill$}
v'_{\un{n},n'}\,=\,\Psi^{-nn'}\,,\label{coc'a}\\
&&\hbox to 4cm{$v'_{n,\un{n}'}\,=\,\Psi^{nn'}c_n\,,$\hfill}
v'_{\un{n},\un{n}'}\,=\,\pm\,\Psi^{-\frac{1}{2}n_0\left(n_0
+\frac{r(r+1)}{2}\right)}\,\Psi^{-n(n_0+n')}\,\,c_n^{-1}\,,
\label{coc'b}
\qqq
where $\,c_n=\Psi^{-\frac{1}{2}\left(n^2+(r+1)n\right)}\,$ satisfies
\qq
c_n\,c_{[n+n']}^{-1}\,c_{n'}\,=\,\Psi^{nn'}\,, \qquad c_{[-n]}\,=\,c_n\,.
\label{ccc}
\qqq 
Note that the lift of the 2-cochain $\,\tilde v\,$ to $\,\G\,$ appears as
an explicit factor in $\,v'$. \,Writing
\qq
u_{\gamma,\gamma',\gamma''}\,=\,u'_{\gamma,\gamma',\gamma''}\,
(\delta v')_{\gamma,\gamma',\gamma''}\,,
\nonumber
\qqq
we obtain the following formulae by a straightforward calculation using  
(\ref{coc'a})-(\ref{ccc}):
\qq
&&u'_{n,n',n''}\,=\,u'_{n,\un{n}',n''}\,=\,u'_{n,n',\un{n}''}\,=\,
u'_{n,\un{n}',\un{n}''}\,=\,1\,,\cr\cr
&&u'_{\un{n},n',n''}\,=\,u'_{\un{n},\un{n}',n''}\,
=\,\Phi^{\,n_0\,\frac{n'+n''-[n'+n'']}{r+1}}\,,\cr\cr
&&u'_{\un{n},n',\un{n}''}\,=\,u'_{\un{n},\un{n}',\un{n}''}\,
=\,\Phi^{\,n_0\,\frac{n''-n'-[n''-n']}{r+1}}\,.
\nonumber
\qqq
If $\,m\,$ is odd then the cocycle $\,u'_{\gamma,\gamma',\gamma''}\,$
may be trivialized by setting
\qq
v''_{n,n'}\,=\,v''_{n,\un{n}'}\,=\,1\,,\qquad
v''_{\un{n},n'}\,=\,v''_{\un{n},\un{n}'}\,=\,\Phi^{\,n_0\,\frac{mn'}{r+1}}.
\label{coc''}
\qqq
Indeed, using the fact that $\,\Phi=\pm1\,$ and that
\qq
\Phi^{\,n_0\,\frac{mn}{r+1}}\,\Phi^{-n_0\,\frac{m[n+n']}{r+1}}\,
\Phi^{\,n_0\,\frac{mn'}{r+1}}\,=\,\Phi^{\,\frac{n+n'-[n+n']}{r+1}},
\nonumber
\qqq
for $\,m\,$ odd, one easily verifies that
\qq
u'_{\gamma,\gamma',\gamma''}\,=\,(\delta v'')_{\gamma,\gamma',\gamma''}\,.
\nonumber
\qqq
If $\,m\,$ is even and $\,\frac{r+1}{m}\,$ is odd then the condition
(\ref{orbcond}) implies that $\,\Phi=1\,$ so that the cocycle 
$\,u'_{\gamma,\gamma',\gamma''}\,$ is trivial. For $\,m\,$  
and $\,\frac{r+1}{m}\,$ even, however, there exists a further 
obstruction to the trivializability of $\,u'_{\gamma,\gamma',\gamma''}\,$
that is related to the choice of the twist element $\,\zeta={n_0}\,$ in 
the action (\ref{tw}). The analysis of the cohomology 
group $\,H^3(\G,U(1)_\epsilon)\,$ done in Sect.\,\ref{sec:cohocon} showed 
that such an obstruction has to lie in $\,\bZ_2\,$ since the 
part of the obstruction related to the orbifold group has already been 
removed by the condition (\ref{orbcond}). To identify it, we note
that the combination $\,X\,$ of (\ref{loeq}) calculated
for the cocycle $\,u'_{\gamma,\gamma',\gamma''}\,$ and $\,n={r+1\over2}\,$ 
is equal to $\,\Phi^{n_0}\,$ since $\,u'_{\un{0},n,n}\,$ contributes 
the only non-trivial factor to it. \,One obtains this way the equality 
$\,\Phi^{n_0}=1\,$ showing that if $\,u'_{\gamma,\gamma',\gamma''}\,$ 
is a coboundary then 
\qq
\sfk\,n_0\quad\tx{is \,even}\quad\tx{if}\quad m\quad\tx{is\ \,even}\quad
\tx{and}\quad{_{r+1}\over^m}\quad\tx{is\ \,even}.
\label{twcond}
\qqq
In that case, $\,v''_{\gamma,\gamma'}\,$ may be taken trivial
or, which amounts to the same, given by (\ref{coc''}).
Note that $\,\sfk\frac{r+1}{m}\,$ is even for $\,m\,$ even
due to the restriction (\ref{orbcond}) so that the condition 
(\ref{twcond}) holds or fails simultaneously for all $\,n_0\,$ in the
same congruence class modulo $\,\frac{r+1}{m}$, \,in agreement
with the equivalence of the $\,\Gamma\,$ actions for the twist
elements in the same $\,Z$-coset that we discussed in 
Sect.\,\ref{sec:orcLg}. 
\vskip 0.2cm

To summarize, for the orientifold group $\,\G=\bZ_2\lx \bZ_{m}$,
\,the triviality of the cohomology class 
$\,[u_{\gamma,\gamma',\gamma''}]\in H^3(\G,U(1)_\epsilon)\,$ imposes
the conditions (\ref{orbcond}) and (\ref{twcond}). If they are
satisfied then the 2-cochain trivializing $\,u_{\gamma,\gamma',\gamma''}\,$ 
may be taken in the form
\qq
v_{\gamma,\gamma'}\,=\,v'_{\gamma,\gamma'}\,v''_{\gamma,\gamma'}\,,
\nonumber
\qqq
where $\,v'\,$ and $\,v''\,$ are given by (\ref{coc'a})-(\ref{coc'b})
and (\ref{coc''}), respectively. For $\,m\,$ odd,
the two choices of the sign in (\ref{coc'b}) 
give two cohomologically inequivalent trivializing cochains  
from which other trivializing cochains differ by 2-coboundaries.
Indeed, the sign change is induced by multiplication of 
$\,v_{\gamma,\gamma'}\,$ by the 2-cocycle $\,v^{(1)}_{\gamma,\gamma'}\,$ 
given by (\ref{v11}). For $\,m\,$ even, further two cohomologically 
inequivalent solutions are obtained by additionally multiplying 
$\,v_{\gamma,\gamma'}\,$
by the 2-cocyle $\,v^{(2)}_{\gamma,\gamma'}\,$ given by (\ref{v22}).
\vskip 0.2cm

Let us illustrate how the above analysis provides concrete information about
the numbers of inequivalent orientifold gerbes on a few examples of 
the $\,A_r\,$ groups of low ranks.
\vskip 0.1cm

For $\,G=SU(2)$, \,if $\,\Gamma=\bZ_2\,$
with its generator acting by (\ref{tw}) then there are two 
inequivalent $\,\G$-equivariant (or Jandl) structures on the 
gerbe of level $\,\sfk\,$ on $\,SU(2)\,$ for each 
$\,\sfk\,$ and each of the two choices of the twist element 
$\,\zeta$. \,For $\,\G=\bZ_2\lx \bZ_2\,$ with the second 
factor being the center of $\,SU(2)$, \,the condition (\ref{orbcond})
imposes that the level $\,\sfk\,$ be even. For each of the
two choices of the twist element $\,\zeta$, \,there are then 4 inequivalent 
$\,\Gamma$-equivariant structures. The different choices of the twist 
element lead to equivalent actions of $\,\Gamma\,$ on $\,SU(2)\,$ 
and there are altogether 4 inequivalent 
Jandl structures on the induced gerbe on $\,SO(3)$, \,see 
the discussion in Sect.\,\ref{sec:orcLg}. These results are in agreement 
with the analysis of refs.\,\cite{SSW} and \cite{PSS95a,PSS95b}.  
\vskip 0.1cm 

For $\,G=SU(3)$, \,there are no obstructions. There are two inequivalent 
$\,\G$-structures on the gerbe on $\,G\,$ for $\,\G=\bZ_2\,$ or
$\,\G=\bZ_2\lx \bZ_3\,$ for each level 
$\,\sfk\,$ and each of the three choices of the twist element. 
For $\,\G=\bZ_2\lx \bZ_3$, \,different choices of the
twist element lead to equivalent $\,\Gamma$-actions. Consequently,
there are, altogether, two inequivalent Jandl structures on 
the induced gerbe on the quotient group $\,SU(3)/\bZ_3\,$ for
each $\,\sfk$.  
\vskip 0.1cm

For $\,G=SU(4)$, \, there are two inequivalent Jandl structures 
on the gerbe on $\,G\,$ for each level $\,\sfk\,$ and each
of the four choices of the twist element. For 
$\,\G=\bZ_2\lx \bZ_2$, \,there are four inequivalent 
$\,\Gamma$-equivariant structures for each $\,\sfk\,$ even 
and each choice of the twist element
in $\,\bZ_4$, \,and for each $\,\sfk\,$ odd and each 
twist element in $\,\bZ_2\subset\bZ_4$. \,There are no
$\,\Gamma$-equivariant structures for $\,\sfk\,$ odd and
twist elements in $\,\bZ_4\setminus\bZ_2$. \,We get
this way eight inequivalent Jandl structures on the induced
gerbe on the quotient group $\,SU(4)/\bZ_2\,$ if
$\,\sfk\,$ is even and four if $\,\sfk\,$ is odd.
\,Finally, if $\,\G=\bZ_2\lx \bZ_4\,$ there are
four inequivalent $\,\Gamma$-equivariant structures for
$\,\sfk\,$ even and each of the four choices of the twist
element. Overall, they give rise to four inequivalent Jandl
structures on the induced gerbe on $\,SU(4)/\bZ_4$. \,There
are no $\,\Gamma$-equivariant structures for $\,\sfk\,$ odd.

\subsection{The case of $\,G=B_r = {Spin}(2r+1)$}\label{sec:Br}
\noindent The Lie algebra $\,\Ng=\gt{spin}(2r+1)\,$ is composed 
of the imaginary antisymmetric $\,(2r+1)\times(2r+1)$-matrices. We shall 
denote by $\,e_i$, $i=1,\dots,r$,  \,the matrices with the matrix elements 
$\,(e_i)_{j,j'}=\sfi(\delta_{j,2i}\delta_{2i-1,j'}-\delta_{j,2i-1}
\delta_{2i,j'})\,$ that span the Cartan algebra $\,\Nt\subset\Ng$.
\,The Killing form is normalized so that $\,\tr\,e_ie_{i'}=\delta_{i,i'}$. 
\,The center is $\,Z(G)\cong\bZ_2\,$ with the generator $\,z=\ee^{-2\pi\sfi\,
\lambda^\vee_1}$, \,where
\qq
\lambda_i^\vee\,=\,\sum\limits_{j=1}^ie_i
\nonumber
\qqq
are the simple coweights corresponding to the simple roots 
$\,\alpha_i=e_i-e_{i+1}$ for $i=1,\dots,r-1\,$ and $\,\alpha_r=e_r$.
\,We have $\,Spin(2r+1)/\bZ_2=SO(2r+1)$. \,The vertices of the positive Weyl
alcove are $\,\tau_0=0$, $\,\tau_1=\lambda_1^\vee\,$ and
$\,\tau_i=\frac{1}{2}\lambda_i^\vee\,$ for $\,i=2,\dots,r$.
\,For $\,\tau\in\CA$, \,the relations (\ref{zact}) and (\ref{kact}) 
hold for $\,w_z\,$ and $\,w_\kappa\,$ that project to
the $\,SO(2r+1)\,$ matrices
\qq
&&(w_z)_{j,j'}\,=\,-(-1)^{\delta_{1j}}\delta_{j,j'}\,,\\ 
&&(w_\kappa)_{j,j'}\,=\,\sum\limits_{i=1}^r(\delta_{j,2i}\delta_{2i-1,j'}
+\delta_{j,2i-1}\delta_{2i,j'})\,+\,(-1)^r\delta_{j,2r+1}\delta_{2r+1,j'}\,,
\nonumber
\qqq
and for the transformations of the positive Weyl alcove acting on the 
vertices of $\,\CA\,$ by
\qq
z\tau_0\,=\,\tau_{1}\,,\quad z\tau_1\,=\,0\,,\quad z\tau_i\,=\,\tau_i
\quad{\rm for}\quad i=2,\dots,r\,,\qquad\kappa\tau_i\,=\,\tau_{i}\quad
\tx{for}\quad i=0,\dots,r\,.
\nonumber
\qqq
The symmetry of the extended Dynkin diagram corresponding to the
index transformations under $\,z\,$ is represented in Fig.\ref{fig:Br}.
The index transformation under 
$\,\kappa\,$ induces a trivial symmetry of the Dynkin diagram.
\vspace{0.6cm}
\FIGURE[!h]{
\epsfxsize = 3in
\epsfysize = 0.8in
\centerline{
\epsfbox{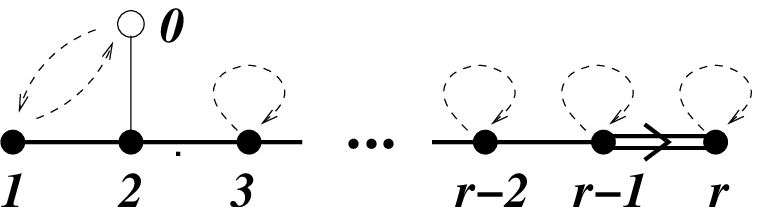}}
\vspace{-0.3cm}
\caption{{\sf The transformation of the extended Dynkin diagram of $B_r$
under $z$.}}
\label{fig:Br}}
It is easy to see, by calculating first the eigenvalues of the projections
of $\,w_z\,$ and $\,w_\kappa\,$ to $\,SO(2r+1)$, \,that 
\qq
w_z\,=\,z^{n_z}\,O_z\,\ee^{\pi\sfi\,\lambda_r^\vee}\,O_z^{-1}\,,
\qquad\quad w_\kappa\,
=\,z^{n_\kappa}\,O_\kappa\,\ee^{\pi\sfi\,\lambda_{r'}^\vee}
\,O_\kappa^{-1}\,,
\nonumber
\qqq
where $\,n_z,\,n_\kappa=0$ or $1$, $\,O_z,\,O_\kappa\in Spin(2r+1)\,$ 
and $\,r'=\frac{r}{2}\,$ for even $\,r\,$ and $\,r'=\frac{r+1}{2}\,$ 
for odd $\,r$. \,The coroot lattice of $\,B_r\,$ is spanned by the 
simple coroots $\,\alpha^\vee_i=e_i-e_{i+1}\,$ for $\,i=1,\dots,r-1$, \,and
$\,\alpha_r^\vee=2e_r$. \,By checking that the coweights 
$\,\lambda_r^\vee\,$ and $\,\lambda_{r'}^\vee\,$ belong to the coroot
lattice if and only if, respectively, $\,r\,$ and $\,r'\,$ 
are even, \,one infers from the above relations that 
\qq
w_z^2\,=\,z^r\,,\quad\qquad w_\kappa^2\,=\,z^{r'}.
\nonumber
\qqq
As far as \,$(w_\kappa w_z)^2\,$ is concerned, we note that it projects
to the same matrix in $\,SO(2r+1)\,$ as $\,\ee^{\pi i\,\lambda_1^\vee}\,$
so that
\qq
(w_\kappa w_z)^2\,=\,\ee^{\pm\pi i\,\lambda_1^\vee}
\nonumber
\qqq
for some choice of the sign. 
\vskip 0.2cm

For the maximal orientifold group $\,\bZ_2\lx\bZ_2$, \,we define
$\,w_n\,$ and $\,w_{\un{n}}\,$ for $\,n=0,1\,$ according 
to (\ref{wnn}). \,One can satisfy the relation 
(\ref{gggb}) by taking 
\qq
&&b_{n,n}\,=\,b_{\un{n},n'}\,=\,m_{n,n'}\lambda_1^\vee,\cr
&&b_{n,\un{n}'}\,=\,(\frac{_1}{^2}\delta_{n,1}+m_{n,\un{n}'})
\lambda_1^\vee,\cr
&&b_{\un{n},\un{n}'}\,=\,(\frac{_1}{^2}\delta_{[n_0+n],1}
+m_{\un{n},\un{n}'})\lambda_1^\vee
\nonumber
\qqq
where $\,m_{n,n'},m_{n,\un{n}'},m_{\un{n},\un{n}'}\,$ are integers.
Since
\qq
\tau_{z^{-n}0}\,=\,\delta_{[n],1}\lambda_1^\vee,\qquad
\tau_{(z_0z^n)^{-1}0}\,=\,\delta_{[n_0+n],1}\lambda_1^\vee,
\nonumber
\qqq
and $\,\tr\,(\lambda_1^\vee)^2=1$, \,one readily sees that the contribution
of the integer multiplicities of $\,\lambda_1^\vee\,$ to
$\,b_{\gamma,\gamma'}\,$ drops out from the expression
(\ref{g3cocyc}) for the obstruction cocycle which, accordingly, takes the
following form:
\qq
&u_{n,n',n''}\,=\,u_{\un{n},n',n''}\,=\,u_{n,\un{n}',n''}\,
=\,u_{\un{n},\un{n}',n''}\,=\,1,\label{br1}&\\
&u_{n,n',\un{n}''}\,=\,(-1)^{\sfk\,n n'},\qquad
u_{\un{n},n',\un{n}''}\,=\,(-1)^{\sfk\,(n_0+n)n'},&\\
&u_{n,\un{n}',\un{n}''}\,=\,(-1)^{\sfk\,n(n_0+n')},\qquad
u_{\un{n},\un{n}',\un{n}''}\,=\,(-1)^{\sfk\,(n_0+n)(n_0+n')}.&
\label{br3}
\qqq
The cohomological equation (\ref{cohoeq}) can be always solved.
Two cohomologically inequivalent solutions are obtained by taking
\qq
&v_{n,n'}\,=\,v_{\un{n},n'}\,=\,(-1)^{\sfk\,nn'},
\qquad v_{n,\un{n}'}\,=\,(-1)^{\sfk\,nn'}\,\ee^{-\frac{3\pi\sfi}{2}\sfk\,n},
&\cr
&v_{\un{n},\un{n}'}\,
=\,\pm(-1)^{\sfk\,n(n_0+n')}\,\ee^{\frac{\pi\sfi}{2}\sfk\,
(n_0+3n)}.&
\nonumber
\qqq
Two further cohomologically inequivalent solutions for
the maximal orientifold group $\,\bZ_2\lx Z(G)\,$ are obtained by 
multiplying $\,v_{\gamma,\gamma'}\,$ by the 2-cocycle 
$\,v^{(2)}_{\gamma,\gamma'}\,$ given by (\ref{v22}).
\vskip 0.2cm

In summary, there are no obstructions to the trivialization of the
3-cocycle (\ref{br1})-(\ref{br3}) on $\,\G=\bZ_2\lx \bZ_2$.
For each $\,\sfk\,$ and each choice of the twist element $\,\zeta\in\bZ_2$,
\,there are four cohomologically inequivalent trivializing
cochains that give rise to inequivalent $\,\G$-equivariant structures 
on the level $\,\sfk\,$ gerbe on $\,Spin(2r+1)$. The latter induce
altogether four inequivalent Jandl structures on the level 
$\,\sfk\,$ gerbe on $\,SO(2r+1)$. \,Restriction to the
inversion group $\,\G=\bZ_2\,$ reduces the number
of inequivalent trivializing cochains to two for each $\,\sfk\,$
and each $\,\zeta$. \,Altogether, they induce four inequivalent
Jandl structures on the level $\,\sfk\,$ gerbe on $\,Spin(2r+1)$.
\vskip 0.7cm

\subsection{The case of $\,G=C_r=Sp(2r)$}\label{sec:Cr}
\noindent The group $\,Sp(2r)\,$ is composed of the unitary 
$\,(2r)\times(2r)$-matrices such that $\,U^T\Omega\,U=\Omega\,$ for
\qq
(\Omega)_{j,j'}\,=\,\sum\limits_{i=1}^r(\delta_{j,2i-1}\delta_{2i,j'}
-\delta_{j,2i}\delta_{2i-1,j'})\,.
\nonumber
\qqq
The Lie algebra $\,\gt{sp}(2r)\,$ of $\,Sp(2r)\,$ is composed
of the hermitian matrices $\,X\,$ such that $\,\Omega X\,$
is symmetric. \,The Cartan subalgebra $\,\Nt\subset\gt{sp}(2r)\,$
is spanned by matrices $\,e_i$, $i=1,\dots,r$, \,with $\,(e_i)_{j,j'}=
\sfi(\delta_{j,2i}\delta_{2i-1,j'}-\delta_{j,2i-1}\delta_{2i,j'})\,$
and the Killing form is normalized so that $\,\tr\,e_ie_{i'}=2\delta_{ij}$.
\,The center $\,Z(G)\cong\bZ_2\,$ with the generator 
$\,z=\ee^{-2\pi\sfi\,\lambda_r^\vee}=-1$, \,where
\qq
\lambda^\vee_i\,=\,\sum\limits_{j=1}^ie_j\quad{\rm for}\quad i=1,\dots,r-1,
\qquad\lambda^\vee_r={_1\over^2}\sum\limits_{j=1}^re_j
\nonumber
\qqq
are the coweights corresponding to the simple roots $\,\alpha_i=
{_1\over^2}(e_i-e_{i+1})\,$ for $\,i=1,\dots,r-1\,$ and $\,\alpha_r=e_r$.
\,The vertices of the positive Weyl alcove $\,\CA\,$ are 
$\,\tau_0=0$, $\,\tau_i={1\over2}\lambda_i^\vee\,$ for $\,i=1,\dots,r-1\,$ 
and $\,\tau_r=\lambda_r^\vee$. \,To satisfy the relations 
(\ref{zact}) and (\ref{kact}) for $\,\tau\in\CA$, \,we may take
for $\,w_z\,$ and $\,w_\kappa\,$ the matrices with the elements
\qq
(w_z)_{j,j'}\,=\,\sfi\,\delta_{j,2r+1-j'}\,,\qquad
(w_\kappa)_{j,j'}\,=\,\sfi\,\sum\limits_{i=1}^r(\delta_{j,2i-1}\delta_{2i,j'}
+\delta_{j,2i}\delta_{2i-1,j'})
\nonumber
\qqq
and the actions of $\,z\,$ and $\,\kappa\,$ on the positive Weyl
alcove reducing to
\qq
z\tau_i\,=\,\tau_{r-i}\,,\ \qquad\kappa\tau_i\,=\,\tau_i
\nonumber
\qqq
on the vertices. The symmetry of the extended Dynkin diagram 
corresponding to the action of $\,z\,$ is depicted
in Fig.\ref{fig:Cr}.
\FIGURE[!h]{
\epsfxsize = 3in
\epsfysize =1.5in
\centerline{
\epsfbox{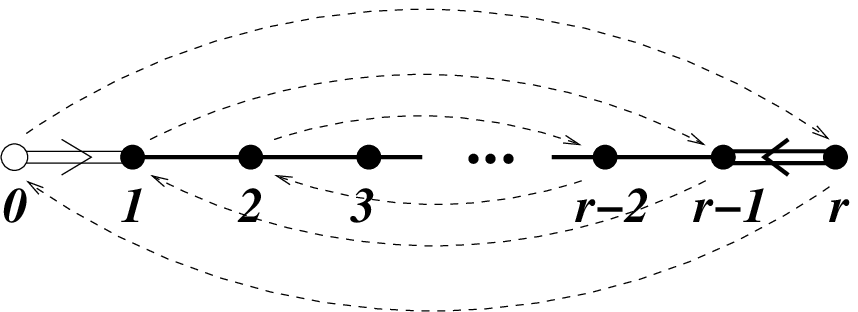}}
\vspace{-0.3cm}
\caption{{\sf The transformation of the extended Dynkin diagram of $C_r$
under $z$.}}
\label{fig:Cr}}
Note that $\,w_z^2=w_\kappa^2=-1=z\,$ and $\,(w_\kappa w_z)^2=1$.
\,Defining $\,w_n\,$ and $\,w_{\un{n}}\,$ for the orientifold
group $\,\G=\bZ_2\lx\bZ_2\,$ by (\ref{wnn}), we may 
satisfy (\ref{gggb}) by taking
\qq
b_{n,n'}\,=\,b_{n,\un{n}'}\,=\,b_{\un{n},n'}\,=\,nn'\lambda_r^\vee,
\qquad b_{\un{n},\un{n}'}\,=\,(1+n_0+nn')\lambda_r^\vee.
\nonumber
\qqq
Since 
\qq
\tau_{z^{-n}0}\,=\,\delta_{[n],1}\lambda_r^\vee,\qquad
\tau_{(z_0z^n)^{-1}0}\,=\,\delta_{[n_0+n],1}\lambda_r^\vee,
\nonumber
\qqq
and $\,\tr\,(\lambda_r^\vee)^2={r\over2}\,$ one infers from 
(\ref{g3cocyc}) that
\qq
&u_{n,n',n''}\,=\,u_{n,\un{n}',n''}\,=\,u_{n,n',\un{n}''}\,=\,(-1)^{\sfk r\,
nn'n''},&\label{uCr1}\\
&u_{\un{n},n',n''}\,=\,u_{\un{n},\un{n}',n''}\,=\,u_{\un{n},n',\un{n}''}\,
=\,(-1)^{\sfk r\,(n_0+n)n'n''},&\\
&u_{n,\un{n}',\un{n}''}\,=\,(-1)^{\sfk r\,n(1+n_0+n'n'')},
\qquad u_{\un{n},\un{n}',\un{n}''}\,=\,(-1)^{\sfk r\,(n_0+n)(1+n_0+n'n'')}.&
\label{uCr3}
\qqq
The restriction $\,u_{n,n',n''}\,$ of the cocycle 
$\,u_{\gamma,\gamma',\gamma''}\,$ to the orbifold subgroup $\,\bZ_2\,$
is trivializable if and only if 
\qq
\sfk\quad\tx{is \,even}\quad\tx{if}\quad r\quad\tx{is \,odd}, 
\nonumber
\qqq
see \cite{GR2}. Under this condition, 
$\,u_{\gamma,\gamma',\gamma''}\equiv 1\,$ and four cohomologically
inequivalent solutions of (\ref{cohoeq}) may be given by the
formulae:
\qq
v_{n,n'}\,=\,v_{n,\un{n}'}\,=\,1,\qquad v_{\un{n},n'}\,=\,
\sigma^{n'},\qquad v_{\un{n},\un{n'}}\,=\,\un{\sigma}\,
\sigma^{n'}
\nonumber
\qqq
with $\,\sigma,\un{\sigma}=\pm1$. \,They lead to four
inequivalent $\,\bZ_2\lx\bZ_2$-equivariant structures on the
level $\,\sfk\,$ gerbe on $\,Sp(2r)\,$ for each choice of the
twist element $\,\zeta=\bZ_2\,$ and, altogether, to four inequivalent
Jandl structures on the quotient gerbe on $\,Sp(2r)/\bZ_2$.
\vskip 0.2cm

The restriction of the 3-cocycle (\ref{uCr1})-(\ref{uCr3})
to the inversion group $\,\Gamma=\bZ_2\,$ is trivial for any level
$\,\sfk\,$ and any choice of the twist element $\,\zeta\in\bZ_2$. 
\,For such a restriction, \,the two cohomologically inequivalent
solutions of (\ref{cohoeq}) are given by 
\qq
v_{0,0}\,=\,v_{0,\un{0}}\,=\,v_{\un{0},0}\,=\,1\,,\qquad v_{\un{0},
\un{0}}=\pm1\,.
\label{coC}
\qqq
Altogether, they lead to four inequivalent Jandl structures 
on the level $\,\sfk\,$ gerbe on $\,Sp(2r)$.

\subsection{The case of $\,G=D_r = Spin(2r)$}\label{sec:Dr}
\noindent The Lie algebra $\,\Ng=\gt{spin}(2r)\,$ is composed
of the imaginary antisymmetric $\,(2r)\times(2r)$-matrices,
with the Cartan subalgebra $\,\Nt\subset\Ng\,$ spanned by the 
matrices $\,e_i$, $i=1,\dots,r$, \,with the matrix elements 
$\,(e_i)_{j,j'}=\sfi(\delta_{j,2i}\delta_{2i-1,j'}-\delta_{j,2i-1}
\delta_{2i,j'})\,$ and $\,\tr\,e_ie_{i'}=\delta_{i,i'}$. \,The vertices of the positive Weyl alcove
are 
\qq
&\tau_0\,=\,0,\ \quad\tau_1\,=\,\lambda_1^\vee,\ \quad 
\tau_i\,=\,{1\over 2}\lambda_i^\vee\quad\tx{for}\quad i=2,\dots,r-2,&\cr\cr
&\tau_{r-1}\,=\,\lambda_{r-1}^\vee,\ \quad\tau_r\,=\,\lambda_r^\vee,&
\nonumber
\qqq
where
\qq
&\lambda_i^\vee\,=\,\sum\limits_{j=1}^ie_i\quad\tx{for}\quad i=1,\dots,r-2,&\cr
&\lambda_{r-1}^\vee\,=\,{_1\over^2}\sum\limits_{j=1}^{r-1} e_j\,
-\,{_1\over^2}e_r\,,
\ \quad \lambda_{r}^\vee\,=\,{_1\over^2}\sum\limits_{j=1}^r e_j&
\nonumber
\qqq
are the simple coweights corresponding to the simple roots $\,\alpha_i=
e_i-e_{i+1}$, $i=1,\dots,r-1$, $\,\alpha_r=e_{r-1}+e_r\,$ that coincide
with the simple coroots. \,The subsequent discussion depends on the
parity of $\,r\,$ and hence will be split into two parts.

\subsubsection{The subcase of $\,r\,$ odd}\label{sec:Dro}
\noindent For $\,r=2s+1$, \,the center $\,Z(G)\cong\bZ_4\,$
is generated by $\,z=\ee^{-2\pi\sfi\,\lambda_r^\vee}$, \,with
$\,Spin(2r)/\{1,z^2\}=SO(2r)$. \,For $\,\tau\in\CA$, 
\,the relations (\ref{zact}) and (\ref{kact}) are satisfied if
we take for $\,w_z\,$ and $\,w_\kappa\,$ the elements
of $\,Spin(2r)\,$ that project to matrices in $\,SO(2r)\,$ with 
the elements
\qq
&(w_z)_{j,j'}\,=\,(-1)^{\delta_{j,2r}}\delta_{j,2r+1-j'}\,,&\\
&(w_\kappa)_{j,j'}\,=\,\sum\limits_{i=1}^{r-1}
(\delta_{j,2i-1}\delta_{2i,j'}+\delta_{j,2i}\delta_{2i-1,j'})\,+\,
\delta_{j,2r-1}\delta_{2r-1,j'}\,+\,\delta_{j,2r}\delta_{2r,j'}\,,&
\nonumber
\qqq
with the actions of $\,z\,$ and $\,\kappa\,$ on the positive Weyl
alcove reducing to
\qq
&z\tau_0\,=\,\tau_{r-1}\,,\qquad z\tau_1\,=\,\tau_{r}\,,\qquad
z\tau_i\,=\,\tau_{r-i}\quad\tx{for}\quad i=2,\dots,r,&\cr\cr
&\kappa\tau_i\,=\,\tau_i\quad\tx{for}\quad i=0,\dots,r-2,\qquad
\kappa\tau_{r-1}\,=\,\tau_r\,,\qquad\kappa\tau_{r}\,=\,\tau_{r-1}\,.
\nonumber
\qqq
on the vertices. The corresponding symmetries of the Dynkin diagrams 
are depicted in Fig.\ref{fig:Dro} and Fig.\ref{fig:DroK}.  
\FIGURE[!h]{
\epsfxsize = 3in
\epsfysize =1.8in
\centerline{
\epsfbox{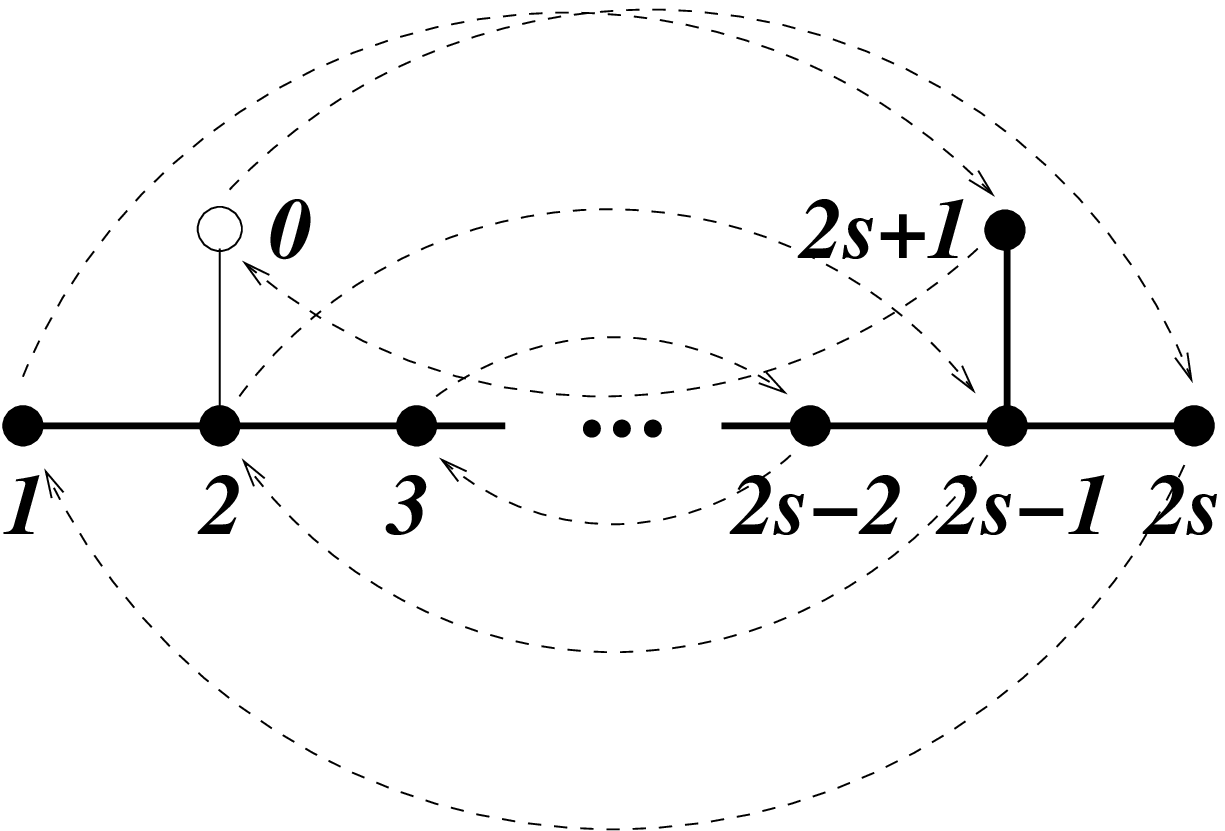}}
\vspace{-0.3cm}
\caption{{\sf 
The transformation of the extended Dynkin diagram of $D_{2s+1}$
under $z$.}}
\label{fig:Dro}}
\FIGURE[!h]{
\epsfxsize = 3in
\epsfysize =0.9in
\centerline{
\epsfbox{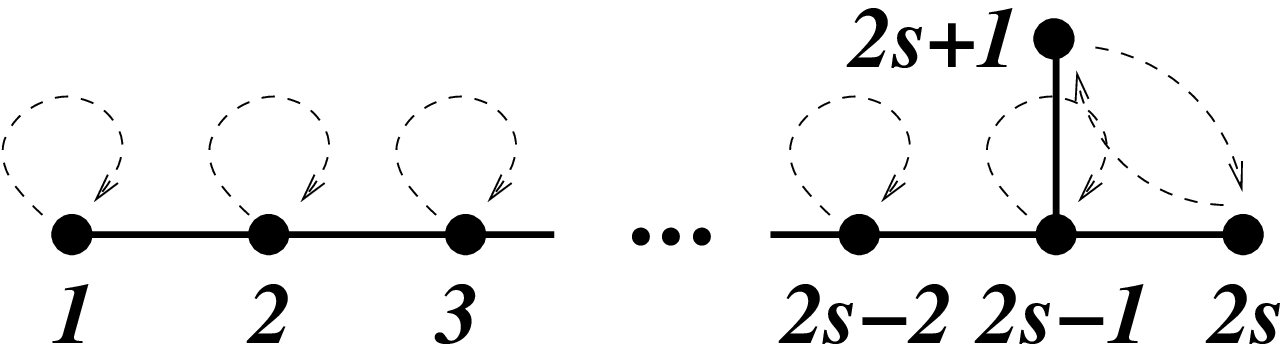}}
\vspace{-0.3cm}
\caption{{\sf The inversion map $\kappa$ flips the last nodes 
of the Dynkin diagram of $D_{2s+1}$.}}
\label{fig:DroK}}
Note the adjoint action
\qq
w_z\,e_i\,w_z^{-1}\,=\,-(-1)^{\delta_{i,1}}\,e_{r+1-i}\,.
\label{actz2}
\qqq
It is easy to see, comparing first the eigenvalues of the
projections of both sides to $\,SO(2r)$, \,that
\qq
&&\hbox to 5.8cm{$w_z\,=\,z^{2n_z}\,O_z\,\ee^{2\pi\sfi\,\tau_z}
\,O_z^{-1}$\hfill}\tx{for}\quad
\tau_z={_1\over^2}\sum\limits_{i=1}^s e_i\,+\,{_1\over^4}e_{s+1}\,,\cr
&&\hbox to 5.8cm{$w_\kappa\,=\,z^{2n_\kappa}\,O_\kappa\,
\ee^{2\pi\sfi\,\tau_\kappa}\,O_z^{-1}$\hfill}\tx{for}\quad\tau_\kappa\,
=\,{_1\over^2}\sum\limits_{i=1}^s e_i\,,\cr
&&\hbox to 5.8cm{$w_\kappa w_z\,=\,z^{2n_{\kappa z}}\,O_{\kappa z}
\,\ee^{2\pi\sfi\,\tau_{\kappa z}}\,O_{\kappa z}^{-1}$\hfill}\tx{for}\quad
\tau_{\kappa z}\,=\,{_1\over^2}\sum\limits_{i=1}^{s-1}e_i\,+\,
{_3\over^8}e_s\,+\,{_1\over^8}e_{s+1}\qquad
\nonumber
\qqq
for some integers $\,n_z,n_\kappa,n_{\kappa_z}\,$ and $\,O_z,O_\kappa,
O_{\kappa z}\in Spin(2r)$. \,The Cartan algebra elements 
$\,\tau_z$, $\,\tau_\kappa\,$ and $\,\tau_{\kappa z}\,$ belong to the 
positive Weyl alcove $\,\CA$. \,It is easy to see that $\,4\tau_z\,$ 
belongs also to the coweight lattice but not to the coroot lattice. 
On the other hand, $\,2\tau_\kappa =\lambda_s^\vee\,$ belongs to 
the coroot lattice if and only if $\,s\,$ is even. Since 
$\,w_z^4\,$ and $\,w_\kappa^2\,$ project to the identity matrix in $\,SO(2r)$,
it follows that
\qq
w_z^4\,=\,z^2,\ \qquad w_\kappa^2\,=\,z^{2s}. 
\label{wz4wk2}
\qqq
We also have 
\qq
&O_{\kappa z}^{-1}\,(w_\kappa w_z)^2\,O_{\kappa z}\,=\,
\ee^{2\pi\sfi(2\tau_{\kappa z})}\,=\,\begin{cases}
\hbox to 5cm{$\ee^{2\pi\sfi(e_1+{3\over 4}e_s+{_1\over 4}e_{s+1})}$\hfill}
\tx{for}\quad s\quad\tx{even},\cr
\hbox to 5cm{$\ee^{2\pi\sfi({3\over 4}e_s+{_1\over 4}e_{s+1})}$\hfill}
\tx{for}\quad s\quad\tx{odd}\end{cases}&\cr\cr
&=\,z^{2(s+1)}\,\ee^{{\pi\sfi\over2}(3e_s+e_{s+1})}\,=\,
z^{2(s+1)}\,O\,\ee^{{\pi\sfi\over2}(3e_1+e_{2})}\,O^{-1}&
\nonumber
\qqq
for $\,O\in Spin(2r)\,$ that is straightforward to construct.
By the relation (\ref{zact}),
\qq
&z^2\,\ee^{{\pi\sfi\over2}(3e_1+e_{2})}\,=\,
z^2\,\ee^{2\pi\sfi({1\over 2}\tau_1+{_1\over 2}\tau_{2})}\,=\,
w_z^{-2}\,\ee^{2\pi\sfi({1\over2}\tau_0+{1\over^2}\tau_2)}\,w_z^2&\cr
&=\,w_z^{-2}\,\ee^{{\pi\sfi\over2}(e_1+e_2)}\,w_z^2\,=\,
O'\,\ee^{{\pi\sfi\over2}(e_1+e_r)}\,{O'}^{-1}.&
\nonumber
\qqq
We infer that $\,(w_\kappa w_z)^2\,$ is in the same conjugacy class
as $\,z^{2s}\,\ee^{{\pi\sfi\over2}(e_1+e_r)}\,$ and that the latter
is different from the conjugacy class of $\,z^{2(s+1)}\,
\ee^{{\pi\sfi\over2}(e_1+e_r)}$. \,On the other hand, it is easy to check 
that $\,(w_\kappa w_z)^2\,$ projects to the same matrix in $\,SO(2r)\,$ as 
$\,\ee^{{\pi\sfi\over2}(e_1+e_r)}$. \,It follows that
\qq
(w_\kappa w_z)^2\,=\,z^{2s}\,\ee^{{\pi\sfi\over2}(e_1+e_r)}
\nonumber
\qqq
which, together with the second equality in (\ref{wz4wk2}), implies that
\qq
w_\kappa w_z w_\kappa^{-1}\,=\,\ee^{{\pi\sfi\over2}(e_1+e_r)}\,w_z^{-1}\,.
\nonumber
\qqq
Using (\ref{actz2}), we obtain the relations:
\qq
w_\kappa w_z^n w_\kappa^{-1}w_z^{n}\,=\,\ee^{2\pi\sfi\,\Delta^+_n},\qquad
w_z^nw_\kappa w_z^n w_\kappa^{-1}\,=\,\ee^{2\pi\sfi\,\Delta^-_n},
\nonumber
\qqq
where
\qq
\Delta^\pm_n\,=\,\begin{cases}
\hbox to 4cm{$\,0$\hfill}\tx{for}\quad n=0,\cr
\hbox to 4cm{$\,\pm{1\over4}(e_1\pm e_r)$\hfill}
\tx{for}\quad n=1,\cr
\hbox to 4cm{$\,\pm{1\over2}e_1$\hfill}\tx{for}\quad n=2,\cr
\hbox to 4cm{$\,\pm{1\over4}(e_1\mp e_r)$\hfill}\tx{for}\quad n=3.\end{cases}
\nonumber
\qqq
Together with (\ref{wz4wk2}), they are all what is needed 
to find $\,b_{\gamma,\gamma'}\,$ for $\,\gamma,\gamma'\,$ in the
maximal orientifold group $\,\G=\bZ_2\lx\bZ_4$. \,We may set 
\qq
&&b_{n,n'}\,=\,b_{\un{n},n'}\,=\,{_{n+n'-[n+n']}\over^4}\,e_1\,,\\
&&b_{n,\un{n}'}\,=\,{_{n'-n-[n'-n]}\over^4}\,e_1\,+\,\Delta^-_n\,,\\
&&b_{\un{n},\un{n}'}\,=\,\Big({_{n'-n-[n'-n]}\over^4}+s\Big)e_1\,
+\,\Delta^+_{[n_0+n]}
\nonumber
\qqq
for $\,n,n'=0,1,2,3$. \,We also have
\qq
\tau_{z^{-1}0}\,=\,\lambda_r^\vee,\qquad\tau_{z^{-2}0}\,
=\,\lambda_1^\vee,\qquad
\tau_{z^{-3}0}\,=\,\lambda_{r-1}^\vee.
\nonumber
\qqq
Rather than displaying the corresponding obstruction 3-cocycle 
(\ref{g3cocyc}) in full, we shall focus on its specific components.
\vskip 0.2cm

First, for the inversion group $\,\G=\bZ_2$, \,the only
entry of the 3-cocycle different from $\,1\,$ is  
\qq
u_{\un{0},\un{0},\un{0}}\,=\,\ee^{{\pi\sfi\over2}\sfk r(n_0+2\delta_{n_0,3})}.
\nonumber
\qqq
The trivializing cochain may be given by the formulae:
\qq
v_{0,0}\,=\,v_{\un{0},0}\,=\,v_{0,\un{0}}\,=\,1\,,\qquad
v_{\un{0},\un{0}}\,=\,\pm\,
\ee^{-{\pi\sfi\over4}\sfk r(n_0+2\delta_{n_0,3})},
\nonumber
\qqq
with the two signs corresponding to cohomologically inequivalent 
solutions. 
Next, we pass to the case of orientifold groups $\,\G=\bZ_2\lx\bZ_m\,$
with $\,m=2,4$. \,The restriction of the obstruction 3-cocycle to the 
orbifold group $\,\bZ_4\,$ is 
\qq
u_{n,n',n''}\,=\,(-1)^{\sfk n\frac{n'+n''-[n'+n'']}{4}}.
\nonumber
\qqq
It is not trivializable if $\,\sfk\,$ is odd, see \cite{GR2}.
On the other hand, its further restriction to $\,\bZ_2\,$ is trivial for all
$\,\sfk$.
\,In order to proceed further, we note that the scalar product
$\,\tr\,\tau_{z^{-n}0}e_1\,$ takes values in integers if $\,n\,$ 
is even and in half-integers if $\,n\,$ is odd. It follows that,  
for even $\,\sfk$, \,only the terms $\,\Delta^\pm\,$ in 
$\,b_{\gamma,\gamma'}\,$ contribute to $\,u_{\gamma,\gamma',\gamma''}\,$
if $\,m=4$. \,This is still the case if
$\,m=2$. \,Indeed, if the twist element 
$\,\zeta\in\bZ_2\subset\bZ_4\,$ then $\,\tr\,\tau_{z^{-n}0} e_1\,$ 
and $\,\tr\,\tau_{(z_0z^n)^{-1}}e_1\,$ take integral values because 
$\,n=0,2$. \,Conversely, if $\,\zeta\in\bZ_4\setminus\bZ_2$, 
\,then a straightforward check shows that the combination $\,X\,$ 
of (\ref{loeq}) is equal to $\,(-1)^\sfk\,$  for $\,n=2$, \,thereby 
contradicting the trivializability of the obstruction cocycle 
for odd $\,\sfk$. \,Summarizing, we obtain the condition
\qq
\sfk\quad\ \tx{is \,even\ \,if}\ \quad m=4\quad\ \tx{or}\quad\ 
m=2\ \quad\tx{and}\ \quad n_0\ \quad\tx{is \,odd}
\nonumber
\qqq
under which only the terms $\,\Delta^\pm\,$ in $\,b_{\gamma,\gamma'}\,$
contribute to the obstruction cocycle $\,u_{\gamma,\gamma',\gamma''}$.
\,With this observation in mind, we obtain, for $\,m=4\,$ 
or for $\,m=2\,$ and $\,n_0\,$ odd, i.e. in both cases 
in which $\,\sfk\,$ has to be even, the following expressions for
the obstruction cocycle: 
\qq
&&u_{n,n',n''}\,=\,u_{\un{n},n',n''}\,=\,u_{n,\un{n}',n''}\,
=\,u_{\un{n},\un{n}',n''}\,=\,1,\\
&&u_{n,n',\un{n}''}\,=\,(-1)^{{1\over2}\sfk(1-\delta_{n,0})
(1-\delta_{n',0})(1-\delta_{n,n'})},\\
&&u_{\un{n},n',\un{n}''}\,=\,(-1)^{{1\over2}\sfk(1-\delta_{[n_0+n],0})
(1-\delta_{n',0})(1-\delta_{[n_0+n],n'})},\\
&&u_{n,\un{n}',\un{n}''}\,=\,(-1)^{{1\over2}\sfk(1-\delta_{n,0})
(1-\delta_{[n_0+n'],0})(1-\delta_{[n_0+n+n'],0})},\\
&&u_{\un{n},\un{n}',\un{n}''}\,=\,(-1)^{{1\over2}\sfk(1-\delta_{[n_0+n],0})
(1-\delta_{[n_0+n'],0})(1-\delta_{[2n_0+n+n'],0})}.
\nonumber
\qqq
Similarly, for $\,m=2\,$ and $\,n_0\,$ even, when $\,\sfk\,$ can be any
integer,
\qq
&&u_{n,n',n''}\,=\,u_{\un{n},n',n''}\,=\,u_{n,\un{n}',n''}\,
=\,u_{\un{n},\un{n}',n''}\,=\,1,\\
&&u_{n,n',\un{n}''}\,=\,(-1)^{\sfk\,\delta_{n,2}\,\delta_{n',2}}
,\quad\ \hspace{0.05cm}\qquad u_{\un{n},n',\un{n}''}\,=\,(-1)^{\sfk\,\delta_{[n_0+n],2}\,
\delta_{n',2}},\\
&&u_{n,\un{n}',\un{n}''}\,=\,(-1)^{\sfk\,\delta_{n,2}\,
\delta_{[n_0+n'],2}},\qquad\hspace{-0.04cm}
u_{\un{n},\un{n}',\un{n}''}\,=\,(-1)^{\sfk\,\delta_{[n_0+n],2}\,
\delta_{[n_0+n'],2}}.
\nonumber
\qqq
In all these cases, there exists a trivializing cochain. It may
be taken in the form:
\qq
&v_{n,n'}\,=\,1,\qquad v_{\un{n},n'}\,=\,\sigma^{{1\over4}mn'},\quad\ 
v_{n,\un{n}'}\,=\,\ee^{{\pi\sfi\over 4}
\sfk(n+2\delta_{n,3})},&\\
&v_{\un{n},\un{n}'}\,=\,\un{\sigma}\,\sigma^{{1\over4}mn'}\,
\ee^{-{\pi\sfi\over 4}\sfk([n_0+n]+2\delta_{[n_0+n],3})}&
\nonumber
\qqq
with different signs $\,\sigma,\un{\sigma}=\pm1\,$ giving 
four cohomologically inequivalent solutions.
\vskip 0.2cm

In summary, for each $\,\sfk\,$ and each choice of the twist
element $\,\zeta\in\bZ_4$, \,there are two inequivalent Jandl structures
on the level $\,\sfk\,$ gerbe on $\,Spin(2r)\,$ with $\,r\,$ odd.
\,For each $\,\sfk\,$ even and each choice of the twist element 
$\,\zeta\in\bZ_4\,$ and for each $\,\sfk\,$ odd and 
$\,\zeta\in\bZ_2\subset\bZ_4$, \,there are four inequivalent 
$\,\bZ_2\lx\bZ_2$-equivariant structures on the level $\,\sfk\,$ gerbe 
on $\,Spin(2r)$, \,giving rise, altogether, to eight inequivalent 
Jandl structures on the induced gerbe on $\,SO(2r)\,$ when  
$\,\sfk\,$ is even and to four ones when $\,\sfk\,$ is odd. \,Finally, 
for each $\,\sfk\,$ even and each choice of $\,\zeta\in\bZ_4$, \,there 
are four inequivalent $\,\bZ_2\lx\bZ_4$-equivariant structures on the
level $\,\sfk\,$ gerbe on $\,Spin(2r)$, \,giving rise, altogether,
to four inequivalent Jandl structures on the induced gerbe
on $\,Spin(2r)/\bZ_4$. \,Note that the count is similar to that for
the group $\,SU(4)$.

\subsubsection{The subcase of $\,r\,$ even}\label{sec:Dre}
\noindent For $\,r=2s$, \,the center 
$\,Z(G)\cong\bZ_2\times\bZ_2\,$ is generated
by $\,z_1=\ee^{-2\pi\sfi\,\lambda_r^\vee}\,$ and $\,z_2=\ee^{-2\pi\sfi
\,\lambda_1^\vee}$, \,with $\,Spin(2r)/\{1,z_2\}=SO(2r)$. 
\,For $\,\tau\in\CA\,$ and $\,z=z_1,z_2$, \,the relations
(\ref{zact}) are  satisfied if
we take for $\,w_{z_1}\,$ and $\,w_{z_2}\,$ the elements
of $\,Spin(2r)\,$ that project to $\,SO(2r)\,$ matrices with the 
elements\footnote{For later convenience, we make a different choice
from that in \cite{GR2}.}
\qq
&&(w_{z_1})_{j,j'}\,=\,\begin{cases}\hbox to 2cm{$
-\delta_{j,2r+1-j'}$\hfill}\tx{for}\quad j=1,\dots,r,\cr
\hbox to 2cm{$\ \delta_{j,2r+1-j'}$\hfill}\tx{for}\quad j=r+1,\dots,2r,
\end{cases}\\
&&(w_{z_2})_{j,j'}\,=\,\delta_{j,1}\delta_{2,j'}+\delta_{j,2}\delta_{1,j'}
+\delta_{j,2r-1}\delta_{j',2r}+\delta_{j,2r}\delta_{2r-1,j'}
+\sum\limits_{i=3}^{2r-2}\delta_{j,i}\delta_{i,j'}\,,
\nonumber
\qqq
with the actions of $\,z_1\,$ and $\,z_2\,$ on the positive Weyl
alcove reducing to
\qq
&z_1\tau_i\,=\,\tau_{r-i}\,,\qquad\quad z_2\tau_0\,=\,\tau_{1}\,,\quad
z_2\tau_{1}\,=\,\tau_0\,,&\cr
&z_2\tau_{i}\,=\,\tau_i\quad\tx{for}
\quad i=2,\dots,r-2\,,\qquad z_2\tau_{r-1}\,=\,\tau_{r}\,,
\quad z_2\tau_r\,=\,\tau_{r-1}&
\nonumber
\qqq
on the vertices.
The corresponding symmetries of the extended Dynkin diagram are depicted
in Fig.\ref{fig:Dre1} and Fig.\ref{fig:Dre2}.  
\FIGURE[!h]{
\epsfxsize = 3in
\epsfysize = 1.8in
\centerline{
\epsfbox{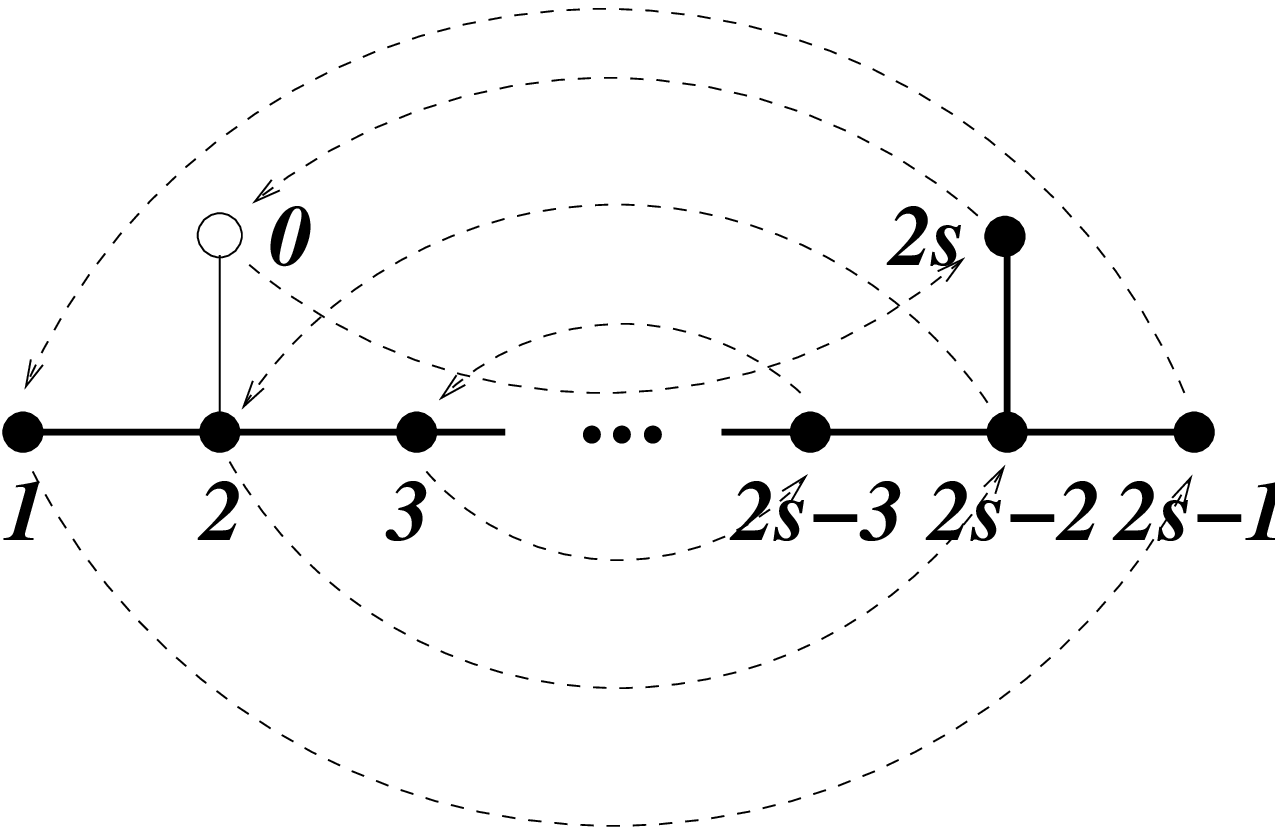}}
\vspace{-0.3cm}
\caption{{\sf The transformation of the extended Dynkin diagram of $D_{2s}$
under $z_1$.}}
\label{fig:Dre1}}
\FIGURE[!h]{
\epsfxsize = 3.1in
\epsfysize = 1.1in
\centerline{
\epsfbox{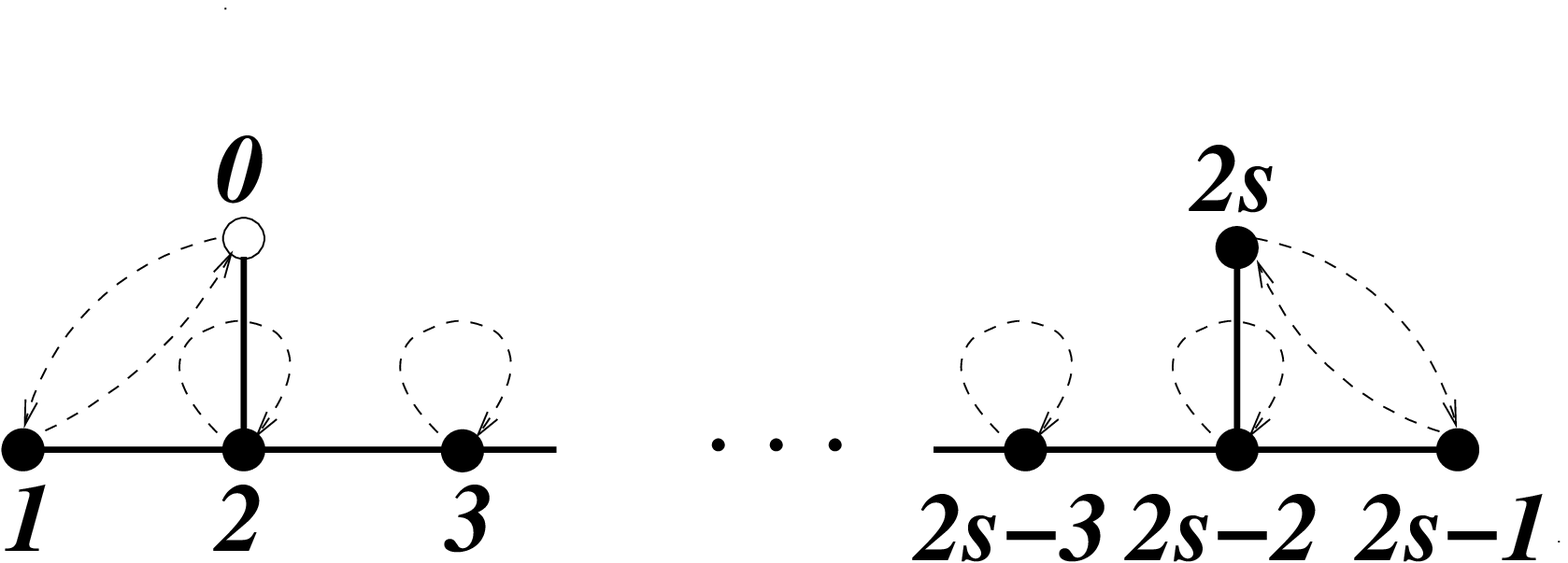}}
\vspace{-0.3cm}
\caption{{\sf The transformation of the extended Dynkin diagram of $D_{2s}$
under $z_2$.}}
\label{fig:Dre2}}
The adjoint action of $\,w_{z_1}\,$ and $\,w_{z_2}\,$
on the Cartan algebra is given by the equations:
\qq
w_{z_1}\,e_i\,w_{z_1}^{-1}\,=\,-e_{r-i+1}\,,\qquad
w_{z_2}\,e_i\,w_{z_2}^{-1}\,=\,(-1)^{\delta_{i,1}+\delta_{i,r}}\,e_i\,.
\nonumber
\qqq
The relations (\ref{kact}) are, in turn, satisfied for $\,\tau\in\CA\,$ if
we take for $\,w_\kappa\,$ the element
of $\,Spin(2r)\,$ that projects to an $\,SO(2r)\,$ matrix with the 
elements
\qq
(w_{\kappa})_{j,j'}\,=\,\sum\limits_{i=1}^r(\delta_{j,2i-1}\delta_{2i,j'}
+\delta_{j,2i}\delta_{2i-1,j'}),
\nonumber
\qqq
with the trivial action of $\,\kappa\,$ on the positive Weyl
alcove. 
We have the relations:
\qq
&&\hbox to 6.8cm{$w_{z_1}\,=\,z_2^{n_{z_1}}\,O_{z_1}\,\ee^{2\pi\sfi\,
\tau_{z_1}}\,O_{z_1}^{-1}$\hfill}\tx{for}\ \qquad\hbox to 0.8cm{$\tau_{z_1}$
\hfill}\,=\,-{_1\over^2}\lambda_r^\vee\,,\\
&&\hbox to 6.8cm{$w_{z_2}\,=\,z_2^{n_{z_2}}\,O_{z_2}\,\ee^{2\pi\sfi\,
\tau_{z_2}}\,O_{z_2}^{-1}$\hfill}\tx{for}\ \qquad\hbox to 0.8cm{$\tau_{z_2}
$\hfill}\,=\,-{_1\over^2}\lambda_1^\vee\,,\\
&&\hbox to 6.8cm{$w_{z_1}w_{z_2}\,=\,z_2^{n_{z_1z_2}}\,O_{z_1z_2}\,
\ee^{2\pi\sfi\,\tau_{z_1z_2}}\,O_{z_1z_2}^{-1}$\hfill}\tx{for}\ 
\qquad\hbox to 0.8cm{$\tau_{z_1z_2}$\hfill}\,=\,\ {_1\over^2}(\lambda_1^\vee
-\lambda_r^\vee)\,,\\
&&\hbox to 6.8cm{$w_\kappa\,=\,z_2^{n_{\kappa}}\,O_{\kappa}\,\ee^{2\pi\sfi\,
\tau_{\kappa}}\,O_{\kappa}^{-1}$\hfill}\tx{for}\ \qquad
\hbox to 0.8cm{$\tau_{\kappa}$\hfill}\,=\,\ {_1\over^2}\lambda_s^\vee
\,,\\
&&\hbox to 6.8cm{$w_\kappa w_{z_1}\,=\,z_2^{n_{\kappa z_1}}\,O_{\kappa z_1}
\,\ee^{2\pi\sfi\,\tau_{\kappa z_1}}\,O_{\kappa z_2}^{-1}$\hfill}\tx{for}\ 
\qquad\hbox to 0.8cm{$\tau_{\kappa z_1}$\hfill}\,=\,\ {_1\over^2}(\lambda_s^\vee-\lambda_r^\vee)\,,\\
&&\hbox to 6.8cm{$w_\kappa w_{z_2}\,=\,z_2^{n_{\kappa z_2}}\,
O_{\kappa z_2}\,\ee^{2\pi\sfi\,
\tau_{\kappa z_2}}\,O_{\kappa z_2}^{-1}$\hfill}\tx{for}\ \qquad
\hbox to 0.8cm{$\tau_{\kappa z_2}$\hfill}\,=\,-{_1\over^2}(\lambda_1^\vee
-\lambda_s^\vee)\qquad
\nonumber
\qqq
from which it follows that
\qq
&w_{z_1}^2\,=\,z_1\,,\ \qquad w_{z_2}^2\,=\,z_2\,,
\ \qquad(w_{z_1}w_{z_2})^2\,
=\,z_1z_2\,,\ \qquad w_\kappa^2\,=\,z_2^s\,,\qquad&\label{sosq1}\\
&(w_\kappa w_{z_1})^2\,=\,z_1z_2^{s}\,,\qquad
(w_\kappa w_{z_2})^2\,=\,z_2^{s+1}\,,
\qquad(w_{\kappa} w_{z_1} w_{z_2})^2\,=\,z_1z_2^{s+1}\,.\qquad&
\label{sosq2}
\qqq
The last equality is a consequence of the previous ones since
\qq
&(w_\kappa w_{z_{1}}w_{z_2})^2\,(w_\kappa w_{z_2})^{-2}\,=\,
w_\kappa w_{z_1} w_{z_2}w_\kappa w_{z_1} w_\kappa^{-1} w_{z_2}^{-1} 
w_\kappa^{-1}\quad&\cr
&=\,z_2^s\,w_\kappa w_{z_1}w_{z_2}(w_\kappa w_{z_1})^2w_{z_1}^{-1}w_{z_2}^{-1}
w_\kappa^{-1}\,=\,z_2\,w_\kappa(w_{z_1} 
w_{z_2})^2 w_\kappa^{-1}\,=\,z_1\,.\quad&
\nonumber
\qqq
Note that the relations (\ref{sosq1}) and (\ref{sosq2}) imply that
$\,w_{z_1}$, $w_{z_2}$ and $w_\kappa\,$ all commute. This will lead
to simple expressions for the obstruction cocycle. 
\vskip 0.2cm

Similarly as in the case of groups with cyclic centers, we shall use 
the abbreviated notation:
\qq
z_1^{n_1}z_2^{n_2}\equiv n_1n_2\,,\qquad\quad z_0z_1^{n_1}z_2^{n_2}\equiv
\un{n_1n_2} 
\nonumber
\qqq
for the elements of the orientifold group $\,\bZ_2\lx(\bZ_2\times\bZ_2)$, 
\,setting 
\qq
w_{n_1n_2}\,=\,w_{z_1}^{n_1}w_{z_2}^{n_2}\,,\qquad
w_{\un{n_1n_2}}\,=\,w_\kappa w_{z_1}^{n_{01}}
w_{z_2}^{n_{02}}w_{z_1}^{n_1}w_{z_2}^{n_2}
\nonumber
\qqq
if $\,n_1,n_2,n_{01},n_{02}=0,1\,$ and if the twist element
$\,\zeta=z_1^{n_{01}}z_2^{n_{02}}\equiv n_{01}n_{02}$.
\,It is easy to show with the help of (\ref{sosq1}) and (\ref{sosq2})
that the Cartan algebra elements $\,b_{\gamma,\gamma'}\,$ 
may be taken in the form:
\qq
&b_{n_1n_2,n'_1n'_2}\,=\,b_{\un{n_1n_2},n'_1n'_2}\,
=\,b_{n_1n_2,\un{n'_1n'_2}}\,=\,n_1n_1'\,\lambda_r^\vee\,+\,
n_2n_2'\,\lambda_1^\vee,&\cr
&b_{\un{n_1n_2},\un{n'_1n'_2}}\,=\,(n_{01}+n_1n'_1)\,\lambda_r^\vee\,+\,
(s+n_{02}+n_2n'_2)\,\lambda_1^\vee.\ \,
\nonumber
\qqq
Employing the relations  
\qq
\tau_{(z_1^{n_1}z_2^{n_2})^{-1}0}\,=\,(1-n_1)n_2\,\lambda_1^\vee\,+\,
n_1n_2\,\lambda_{r-1}^\vee\,+\,n_1(1-n_2)\,\lambda_r^\vee,
\qqq
together with
\qq
&\tr\,(\lambda_1^\vee)^2\,=\,1,\qquad\tr\,\lambda_1^\vee
\lambda_{r-1}^\vee\,=\,\tr\lambda_1^\vee\lambda_r^\vee\,=\,{1\over2}\,,&\cr
&\tr\,\lambda_{r-1}^\vee\lambda_r^\vee\,=\,{{s-1}\over2}\,,
\qquad\tr\,(\lambda_r^\vee)^2\,=\,{s\over2}\,,&
\nonumber
\qqq
we obtain from the definition (\ref{g3cocyc}) the explicit
expressions for the obstruction cocycle 
\qq
u_{n_1n_2,n_1'n_2',n_1''n_2''}&=&u_{n_1n_2,\un{n_1'n_2'},n_1''n_2''}\,=\,
u_{n_1n_2,n_1'n_2',\un{n_1''n_2''}}\,\cr
&=&(-1)^{\sfk\left(s\,n_1n'_1n''_1
+n_1n'_2n''_2+n_2n'_1n''_1\right)}\,,\label{Deobst1}\\
u_{n_1n_2,\un{n_1'n_2'},\un{n_1''n_2''}}
&=&(-1)^{\sfk\left(s\,n_1(1+n_{01}+n'_1n''_1)+n_1(n_{02}+n'_2n''_2)
+n_2(n_{01}+n'_1n''_1)\right)}\,,\\
u_{\un{n_1n_2},n_1'n_2',n_1''n_2''}&=&u_{\un{n_1n_2},\un{n_1'n_2'},
n_1''n_2''}\,=\,
u_{\un{n_1n_2},n_1'n_2',\un{n_1''n_2''}}\,\cr
&=&(-1)^{\sfk\left(s(n_{01}+n_1)n'_1n''_1
+(n_{01}+n_1)n'_2n''_2+(n_{02}+n_2)n'_1n''_1\right)}\,,\\
u_{\un{n_1n_2},\un{n_1'n_2'},\un{n_1''n_2''}}
&&\cr
&&\hspace*{-2.5cm}=\ (-1)^{\sfk\left(s(n_{01}+n_1)(1+n_{01}+n'_1n''_1)+(n_{01}+n_1)
(n_{02}+n'_2n''_2)+(n_{02}+n_2)(n_{01}+n'_1n''_1)\right)}\qquad
\label{Deobst2}
\qqq
that can easily be analyzed.
\vskip 0.2cm

First, we note that the restriction of $\,u_{\gamma,\gamma',\gamma''}\,$
to the inversion group $\,\G=\bZ_2\,$ is trivial, with the formulae
\qq
v_{00,00}\,=\,v_{\un{00},00}\,=\,v_{00,\un{00}}\,=\,1\,,\qquad
v_{\un{00},\un{00}}\,=\,\pm1
\nonumber
\qqq
providing two cohomologically inequivalent trivializing cochains.
\vskip 0.2cm 

As the next case, let us consider the restriction of the
obstruction cocycle to the orientifold subgroup $\,\G=\bZ_2\lx Z\,$
with $\,Z=\{1,z_2\}$. \,Since the combination $\,X\,$ of 
(\ref{loeq}) with $\,n=01\,$ is easily calculated to be
equal to $\,(-1)^{\sfk\,n_{01}}$,  \,we infer that the obstruction
cocycle restricted to $\,\G\,$ may be trivialized only if
\qq
\sfk\ \quad\tx{is\ \,even\ \,if}\quad\ n_{01}=1.
\nonumber
\qqq
Under this condition, the restricted cocycle becomes trivial 
for all choices of the twist element.  
\vskip 0.2cm

Passing to the the orientifold groups $\,\G=\bZ_2\lx Z\,$
with $\,Z=\{1,z_1\}$, $\,Z=\{1,z_1z_2\}\,$ or $\,Z=\bZ_2\times\bZ_2$,
\,we recall from \cite{GR2} that, in all these three cases,
the restriction of the obstruction cocycle to the
orbifold group $\,Z\,$ may be trivialized only under the
condition that
\qq
\sfk\quad\ \tx{is\ \,even}\quad\ \tx{if}\ \quad s={_r\over^2}\ \quad
\tx{is\ \,odd}.
\label{Decond}
\qqq
If $\,\sfk\,$ is even the whole obstruction cocycle becomes trivial.
Suppose then that $\,\sfk\,$ is odd but $\,s\,$ is even so that
the terms multiplied by $\,s\,$ may be dropped in the explicit 
expression for the cocycle. The combinations $\,X\,$ of (\ref{loeq}) 
with $\,n=10\,$ and $\,n=11\,$ are now easily calculated to take the values
$\,(-1)^{\sfk\,n_{02}}\,$ and $\,(-1)^{\sfk(n_{01}+n_{02})}$,
\,respectively. For $\,Z=\{1,z_1\}$, \,we then obtain the condition
\qq
\sfk\quad\ \tx{is\ \,even}\quad\ \tx{if}\ \quad n_{02}=1
\label{Decond1}
\qqq
and, for $\,Z=\{1,z_1z_2\}$, \,the condition
\qq
\sfk\quad\ \tx{is\ \,even}\quad\ \tx{if}\ \quad n_{01}+n_{02}
\quad\ \tx{is\ \,odd}.
\nonumber
\qqq
The obstruction cocycle restricted to $\,\G=\bZ_2\lx Z\,$ with 
$\,Z=\{1,z_1\}\,$ or $\,Z=\{1,z_1z_2\}\,$ becomes trivial
under the conditions (\ref{Decond}) and (\ref{Decond1}) or
(\ref{Decond}) and (\ref{Decond1}), respectively.
\vskip 0.2cm

Finally, for the maximal orientifold group, the conditions
(\ref{Decond}), (\ref{Decond1}) and (\ref{Decond}) must hold
simultaneously, implying that if the twist element $\,\zeta\not=1\,$
then the obstruction cocycle can be trivialized only if 
$\,\sfk\,$ is even. On the other hand, the trivializability
of $\,u_{\gamma,\gamma',\gamma''}\,$ cannot depend on 
the choice of the twist element in this case so that
for $\,\G=\bZ_2\lx(\bZ_2\times\bZ_2)\,$ the  cohomological 
equation (\ref{cohoeq}) has a solution only if
\qq
\sfk\quad\ \tx{is\ \,even}
\label{DecondZ}
\qqq
whatever the choice of the twist element. Indeed, if
$\,\zeta=1\,$ then the combination $\,X\,$ of (\ref{loeq}) 
with $\,n=10\,$ takes the value $\,(-1)^\sfk\,$
for the cocycle $\,u'_{\gamma,\gamma',\gamma''}\,$ 
obtained by composing $\,u_{\gamma,\gamma',\gamma''}\,$ with
the automorphism $\,h_{z_1}\,$ of $\,\Gamma$, \,see (\ref{twh}).
\,Since $\,u'_{\gamma,\gamma',\gamma''}\,$
is trivializable if and only if $\,u_{\gamma,\gamma',\gamma''}\,$ is,
the condition (\ref{DecondZ}) for the trivial twist element follows.
\vskip 0.2cm

For all orientifold groups $\,\G=\bZ_2\lx Z\,$ with a non-trivial 
orbifold subgroup $\,Z$, \,the obstruction 
cocycle $\,u_{\gamma,\gamma,\gamma''}\,$ of 
(\ref{Deobst1})-(\ref{Deobst2}) is then trivial whenever it may
be trivialized. Sixteen cohomologically inequivalent trivializing 2-cocycles
$\,v_{\gamma,\gamma'}\,$ on $\,\G=\bZ_2\lx(\bZ_2\times\bZ_2)\,$ are 
given by the formulae
\qq
&&v_{n_1n_2,n'_1n'_2}\,=\,v_{n_1n_2,\un{n'_1n'_2}}\,=\,\sigma^{n_2n'_1},\cr
&&v_{\un{n_1n_2},n'_1n'_2}\,=\,\sigma^{n_2n'_1}\,\sigma_1^{n'_1}
\,\sigma_2^{n'_2},\cr
&&v_{\un{n_1n_2},\un{n'_1n'_2}}\,=\,\un{\sigma}\,\sigma^{n_2n'_1}
\,\sigma_1^{n'_1}\,\sigma_2^{n'_2}
\nonumber
\qqq
with $\,\sigma,\sigma_1,\sigma_2,\un{\sigma}=\pm1$. 
\,In particular, the choice of $\,\sigma\,$ distinguishes two
inequivalent restrictions of the 2-cocycle $\,v_{\gamma,\gamma'}\,$
to the orbifold group $\,\bZ_2\times\bZ_2\,$ that give rise to two 
inequivalent gerbes on $\,Spin(2r)/(\bZ_2\times\bZ_2)$, \,see \cite{GR2}.  
For $\,\Gamma=\bZ_2\lx Z\,$ with $\,Z\cong\bZ_2$, \,four inequivalent
cohomologically non-trivial 2-cocycles are obtained from the above
expressions (with, say, $\,\sigma=1$) \,by restriction.
\vskip 0.2cm

Let us summarize the results for the $\,Spin(2r)\,$ group with
even $\,r$.  \,First, for each $\,\sfk\,$ and each of the four choices 
of the twist element, there are two inequivalent Jandl structures on 
the level $\,\sfk\,$ gerbe on $\,Spin(2r)$. \,Next, for each $\,\sfk\,$
even and each choice of the twist element, there are four inequivalent 
$\,\G$-equivariant structures on the level $\,\sfk\,$ gerbe on $\,Spin(2r)$
for $\,\G=\bZ_2\lx Z\,$ with $\,Z\cong\bZ_2$. \,They give rise to eight 
inequivalent Jandl structures on the induced gerbe on $\,Spin(2r)/Z$.
\,For such orientifold groups and $\,\sfk\,$ odd, there exist four 
inequivalent $\,\G$-equivariant structures only if the twist belongs 
to $\,Z\,$ and, for $\,Z=\{1,z_1\}\,$
or $\,Z=\{1,z_1z_2\}$, \,if, additionally, $\,s={r\over2}\,$ is
even. For fixed $\,Z$, \,we thus obtain four inequivalent Jandl
structures on the induced gerbe on $\,Spin(2r)/Z$. \,Finally,
for the maximal orientifold group $\,\G=\bZ_2\lx(\bZ_2\times\bZ_2)\,$ and
each $\,\sfk\,$ even, there exist sixteen inequivalent $\,\G$-equivariant
structures on the level $\,\sfk\,$ gerbe on $\,Spin(2r)\,$ for each
choice of the twist element. They give rise to, altogether, eight 
inequivalent Jandl structures on each of the two inequivalent
gerbes induced on $\,Spin(2r)/(\bZ_2\times\bZ_2)$.

\subsection{The case of $\,G=E_6$}\label{sec:E6}
\noindent As in Sect.\,4.7 of \cite{GR2}, we identify the
Cartan algebra $\,\Nt\,$ of $\,E_6\,$ with the subspace of
$\,\bR^7\,$ with the first six coordinates summing to zero.
The Killing form is inherited from the scalar product
in $\,\bR^7$. \,The vertices of the positive Weyl alcove 
$\,\CA\,$ are
\qq
\tau_0=0,\ \,\tau_1=\lambda_1^\vee,\ \,\tau_2={_1\over^2}
\lambda_2^\vee,\ \,\tau_3={_1\over^3}\lambda_3^\vee,
\ \,\tau_4={_1\over^2}\lambda_4^\vee,\ \,\tau_5=\lambda_5^\vee, 
\ \,\tau_6={_1\over^2}\lambda_6^\vee\qquad
\nonumber
\qqq
for the simple coweights $\,\lambda_i^\vee\,$ corresponding to the
simple roots
\qq
&\alpha_i\,=\,e_i-e_{i+1}\quad\ \tx{for}\ \quad i=1,\dots,5\,,&\cr
&\alpha_6\,=\,{_1\over^2}(-e_1-e_2-e_3+e_4+e_5+e_6)+{_1\over^{\sqrt{2}}}\,
e_7\,,&
\nonumber
\qqq
where $\,e_i\,$ are the vectors of the canonical basis of $\,\bR^7$. 
\,The positive roots have the form $\,e_i-e_j\,$ for $\,1\leq i<j\leq6$,
$\,{1\over2}(\pm e_1\pm e_2\pm e_3\pm e_4\pm e_5\pm e_6)+{1\over\sqrt{2}}
e_7\,$ with three signs $+$ and three signs $-$, \,and 
$\,\phi=\sqrt{2}\,e_7\,$ (the highest root). 
The center $\,Z(E_6)\cong\bZ_3\,$ is generated by 
$\,z=\ee^{-2\pi\sfi\,\lambda_5^\vee}$.
\,We shall construct the elements $\,w_z\,$ and $\,w_\kappa\,$
entering the relations (\ref{zact}) and (\ref{kact}) in terms
of group elements $\,w_\alpha=\ee^{{\pi\sfi\over2}(e_\alpha+e_{-\alpha})}\,$
that implement the Weyl reflections $\,r_\alpha\,$ in roots $\,\alpha$,
\,acting on the Cartan algebra by
\qq
\tau\ \longrightarrow\ w_\alpha\,\tau\,w_\alpha^{-1}\,
=\,\tau\,-\,\alpha^\vee\,\tr\,\tau\alpha\,
\equiv\,r_\alpha(\tau)\,,
\nonumber
\qqq
where $\,e_{\pm\alpha}\,$ and $\,\alpha^\vee\,$ stand for the step 
generators and the coroot associated to $\,\alpha$, \,respectively. One has
\qq
w_\alpha^2\,=\,\ee^{\pi\sfi\,\alpha^\vee}.
\nonumber
\qqq
Besides, since $\,[e_\alpha,e_\beta]\,$ does not vanish only if 
$\,\alpha+\beta\,$ is a root, $\,w_\alpha\,$ and $\,w_\beta\,$ commute 
if neither $\,\alpha+\beta\,$ nor $\,\alpha-\beta\,$ is a root.
The relation (\ref{kact}) is satisfied for $\,\tau\in\CA\,$ if we take
\qq
w_\kappa\,=\,w_{\alpha_3}\,w_{\alpha_2+\alpha_3+\alpha_4}\,
w_{\alpha_1+\alpha_2+\alpha_3+\alpha_4+\alpha_5}\,w_{\phi}\,, 
\label{wk}
\qqq
with the action of $\,\kappa\,$ on $\,\CA\,$
reducing to
\qq
\kappa\tau_0\,=\,\tau_0,\ \quad\kappa\tau_i\,=\,\tau_{6-i}\quad\tx{for}
\quad i=1,\dots,5,\ \quad\kappa\tau_6\,=\tau_6
\nonumber
\qqq
on the vertices and thereby giving rise to the symmetry of the Dynkin 
diagram represented in Fig.\ref{fig:E6K}.
\FIGURE[!h]{
\epsfxsize = 2.5in
\epsfysize = 1.9in
\centerline{
\epsfbox{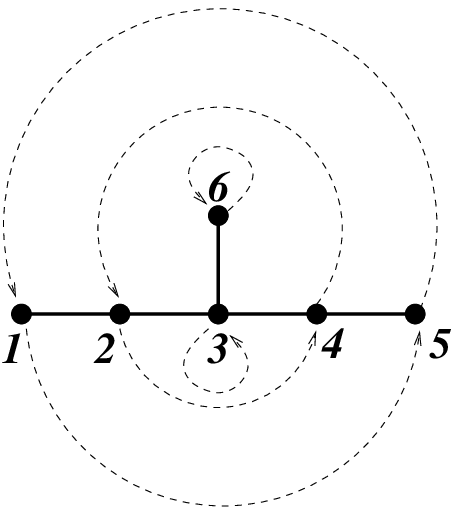}}
\vspace{-0.3cm}
\caption{{\sf The Weyl reflection of the Dynkin diagram of $E_6$ under 
$\kappa$.}}
\label{fig:E6K}}
It is easy to check that all the factors on the right hand side
of (\ref{wk}) commute so that
\qq
w_\kappa^2&=&w_{\alpha_3}^2\,w_{\alpha_2+\alpha_3+\alpha_4}^2\,
w_{\alpha_1+\alpha_2+\alpha_3+\alpha_4+\alpha_5}^2\,w_{\phi}^2&\cr\cr
&=&\ee^{\pi\sfi(\alpha_3^\vee+\alpha_2^\vee+\alpha_3^\vee+\alpha_4^\vee
+\alpha_1^\vee+\alpha_2^\vee+\alpha_3^\vee+\alpha_4^\vee
+\alpha_5^\vee+\phi^\vee)}\,=\,1\,.
\nonumber
\qqq 
As observed in \cite{GR2}, there is another set of simple roots
\qq
&&\beta_1\,=\,\alpha_1+\alpha_2+\alpha_3+\alpha_4\,,\qquad
\beta_2\,=\,\alpha_3+\alpha_4+\alpha_5+\alpha_6\,,\cr
&&\beta_3\,=\,-\alpha_1-\alpha_2-2\alpha_3-\alpha_4-\alpha_5-\alpha_6\,,\cr
&&\beta_4\,=\,\alpha_1+\alpha_2+\alpha_3+\alpha_6\,,\qquad
\beta_5\,=\,\alpha_2+\alpha_3+\alpha_4+\alpha_5\,,\cr
&&\beta_6\,=\,\alpha_3
\nonumber
\qqq
such that (\ref{zact}) may be satisfied for $\,\tau\in\CA\,$
if the adjoint action of $\,w_z\,$ on the Cartan algebra is given by the
product of the Weyl reflections
\qq
w_z\,\tau\,w_z^{-1}\,=\,r_{\beta_1}r_{\beta_4}r_{\beta_5}r_{\beta_2}(\tau)\,,
\nonumber
\qqq
with the action of $\,z\,$ on $\,\CA\,$ reducing to
\qq
&z\tau_0\,=\,\tau_1,\quad z\tau_1\,=\,\tau_5,\quad z\tau_2\,=\,\tau_4,&\cr 
&z\tau_3\,=\,\tau_3,\quad z\tau_4\,=\,\tau_6,\quad z\tau_5\,=\,\tau_0,\quad
z\tau_6\,=\,\tau_2&
\nonumber
\qqq
on the vertices and thus giving rise to the symmetry of the extended 
Dynkin diagram depicted in Fig.\ref{fig:E6}. 
\FIGURE[!h]{
\epsfxsize = 2.5in
\epsfysize = 1.9in
\centerline{
\epsfbox{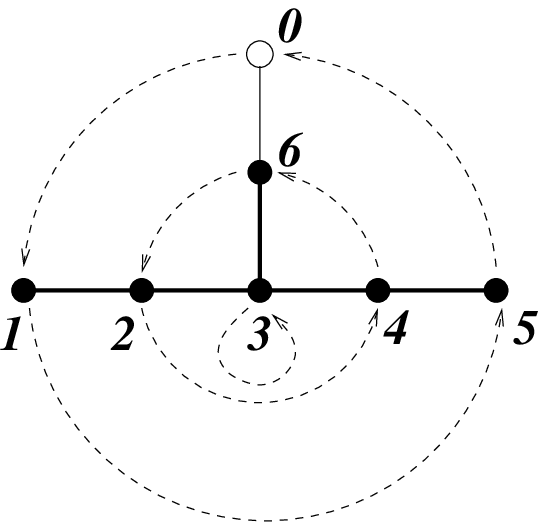}}
\vspace{-0.3cm}
\caption{{\sf The transformation of the extended Dynkin diagram of $E_6$
under $z$.}}
\label{fig:E6}}
\noindent Note that $\,w_\k\beta_1^\vee w_\k^{-1}=-\beta_5^\vee\,$  and 
$\,w_\k\beta_2^\vee w_\k^{-1}=-\beta_4^\vee$. \,It follows that
\qq
w_\k\,e_{\pm\beta_1}\,w_\k^{-1}\,=\,\mu_1^{\mp1}e_{\mp\beta_5},\qquad
w_\k\,e_{\pm\beta_2}\,w_\k^{-1}\,=\,\mu_2^{\mp1}e_{\mp\beta_4}
\nonumber
\qqq
for some $\,\mu_1\,$ and $\,\mu_2\,$ of absolute value $1$.
\,Hence 
\qq
w_\k\,w_{\beta_1}\,w_\k^{-1}\,=\,\ee^{{\pi\sfi\over2}(\mu_1e_{\beta_5}+
\bar\mu_1e_{-\beta_5})}\,,\qquad
w_\k\,w_{\beta_2}\,w_\k^{-1}\,=\,\ee^{{\pi\sfi\over2}(\mu_2e_{\beta_4}+
\bar\mu_2e_{-\beta_4})}\,.
\nonumber
\qqq
Since conjugation with $\,\ee^{{\pi\sfi\over2}
(\mu e_\alpha+\bar\mu e_{-\alpha})}\,$ induces the Weyl reflection 
$\,r_\alpha\,$ on the Cartan algebra for all $\,\mu\,$ with $\,|\mu|=1$,
we may set
\qq
w_z\,=\,w_{\beta_1}w_\k w_{\beta_2}w_\k^{-1}w_\k w_{\beta_1}w_\k^{-1}
w_{\beta_2}\,.
\nonumber
\qqq
The elements $\,e_{\pm\beta_i}\,$ with $\,i=1,\dots,5$ \,generate 
an $\,\gt{su}(6)\,$ subalgebra of the Lie algebra of $\,E_6$.
The coroots $\,\beta_i^\vee$ may be taken as its simple 
coroots and $\,e_{\pm\beta_i}\,$ as its step generators.
Clearly, $\,w_z\,$ belongs to the $\,SU(6)\,$ subgroup of $\,E_6\,$ 
corresponding to this subalgebra and, with the standard identification
of the simple roots and the step generators of $\,\gt{su}(6)\,$ in terms
of matrices,
\qq
w_z\ =\ \left(\begin{matrix}\ _0&\ \ _0&\,\ _{-1}&_0&_0&_0\cr\ _\sfi&\ \ _0
&\,\ _0&_0&_0&_0\cr
\ _0&\ \ _\sfi&\,\ _0&_0&_0&_0\cr\ _0&\ \ _0&\,\ _0&_0&_0&_{-\mu_1\mu_2}
\cr\ _0&\ \ _0&\,\ _0&_{\sfi\bar\mu_2}&_0&_0\cr\ _0&\ \ _0&\,\ _0&_0
&_{\sfi\bar\mu_1}&_0\end{matrix}\right)\ \in SU(6)\,\subset\,E_6\,.
\nonumber
\qqq
The relation 
\qq
w_z^3\ =\ 1
\label{wz3=1}
\nonumber
\qqq 
follows by raising the above matrix to the third power. 
Let us further note that, since $\,[e_{\beta_1},e_{\pm\beta_4}]=0\,$ 
and $\,[e_{\beta_2},e_{\pm\beta_5}]=0$, we have the commutation
relations:
\qq
w_{\beta_1}w_\k w_{\beta_2}w_\k^{-1}\,=\,w_\k w_{\beta_2}w_\k^{-1}
w_{\beta_1}\,,
\qquad
w_{\beta_2}w_\k w_{\beta_1}w_\k^{-1}\,=\,w_\k w_{\beta_1}w_\k^{-1}
w_{\beta_2}\,.\quad
\nonumber
\qqq
Using these identities and the equality $\,w_\k=w_\k^{-1}$, \,we infer that
\qq
&&(w_\k w_z)^2\,=\,
w_\k w_{\beta_1}w_\k^{-1}w_{\beta_2}w_{\beta_1}^2w_\k
w_{\beta_2}^2w_\k^{-1}w_\k w_{\beta_1}w_\k^{-1}w_{\beta_2}\cr
&&=\,w_\k w_{\beta_1}w_\k^{-1}w_{\beta_2}
\ee^{\pi\sfi(\beta_1^\vee+\beta_4^\vee)}w_{\beta_2}w_\k 
w_{\beta_1}w_\k^{-1}\,.
\nonumber
\qqq
Next, the relations $\,[\beta_1^\vee+\beta_4^\vee,e_{\pm\beta_2}]
=\mp e_{\pm\beta_2}\,$ imply that
\qq
\ee^{\pi\sfi(\beta_1^\vee+\beta_4^\vee)}w_{\beta_2}
\ee^{-\pi\sfi(\beta_1^\vee+\beta_4^\vee)}\,=\,
w_{\beta_2}^{-1}\,.
\nonumber
\qqq
Similarly,
\qq
\ee^{\pi\sfi(\beta_1^\vee+\beta_4^\vee)}w_kw_{\beta_1}w_k^{-1}
\ee^{-\pi\sfi(\beta_1^\vee+\beta_4^\vee)}\,=\,
w_kw_{\beta_1}^{-1}w_k^{-1}\,
\nonumber
\qqq
so that we obtain the identities:
\qq
&&(w_kw_z)^2\,=\,\ee^{\pi\sfi(\beta_1^\vee+\beta_4^\vee)}\,=\,
\ee^{\pi\sfi(\alpha_4^\vee+\alpha_6^\vee)}\,,\cr
&&(w_zw_k)^2\,=\,w_k(w_kw_z)^2w_k^{-1}\,=\,
\ee^{\pi i(\alpha_2^\vee+\alpha_6^\vee)}\,.
\nonumber
\qqq
It follows easily that we may choose:
\qq
&&b_{n,n'}\,=\,b_{\un{n},n'}\,=\,0,\cr\cr
&&b_{n,\un{n}'}\,=\,\begin{cases}\,\hbox to 3cm{$0$\hfill}\tx{for}
\ \quad n=0,\cr
\,\hbox to 3cm{${1\over2}(\alpha_2^\vee+\alpha_6^\vee)$\hfill}\tx{for}
\ \quad n=1,\cr
\,\hbox to 3cm{${1\over2}(\alpha_4^\vee+\alpha_6^\vee)$\hfill}\tx{for}
\ \quad n=2,\end{cases}\cr\cr
&&b_{\un{n},\un{n}'}\,=\,\begin{cases}\,\hbox to 3cm{$0$\hfill}\tx{for}
\ \quad [n_0+n]=0,\cr
\,\hbox to 3cm{${1\over2}(\alpha_4^\vee+\alpha_6^\vee)$\hfill}\tx{for}
\ \quad [n_0+n]=1,\cr
\,\hbox to 3cm{${1\over2}(\alpha_2^\vee+\alpha_6^\vee)$\hfill}\tx{for}
\ \quad [n_0+n]=2.\end{cases}\cr
\nonumber
\qqq
Since $\,\tau_{z^{-1}0}=\lambda_5^\vee\,$ and $\,\tau_{z^{-2}0}
=\lambda_1^\vee$, \,it follows from the definition (\ref{g3cocyc}) that
the obstruction cocycle $\,u_{\gamma,\gamma',\gamma''}\,$
is trivial on both orientifold groups $\,\G=\bZ_2\lx\bZ_3\,$ and 
$\,\G=\bZ_2\,$ so that two cohomologically inequivalent cocycles
may be taken in the form
\qq
v_{n,n'}\,=\,v_{\un{n},n'}\,=\,v_{n,\un{n}'}\,=\,1,\qquad
v_{\un{n},\un{n}'}\,=\,\pm1.
\nonumber
\qqq
\vskip 0.2cm

In short, for each orientifold group, each $\,\sfk\,$ and each 
of the three choices of the twist element, there are two inequivalent 
$\,\G$-equivariant structures on the level $\,\sfk\,$ gerbe on $\,E_6$. 
\,They give rise to six inequivalent Jandl structures on that gerbe 
and to two inequivalent Jandl structures on the induced gerbe 
on $\,E_6/\bZ_3$.

\subsection{The case of $\,G=E_7$}\label{sec:E7}
\noindent As in Sect.\,4.8 of \cite{GR2}, we identify the Cartan algebra 
of $\,E_7\,$ with the subspace of the vectors in $\,\bR^8\,$ whose
coordinates sum to zero, with the Killing form inherited from
the scalar product in $\,\bR^8$. \,The vertices of the positive
Weyl alcove $\,\CA\,$ are
\qq
&\tau_0=0,\quad\tau_1=\lambda_1^\vee,\quad\tau_2={_1\over^2}
\lambda_2^\vee,\quad\tau_3={_1\over^3}\lambda_3^\vee,&\cr
&\tau_4={_1\over^4}\lambda_4^\vee,\quad\tau_5={1\over^3}\lambda_5^\vee, 
\quad\tau_6={_1\over^2}\lambda_6^\vee,\quad\tau_7={_1\over^2}\lambda_7^\vee&
\nonumber
\qqq
for the simple coweights $\,\lambda_i^\vee\,$ corresponding to the
simple roots
\qq
&\alpha_i\,=\,e_i-e_{i+1}\quad\ \tx{for}\ \quad i=1,\dots,6\,,&\cr
&\alpha_7\,=\,{_1\over^2}(-e_1-e_2-e_3-e_4+e_5+e_6+e_7+e_8),&
\nonumber
\qqq
where $\,e_i\,$ are the vectors of the canonical basis of $\,\bR^8$. 
\,Roots have the form $\,e_i-e_j\,$ for $\,i\not=j\,$ and
$\,{1\over2}(\pm e_1\pm e_2\pm e_3\pm e_4\pm e_5\pm e_6\pm e_7\pm e_8)\,$ 
with four signs $+$ and four signs $-$. \,The highest 
root is $\,\phi=-e_7+e_8$. 
\,The center $\,Z(E_7)\cong\bZ_2\,$ is generated by 
$\,z=\ee^{-2\pi\sfi\,\lambda_1^\vee}$, \,with $\,\lambda_1^\vee=
{1\over4}(3e_1-e_2-e_3-e_4-e_5-e_6-e_7+3e_8)$. \,The relation (\ref{kact})
may be satisfied for $\,\tau\in\CA\,$ if
\qq
w_\kappa\,=\,w_{\alpha_1}\,w_{\alpha_3}\,w_{\alpha_5}\,w_{\alpha_7}
\,w_{\alpha_3+2\alpha_4+\alpha_5+\alpha_7}\,w_{\alpha_1+2\alpha_2+2\alpha_3
+2\alpha_4+\alpha_5+\alpha_7}\,w_{\phi}\,,
\label{wke7}
\qqq
with the trivial action of $\,\kappa\,$ on $\,\CA$.
\,All the factors on the right hand side of (\ref{wke7}) commute so that
\qq
w_\kappa^2\,=\,\ee^{\pi\sfi(\alpha_1^\vee+\alpha_3^\vee+\alpha_7^\vee)}\,=\,
z\,.
\nonumber
\qqq
The roots
\qq
\b_1 &=& \a_1 + 2 \a_2 + 2\a_3 + 2\a_4 + \a_5 + \a_7, \cr
\b_2 &=& - (\a_1 + \a_2 + 2 \a_3 + 2\a_4 + \a_5 + \a_7 ), \cr
\b_3 &=& \a_1 + \a_2 + 2\a_3 + 2\a_4 + \a_5 + \a_6 + \a_7, \cr
\b_4 &=& - (\a_1 + \a_2 + \a_3 + 2 \a_4 + \a_5 + \a_6 + \a_7), \cr 
\b_5 &=& \a_4, \cr
\b_6 &=& \a_7,\cr
\b_7 &=& \a_1 + \a_2 + \a_3 + 2\a_4 + 2\a_5 + \a_6 + \a_7,
\nonumber
\qqq
form another system of simple roots such that (\ref{zact}) 
may be satisfied for $\,\tau\in\CA\,$ if the adjoint action 
of $\,w_z\,$ on the Cartan algebra is given by the
product of the Weyl reflections
\qq
w_z\,\tau\,w_z^{-1}\,=\,r_{\beta_1}r_{\beta_3}r_{\beta_7}(\tau)\,,
\nonumber
\qqq
with the action of $\,z\,$ on $\,\CA\,$ reducing to
\qq
z\tau_0\,=\,\tau_1,\quad z\tau_1\,=\,\tau_0,\quad z\tau_i\,
=\,\tau_{8-i}\quad\tx{for}\quad i=2,\dots,6,\quad z\tau_7=\tau_7
\nonumber
\qqq
 on the vertices, as illustrated in Fig.\ref{fig:E7}. 
\FIGURE[!h]{
\epsfxsize = 2.5in
\epsfysize = 1.8in
\centerline{
\epsfbox{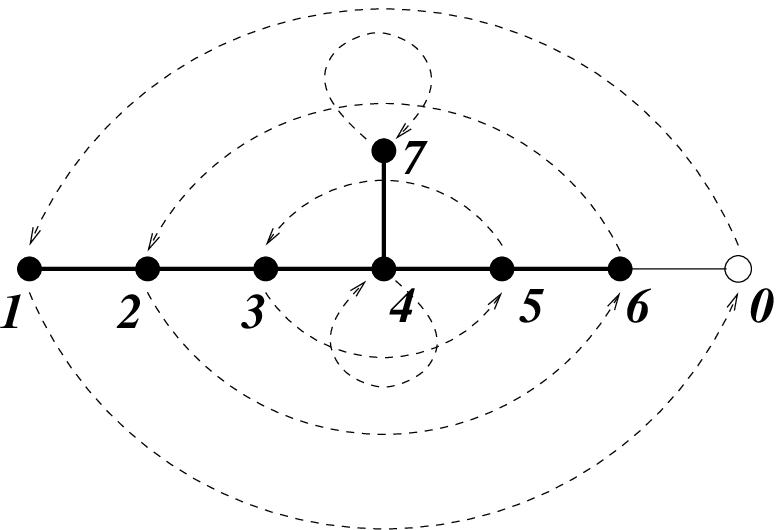}}
\vspace{-0.3cm}
\caption{{\sf The transformation of the extended Dynkin diagram of $E_7$
under $z$.}}
\label{fig:E7}}
Since $\,w_\kappa \beta_i^\vee w_\kappa^{-1}=-\beta_i^\vee$, \,we must have
\qq
w_\kappa\,e_{\pm\beta_i}\,w_\kappa^{-1}\,=\,\mu_i^{\mp 2}e_{\mp\beta_i}
\nonumber
\qqq
for some $\,\mu_i\,$ with $\,|\mu_i|=1$. \,Let 
\qq
\tilde w_{\beta_i}\,=\,\ee^{{\pi\sfi\over2}(\mu_i e_{\beta_i}
+\bar\mu_i e_{-\beta_i})}.
\nonumber
\qqq
Conjugation with $\,\tilde w_{\beta_i}\,$ still induces the Weyl 
reflections $\,r_{\beta_i}\,$ on the Cartan algebra and,
similarly as for $\,w_{\beta_i}$, $\ \tilde w_{\beta_i}^2
=\ee^{\pi\sfi\,\beta_i^\vee}$. \,We may then take
\qq
w_z\,=\tilde w_{\beta_1}\,\tilde w_{\beta_3}\,\tilde w_{\beta_7}\,.
\nonumber
\qqq
Since $\,\tilde w_{\beta_1}$, $\,\tilde w_{\beta_3}\,$ and
$\,\tilde w_{\beta_7}\,$ commute, the relation
\qq
w_z^2\,=\,\ee^{\pi\sfi(\beta_1^\vee+\beta_3^\vee+\beta_7^\vee)}
\,=\,\ee^{\pi\sfi(\alpha_1^\vee+\alpha_3^\vee+\alpha_7^\vee)}\,=\,z\,
\nonumber
\qqq
holds. By construction, $\,w_\kappa\tilde w_{\beta_i}w_\kappa^{-1}=
\tilde w_{\beta_i}$. \,Hence
\qq
(w_\kappa w_z)^2\,=\,w_\kappa w_z w_\kappa^{-1}w_z^{-1}\,=\,1\,.
\nonumber
\qqq
It follows that we may set:
\qq
b_{n,n'}\,=\,b_{\un{n},n'}\,=\,b_{n,\un{n}'}\,=\,nn'\lambda_1^\vee,
\qquad b_{\un{n},\un{n}'}\,=\,(1+n_0+nn')\lambda_1^\vee
\nonumber
\qqq
which, with the help of the relations $\,\tau_{z^{-n}0}
=\delta_{[n],1}\lambda_1^\vee\,$ and $\,\tr(\lambda_1^\vee)^2={3\over2}$, 
\,gives rise to the obstruction cocycle (\ref{g3cocyc}) on 
$\,\G=\bZ_2\lx\bZ_2\,$ of the form:
\qq
&&u_{n,n',n''}\,=\,u_{n,\un{n}',n''}\,=\,u_{n,n',\un{n}''}\,=\,
(-1)^{\sfk\,nn'n''},\cr
&&u_{n,\un{n}',\un{n}''}\,=\,(-1)^{\sfk\,n(1+n_0+n'n'')},\cr
&&u_{\un{n},n',n''}\,=\,u_{\un{n},\un{n}',n''}\,=\,u_{\un{n},n',\un{n}''}
\,=\,(-1)^{\sfk\,(n_0+n)n'n''},\cr
&&u_{\un{n},\un{n}',\un{n}''}\,=\,(-1)^{\sfk\,(n+n_0)(1+n_0+n'n'')}.
\nonumber
\qqq
The restriction of this cocycle to the orbifold subgroup $\,Z(E_7)\,$
may be trivialized if and only if
\qq
\sfk\ \quad\tx{is\ \,even,}
\nonumber
\qqq
in which case the whole cocycle becomes trivial. As four cohomologically 
inequivalent trivializing cochains we may take the cocycles
\qq
v_{n,n'}\,=\,1,\qquad v_{n,\un{n}'}\,=\,1,\quad\ 
v_{\un{n},n'}\,=\,\sigma^{n'},\quad\ v_{\un{n},\un{n}'}\,=\,
\un{\sigma}\,\sigma^{n'}
\label{trcoE7}
\qqq
for $\,\sigma,\un{\sigma}=\pm1$.
On the other hand, the restriction of the obstruction cocycle to 
the inversion group $\,\bZ_2\,$ is trivial for all $\,\sfk$.
Two cohomologically nonequivalent trivializing cocycles
may be obtained by restriction of (\ref{trcoE7}) to $\,n=n'=0$.
\vskip 0.2cm

To summarize, for each $\,\sfk\,$ and each of the two choices 
of the twist element,
there are two inequivalent Jandl structures on the level $\,\sfk\,$ gerbe 
on $\,E_7$. \,For $\,\sfk\,$ even and each choice of the twist element, there
are four inequivalent $\,\bZ_2\lx\bZ_2$-equivariant structures on the
level $\,\sfk\,$ gerbe on $\,E_7$, \,giving rise to, altogether, four 
Jandl structures on the induced gerbe on $\,E_7/\bZ_2$. \,There are no
$\,\bZ_2\lx\bZ_2$-equivariant structures for $\,\sfk\,$ odd.

\subsection{The cases of $G=E_8,\,F_4,\,G_2$}\label{sec:E8F4G2}
\noindent These are the simple groups with a trivial center 
and no non-trivial Dynkin diagram symmetries. The only possible orientifold 
group is the inversion group $\,\G=\bZ_2\,$ and whatever the values 
of $\,b_{\gamma,\gamma'}\,$ the obstruction 3-cocycle (\ref{g3cocyc}) 
is trivial since $\,\tau_{\gamma^{-1}0}=\tau_0=0\,$ for all 
$\,\gamma\in\G$. Two cohomologically inequivalent trivializing cochains 
are given by the 2-cocycles of (\ref{coC}). They give rise to 
two inequivalent Jandl structures on the level $\,\sfk\,$ gerbe 
on $\,G\,$ for each $\,\sfk$.

\section{Conclusions}\label{sec:concl}
\noindent We have studied orientifolds of the WZW theories with 
simple compact simply connected groups $\,G\,$ as targets. \,For orientifold 
groups $\,\Gamma=\bZ_2\lx Z$, \,where the generator of $\,\bZ_2\,$ acts 
by a twisted inversion $\,g\mapsto(\zeta g)^{-1}\,$ on $\,G\,$ and 
$\,Z\,$ is a subgroup of the center of $\,G$, \,we have classified 
all inequivalent $\,\Gamma$-equivariant structures on the level 
$\,\sfk\,$ gerbes on groups $\,G$. \,Such structures are required 
to unambiguously define Feynman amplitudes of classical fields 
of the orientifold theory. For $\,Z\,$ of even order, there may 
be obstructions to existence of the orientifold theory with
a given twist $\,\zeta\,$ even if the $\,Z$-orbifold theory exists. 
The classification of the 
$\,\Gamma$-equivariant structures on the level $\,\sfk\,$ gerbe 
on $\,G\,$ descends to the classification of the Jandl structures 
\cite{SSW} on the induced gerbe on the quotient group $\,G/Z$.
There exists an even number, at least two, of such induced Jandl 
structures, giving rise to different orientifold extensions 
of the $\,Z$-orbifold theory, i.e. to different unoriented 
closed string theories with the $\,G/Z\,$ target space.
Our results also show that, in all cases 
except for $\,G=Spin(8n)\,$ and $\,Z=\bZ_2\times\bZ_2$, \,the only 
obstructions to existence of a $\,\Gamma$-equivariant structure
with the trivial twist element $\,\zeta=1\,$ are the ones that obstruct 
existence of a $\,Z$-equivariant structure. \,In the exceptional 
case, $\,Z$-equivariant structures exist (two inequivalent ones
\cite{GR2}) for all integer levels $\,\sfk$, \,whereas 
$\,\Gamma$-equivariant ones with the trivial twist element exist only 
for $\,\sfk\,$ even. In \cite{FGK}, an additional 
condition was imposed on the $\,Z$-orbifold theory, see (2.15) therein, 
that is equivalent to existence of a $\,\Gamma$-equivariant 
structure with the trivial twist element. This condition, that was unjustly 
related to unitarity of the $\,Z$-orbifold theory, eliminated 
odd levels $\sfk\,$ for the $\,SO(8n)/\bZ_2\,$ WZW theory 
(in fact, the unitarity holds also for odd $\,\sfk\,$ theories; 
what fails is the left-right symmetry of the toroidal partition 
functions).
 
\,As we shall discuss in \cite{GSW2}, our results, based on a 
systematic geometric approach to the classical orientifold theory, 
are in agreement with the ones obtained in \cite{BH} by studying 
the sewing and modular invariance constraints for the crosscap 
states in the simple-current orbifolds of the WZW theory.
\vfill
\eject

\section{Appendix}\label{sec:append}
\noindent Here is a short list of results, with the signs $\,\sigma=\pm1$,
$\,\sigma_1=\pm1$, $\,\sigma_2\pm1\,$ and $\,\un{\sigma}=\pm1\,$ describing
different choices of trivializing cochains.
\qq
&&=================================================\cr
&&\hbox to 4.8cm{\bf Group\hfill}{\bf A_r}\cr
&&\hbox to 4.8cm{center\hfill}\bZ_{r+1}\cr
&&\hbox to 4.8cm{twist element\hfill}n_0=0,1,\dots,r\cr
&&--------------------------------------\cr\cr
&&\hbox to 4.8cm{orientifold group\hfill}\bZ_2\lx\bZ_m, \ \,m\,\ \tx{odd}\,\cr
&&\hbox to 4.8cm{level\hfill}\sfk\in\bZ\cr
&&\hbox to 4.8cm{trivializing cochain for\hfill}n,n'=0,{_{r+1}\over^m},
\dots,{_{r+1}\over^m}(m-1)\cr\cr
&&\hspace{0.4cm}\hbox to 5.2cm{$v_{n,n'}=\ee^{{2\pi\sfi\sfk\over r+1}nn'},$
\hfill}
v_{\un{n},n'}=(-1)^{\sfk rn_0{mn'\over r+1}}\,
\ee^{-{2\pi\sfi\sfk\over r+1}nn'},\cr\cr
&&\hspace{0.4cm}\hbox to 5.2cm{$v_{n,\un{n}'}=\ee^{{\pi\sfi\sfk\over r+1}
(2nn'-n^2-(r+1)n)},$\hfill}
v_{\un{n},\un{n}'}=\un{\sigma}\,\ee^{{\pi\sfi\sfk\over r+1}
\left(-n_0(n_0+{r(r+1)\over2})-2n(n_0+n')+n^2+(r+1)n\right)}\cr\cr
&&--------------------------------------\cr\cr
&&\hbox to 4.8cm{orientifold group\hfill}\bZ_2\lx\bZ_m, \ \,m\,\ \tx{even}\,\cr
&&\hbox to 4.8cm{level\hfill}\sfk\in\bZ\quad\tx{if}\ \ {_{r+1}\over^m}\ \ 
\tx{and}\ \ n_0\quad\tx{are\ \ even}, \ \ \sfk\in2\bZ\ \ \tx{otherwise}\cr
&&\hbox to 4.8cm{trivializing cochain for\hfill}n,n'=0,{_{r+1}\over^m},
\dots,{_{r+1}\over^m}(m-1)\cr\cr
&&\hspace{0.4cm}\hbox to 5.2cm{$v_{n,n'}=\ee^{{2\pi\sfi\sfk\over r+1}nn'},$
\hfill}
v_{\un{n},n'}=\sigma^{mn'\over r+1}\,(-1)^{\sfk rn_0{mn'\over r+1}}\,
\ee^{-{2\pi\sfi\sfk\over r+1}nn'},\cr\cr
&&\hspace{0.4cm}\hbox to 5.2cm{$v_{n,\un{n}'}=\ee^{{\pi\sfi\sfk\over r+1}
(2nn'-n^2-(r+1)n)},$\hfill}
v_{\un{n},\un{n}'}=\un{\sigma}\,\sigma^{mn'\over r+1}\,
\ee^{{\pi\sfi\sfk\over r+1}
\left(-n_0(n_0+{r(r+1)\over2})-2n(n_0+n')+n^2+(r+1)n\right)}\qquad
\cr\cr
&&=================================================\cr
&&\hbox to 4.8cm{\bf Group\hfill}{\bf B_r}\cr
&&\hbox to 4.8cm{center\hfill}\bZ_2\cr
&&\hbox to 4.8cm{twist element\hfill}n_0=0,1\cr
&&--------------------------------------\cr\cr
&&\hbox to 4.8cm{orientifold group\hfill}\bZ_2\cr
&&\hbox to 4.8cm{level\hfill}\sfk\in\bZ\cr
&&\hbox to 4.8cm{trivializing cochain\hfill}\cr\cr
&&\hspace{0.4cm}\hbox to 5.2cm{$v_{0,0}=v_{\un{0},0}=v_{0,\un{0}}=1,$\hfill}
v_{\un{0},\un{0}}=\un{\sigma}\,\ee^{{\pi\sfi\sfk\over2}n_0}\cr\cr
&&--------------------------------------\cr\cr
&&\hbox to 4.8cm{orientifold group\hfill}\bZ_2\lx\bZ_2\cr
&&\hbox to 4.8cm{level\hfill}\sfk\in\bZ\cr
&&\hbox to 4.8cm{trivializing cochain for\hfill}n,n'=0,1\cr\cr
&&\hspace{0.4cm}\hbox to 5.2cm{$v_{n,n'}=(-1)^{\sfk\,nn'},$\hfill}
v_{\un{n},n'}=\sigma^{n'}\,(-1)^{\sfk\,nn'},\cr
&&\hspace{0.4cm}\hbox to 5.2cm{$v_{n,\un{n}'}
=(-1)^{\sfk\,nn'}\,\ee^{{-3\pi\sfi\sfk\over2}n},$\hfill}
v_{\un{n},\un{n}'}=\un{\sigma}\,\sigma^{n'}\,(-1)^{\sfk\,n(n_0+n')}\,
\ee^{{\pi\sfi\sfk\over2}(n_0+3n)}\cr\cr
&&=================================================\cr
&&\hbox to 4.8cm{\bf Group\hfill}{\bf C_r}\cr
&&\hbox to 4.8cm{center\hfill}\bZ_2\cr
&&\hbox to 4.8cm{twist element\hfill}n_0=0,1\cr
&&--------------------------------------\cr\cr
&&\hbox to 4.8cm{orientifold group\hfill}\bZ_2\cr
&&\hbox to 4.8cm{level\hfill}\sfk\in\bZ\cr
&&\hbox to 4.8cm{trivializing cochain\hfill}\cr\cr
&&\hspace{0.4cm}\hbox to 5.2cm{$v_{0,0}=v_{\un{0},0}=v_{0,\un{0}}=1,$\hfill}
v_{\un{0},\un{0}}=\un{\sigma}\cr\cr
&&--------------------------------------\cr\cr
&&\hbox to 4.8cm{orientifold group\hfill}\bZ_2\lx\bZ_2\cr
&&\hbox to 4.8cm{level\hfill}\sfk\in\bZ\ \ \tx{if}\ \ r\ \ \tx{is\ \ even},
\ \ \ \sfk\in2\bZ\ \ \tx{otherwise}\cr
&&\hbox to 4.8cm{trivializing cochain for\hfill}n,n'=0,1\cr\cr
&&\hspace{0.4cm}\hbox to 5.2cm{$v_{n,n'}=1,$\hfill}
v_{\un{n},n'}=\sigma^{n'},\cr
&&\hspace{0.4cm}\hbox to 5.2cm{$v_{n,\un{n}'}
=1,$\hfill}
v_{\un{n},\un{n}'}=\un{\sigma}\,\sigma^{n'}\cr\cr
&&=================================================\cr
&&\hbox to 4.8cm{\bf Group\hfill}{\bf D_{r}}\ \ \tx{for}\ \ r\ \ \tx{odd}\cr
&&\hbox to 4.8cm{center\hfill}\bZ_4\cr
&&\hbox to 4.8cm{twist element\hfill}n_0=0,1,2,3\cr
&&--------------------------------------\cr\cr
&&\hbox to 4.8cm{orientifold group\hfill}\bZ_2\cr
&&\hbox to 4.8cm{level\hfill}\sfk\in\bZ\cr
&&\hbox to 4.8cm{trivializing cochain\hfill}\cr\cr
&&\hspace{0.4cm}\hbox to 5.2cm{$v_{0,0}=v_{\un{0},0}=v_{0,\un{0}}=1,$\hfill}
v_{\un{0},\un{0}}=\un{\sigma}\,\ee^{-{\pi\sfi\over4}\sfk r(n_0
+2\delta_{n_0,3})}\cr\cr
&&--------------------------------------\cr\cr
&&\hbox to 4.8cm{orientifold group\hfill}\bZ_2\lx\bZ_2\cr
&&\hbox to 4.8cm{level\hfill}\sfk\in\bZ\ \ \tx{if}\ \ n_0\ \ \tx{is\ \ even},
\ \ \ \sfk\in2\bZ\ \ \tx{otherwise}\cr
&&\hbox to 4.8cm{trivializing cochain for\hfill}n,n'=0,2\cr\cr
&&\hspace{0.4cm}\hbox to 5.2cm{$v_{n,n'}=1,$\hfill}
v_{\un{n},n'}=\sigma^{{1\over2}n'},\cr
&&\hspace{0.4cm}\hbox to 5.2cm{$v_{n,\un{n}'}
=\ee^{{\pi\sfi\sfk\over4}n},$\hfill}
v_{\un{n},\un{n}'}=\un{\sigma}\,\sigma^{{1\over2}n'}\,
\ee^{-{\pi\sfi\sfk\over4}([n_0+n]+2\delta_{[n_0+n],3})}\cr\cr
&&--------------------------------------\cr\cr
&&\hbox to 4.8cm{orientifold group\hfill}\bZ_2\lx\bZ_4\cr
&&\hbox to 4.8cm{level\hfill}\sfk\in2\bZ\cr
&&\hbox to 4.8cm{trivializing cochain for\hfill}n,n'=0,1,2,3\cr\cr
&&\hspace{0.4cm}\hbox to 5.2cm{$v_{n,n'}=1,$\hfill}
v_{\un{n},n'}=\sigma^{n'},\cr
&&\hspace{0.4cm}\hbox to 5.2cm{$v_{n,\un{n}'}
=\ee^{{\pi\sfi\sfk\over4}(n+2\delta_{n,3})},$\hfill}
v_{\un{n},\un{n}'}=\un{\sigma}\,\sigma^{n'}\,
\ee^{-{\pi\sfi\sfk\over4}([n_0+n]+2\delta_{[n_0+n],3})}\cr\cr
&&=================================================\cr
&&\hbox to 4.8cm{\bf Group\hfill}{\bf D_{r}}\ \ \tx{for}\ \ r\ \ \tx{even}\cr
&&\hbox to 4.8cm{center\hfill}\bZ_2\times\bZ_2=\{n_1n_2\,|\,n_1,n_2=0,1\}\cr
&&\hbox to 4.8cm{twist element\hfill}n_{01}n_{02}=00,10,01,11\cr
&&--------------------------------------\cr\cr
&&\hbox to 4.8cm{orientifold group\hfill}\bZ_2\cr
&&\hbox to 4.8cm{level\hfill}\sfk\in\bZ\cr
&&\hbox to 4.8cm{trivializing cochain\hfill}\cr\cr
&&\hspace{0.4cm}\hbox to 5.2cm{$v_{00,00}=v_{\un{00},00}
=v_{00,\un{00}}=1,$\hfill}
v_{\un{00},\un{00}}=\un{\sigma}\cr\cr
&&--------------------------------------\cr\cr
&&\hbox to 4.8cm{orientifold group\hfill}\bZ_2\lx\{n0\,|\,n=0,1\}\cr
&&\hbox to 4.8cm{level\hfill}\sfk\in\bZ\ \ \tx{if}\ \ {_r\over^2}
\ \ \tx{is\ \ even\ \ and}\ \ n_{02}=0, 
\ \ \ \sfk\in2\bZ\ \ \tx{otherwise}\cr
&&\hbox to 4.8cm{trivializing cochain for\hfill}n,n'=0,1\cr\cr
&&\hspace{0.4cm}\hbox to 5.2cm{$v_{n0,n'0}=1,$\hfill}
v_{\un{n0},n'0}=\sigma^{n'},\cr
&&\hspace{0.4cm}\hbox to 5.2cm{$v_{n0,\un{n'0}}
=1,$\hfill}
v_{\un{n0},\un{n0}'}=\un{\sigma}\,\sigma^{n'}\cr\cr
&&--------------------------------------\cr\cr
&&\hbox to 4.8cm{orientifold group\hfill}\bZ_2\lx\{0n\,|\,n=0,1\}\cr
&&\hbox to 4.8cm{level\hfill}\sfk\in\bZ\ \ \tx{if}\ \ n_{01}=0,
\ \ \ \sfk\in2\bZ\ \ \tx{otherwise}\cr
&&\hbox to 4.8cm{trivializing cochain for\hfill}n,n'=0,1\cr\cr
&&\hspace{0.4cm}\hbox to 5.2cm{$v_{0n,0n'}=1,$\hfill}
v_{\un{0n},0n'}=\sigma^{n'},\cr
&&\hspace{0.4cm}\hbox to 5.2cm{$v_{0n,\un{0n'}}
=1,$\hfill}
v_{\un{0n},\un{0n}'}=\un{\sigma}\,\sigma^{n'}\cr\cr
&&--------------------------------------\cr\cr
&&\hbox to 4.8cm{orientifold group\hfill}\bZ_2\lx\{nn\,|\,n=0,1\}\cr
&&\hbox to 4.8cm{level\hfill}\sfk\in\bZ\ \,\tx{if}\ \,{_r\over^2}\ \,  
\tx{and}\ \,n_{01}+n_{02}\ \,\tx{are\ \,even}, 
\ \ \ \sfk\in2\bZ\ \,\tx{otherwise}\cr
&&\hbox to 4.8cm{trivializing cochain for\hfill}n,n'=0,1\cr\cr
&&\hspace{0.4cm}\hbox to 5.2cm{$v_{nn,n'n'}=1,$\hfill}
v_{\un{nn},n'n'}=\sigma^{n'},\cr
&&\hspace{0.4cm}\hbox to 5.2cm{$v_{nn,\un{n'n'}}=1,$\hfill}
v_{\un{nn},\un{n'n'}}=\un{\sigma}\,\sigma^{n'}\cr\cr
&&--------------------------------------\cr\cr
&&\hbox to 4.8cm{orientifold group\hfill}\bZ_2\lx(\bZ_2\times\bZ_2)\cr
&&\hbox to 4.8cm{level\hfill}\sfk\in2\bZ\cr
&&\hbox to 4.8cm{trivializing cochain for\hfill}n_1,n_2,n'_1,n'_2=0,1\cr\cr
&&\hspace{0.4cm}\hbox to 5.2cm{$v_{n_1n_2,n'_1n'_2}=\sigma^{n_2n'_1},$\hfill}
v_{\un{n_1n_2},n'_1n'_2}=\sigma^{n_2n'_1}\,\sigma_1^{n'_1}\,\sigma_2^{n'_2},\cr
&&\hspace{0.4cm}\hbox to 5.2cm{$v_{n_1n_2,\un{n'_1n'_2}}
=\sigma^{n_2n'_1},$\hfill}
v_{\un{n_1n_2},\un{n'_1n'_2}}=\un{\sigma}\,\sigma^{n_2n'_1}\,\sigma_1^{n'_1}\,
\sigma_2^{n'_2}\cr\cr
&&=================================================\cr
&&\hbox to 4.8cm{\bf Group\hfill}{\bf E_6}\cr
&&\hbox to 4.8cm{center\hfill}\bZ_3\cr
&&\hbox to 4.8cm{twist element\hfill}n_0=0,1,2\cr
&&--------------------------------------\cr\cr
&&\hbox to 4.8cm{orientifold group\hfill}\bZ_2\cr
&&\hbox to 4.8cm{level\hfill}\sfk\in\bZ\cr
&&\hbox to 4.8cm{trivializing cochain\hfill}\cr\cr
&&\hspace{0.4cm}\hbox to 5.2cm{$v_{0,0}=v_{\un{0},0}=v_{0,\un{0}}=1,$\hfill}
v_{\un{0},\un{0}}=\un{\sigma}\cr\cr
&&--------------------------------------\cr\cr
&&\hbox to 4.8cm{orientifold group\hfill}\bZ_2\lx\bZ_3\cr
&&\hbox to 4.8cm{level\hfill}\sfk\in\bZ\cr
&&\hbox to 4.8cm{trivializing cochain for\hfill}n,n'=0,1,2\cr\cr
&&\hspace{0.4cm}\hbox to 5.2cm{$v_{n,n'}=v_{\un{n},n'}=v_{n,\un{n}'}=1,$\hfill}
v_{\un{n},\un{n}'}=\un{\sigma}\cr\cr
&&=================================================\cr
&&\hbox to 4.8cm{\bf Group\hfill}{\bf E_7}\cr
&&\hbox to 4.8cm{center\hfill}\bZ_2\cr
&&\hbox to 4.8cm{twist element\hfill}n_0=0,1\cr
&&--------------------------------------\cr\cr
&&\hbox to 4.8cm{orientifold group\hfill}\bZ_2\cr
&&\hbox to 4.8cm{level\hfill}\sfk\in\bZ\cr
&&\hbox to 4.8cm{trivializing cochain\hfill}\cr\cr
&&\hspace{0.4cm}\hbox to 5.2cm{$v_{0,0}=v_{\un{0},0}=v_{0,\un{0}}=1,$\hfill}
v_{\un{0},\un{0}}=\un{\sigma}\cr\cr
&&--------------------------------------\cr\cr
&&\hbox to 4.8cm{orientifold group\hfill}\bZ_2\lx\bZ_2\cr
&&\hbox to 4.8cm{level\hfill}\sfk\in2\bZ\cr
&&\hbox to 4.8cm{trivializing cochain for\hfill}n,n'=0,1,2\cr\cr
&&\hspace{0.4cm}\hbox to 5.2cm{$v_{n,n'}=1,$\hfill}v_{\un{n},n'}
=\sigma^{n'},\cr
&&\hspace{0.4cm}\hbox to 5.2cm{$v_{n,\un{n}'}=1,$\hfill}
v_{\un{n},\un{n}'}=\un{\sigma}\,\sigma^{n'}\cr\cr
&&=================================================\cr
&&\hbox to 4.8cm{\bf Group\hfill}{\bf E_8}\cr
&&\hbox to 4.8cm{center\hfill}\bZ_1\cr
&&\hbox to 4.8cm{twist element\hfill}n_0=0\cr
&&--------------------------------------\cr\cr
&&\hbox to 4.8cm{orientifold group\hfill}\bZ_2\cr
&&\hbox to 4.8cm{level\hfill}\sfk\in\bZ\cr
&&\hbox to 4.8cm{trivializing cochain\hfill}\cr\cr
&&\hspace{0.4cm}\hbox to 5.2cm{$v_{0,0}=v_{\un{0},0}=v_{0,\un{0}}=1,$\hfill}
v_{\un{0},\un{0}}=\un{\sigma}\cr\cr
&&=================================================\cr
&&\hbox to 4.8cm{\bf Group\hfill}{\bf F_4}\cr
&&\hbox to 4.8cm{center\hfill}\bZ_1\cr
&&\hbox to 4.8cm{twist element\hfill}n_0=0\cr
&&--------------------------------------\cr\cr
&&\hbox to 4.8cm{orientifold group\hfill}\bZ_2\cr
&&\hbox to 4.8cm{level\hfill}\sfk\in\bZ\cr
&&\hbox to 4.8cm{trivializing cochain\hfill}\cr\cr
&&\hspace{0.4cm}\hbox to 5.2cm{$v_{0,0}=v_{\un{0},0}=v_{0,\un{0}}=1,$\hfill}
v_{\un{0},\un{0}}=\un{\sigma}\cr\cr
&&=================================================\cr
&&\hbox to 4.8cm{\bf Group\hfill}{\bf G_2}\cr
&&\hbox to 4.8cm{center\hfill}\bZ_1\cr
&&\hbox to 4.8cm{twist element\hfill}n_0=0\cr
&&--------------------------------------\cr\cr
&&\hbox to 4.8cm{orientifold group\hfill}\bZ_2\cr
&&\hbox to 4.8cm{level\hfill}\sfk\in\bZ\cr
&&\hbox to 4.8cm{trivializing cochain\hfill}\cr\cr
&&\hspace{0.4cm}\hbox to 5.2cm{$v_{0,0}=v_{\un{0},0}=v_{0,\un{0}}=1,$\hfill}
v_{\un{0},\un{0}}=\un{\sigma}
\nonumber
\qqq

\end{document}